\documentclass[12pt,preprint]{aastex}
\usepackage{multirow}
\usepackage{tikz}
\usepackage{ulem}
\usepackage{color, soul}

\shorttitle{Turbulence and Depolarization of Optical Blazars}
\shortauthors{Guo et al.}
\begin{document}
  \title{Can Turbulence Dominate Depolarization of Optical Blazars?}
  \author{Xiaotong Guo\altaffilmark{1,2,3,4},
          Jirong Mao\altaffilmark{1,2,3},
          Jiancheng Wang\altaffilmark{1,2,3}
            }

\altaffiltext{1} {Yunnan Observatories, Chinese Academy of Sciences, 650011 Kunming, Yunnan Province, China}
\altaffiltext{2} {Center for Astronomical Mega-Science, Chinese Academy of Sciences, 20A Datun Road, Chaoyang District, 100012 Beijing, China}
\altaffiltext{3} {Key Laboratory for the Structure and Evolution of Celestial Objects, Chinese Academy of Sciences, 650011 Kunming, China}
\altaffiltext{4} {University of Chinese Academy of Sciences, 100049 Beijing, China}
\email{jirongmao@mail.ynao.ac.cn}
\begin{abstract}
We carefully examine the depolarization feature of blazars in the optical and near-infrared bands using the sample of Mead et al.
Magnetohydrodynamics (MHD) turbulence could be one possible reason for the depolarization of optical/infrared blazars when we apply the theoretical analysis 
of Lazarian \& Pogosyan. We further identify in the sample that the depolarization results shown in most blazars roughly obey the form of
the three-dimensional anisotropic Kolmogorov scaling. 
The effective Faraday rotation window length scale is not small enough to resolve the polarization correlation length scale in the blazar sample.
The depolarization and the related turbulent features show diversities in different blazar sources.
We suggest more simultaneous observations in both the optical/infrared and the high-energy bands for the study of the blazar polarization.
\end{abstract}


\section{Introduction}\label{Introduction}
Blazars, as a subclass of active galactic nuclei (AGNs) with the radiation aligned to the line of sight of observers, are dominated by relativistic
jets \citep{1995PASP..107..803U}. The relativistic jet with synchrotron emission has strong linear polarization. The linear polarization degree can
be calculated as $p=(\alpha+1)/(\alpha+5/3)$, where $\alpha$ is the spectral index of the synchrotron radiation, and the magnetic field distribution can 
slightly modify the spectral polarization \citep{bjor82}.

Numerous studies have been carried out for the blazar polarization. Here, we focus on the optical/infrared depolarization feature shown in blazar jets.
Although astronomers have performed many polarization detections of blazars in the optical/infrared band (e.g., Heidt \& Nilsson 2011; Ikejiri et al. 2011; Covino et al. 2015; Blinov et al. 2016a), in order to analyze the wavelength-dependent polarization properties of blzars, the observations of multi-band polarization are necessary.
For example, \citet{kulsh87} presented the optical polarization measurements for three blazars. In our work, we utilize the observational results given
by \cite{1990A&AS...83..183M}. They presented the multi-band (UBVRIJHK) polarizations for 44 blazars. The observations were carried out on the 3.5 m United Kingdom
Infrared Telescope (UKIRT). In their results, the polarization degrees of some blazars are wavelength dependent. This feature was called frequency dependence
polarization (FDP). This unique dataset provides us with a good possibility to investigate the depolarization properties of optical /infrared blazars.

The depolarization takes effect on the polarization degree of synchrotron sources due to Faraday dispersion. \cite{1966MNRAS.133...67B} proposed that the
polarization has a form of $p\propto exp\{-\lambda^4\}$, where $\lambda$ is the observational wavelength.
The Faraday rotation measure (RM) can be related to the magnetic field topology \citep{kigure04,uchida04,horellou14}, and a special magnetic field twist has been
proposed \citep{sokoloff1998,con09}. Moreover, some studies on the Faraday RM synthesis have also been presented \citep{bre05,andre12,frick10}. On the 
other hand, \cite{1991MNRAS.250..726T} suggested a turbulent feature in the
Faraday RM, a form of $p\propto \lambda^{-4/\zeta}$ was derived, where $\zeta$ is related to the index of the turbulent energy cascade.
Some simulations suggested that the turbulence is a possible reason for the polarization variability in blazars \citep{marscher14}.
Recently, \cite{2016ApJ...818..178L} comprehensively investigated the three-dimensional anisotropic magnetohydrodynamic (MHD) turbulence to the wavelength-dependent polarization by
using the statistic correlation function. It is important to see that they clearly presented
the different length scales of the turbulent Faraday rotation on the depolarization feature. This indicates that we can apply the modeling results of MHD turbulence given by Lazarian \& Pogosyan (2016) to analyze the depolarization measurements of Mead et al. (1990).

The aim of this paper is to examine the depolarization feature in optical/infrared blazars and reveal the physical reasons of the wavelength-dependent polarization.
In Section 2, we collect the polarization
data from \cite{1990A&AS...83..183M} and fit the multi-band polarization degree by a power law of $p\propto \lambda^{-b}$, where $b$ is the fitting parameter.
In Section 3, in the case of the blazar population, we first examine the physical conditions of the turbulent Faraday rotation given
by \cite{2016ApJ...818..178L}. Then, we determine the physical reason for the the wavelength-dependent depolarization features derived in Section 2. We propose other
possibilities to explain the complicated polarization features in Section 4. Finally, the conclusions are given in Section 5.

\section{Sample Selection and Fitting Results} \label{data}

We utilize the blazar polarization data provided by \cite{1990A&AS...83..183M}.
The observations were performed on the UKIRT in the dates of 1986 July 31 - August 7, 1987 July 27-30, 1987 September 18-21, and 1988 February 15-18.
44 blazars were observed. In each observation, multi-band polarization measurements were performed for a blazar, and each blazar was
observed at least once. 
In particular, the optical beam setting on the Mark II Hatfield polarimeter of UKIRT made the polarization measurements in the optical and infrared bands simultaneously for each blazar in each observational time. The simultaneous observations in multi-bands for each blazar provided the proper depolarization measurements. Hence, we can use these data to compare with the theoretical polarization scaling of Lazarian \& Pogosyan (2016). Here, we claim that the scaling 
relations given by Lazarian \& Pogosyan (2016) are time averaged.

We carefully count the polarization results for each blazar. 
Among 178 photometric data sets, 46 sets have the feature of $dp/d\lambda<0$, 7 sets have the feature of $dp/d\lambda>0$, 76 sets have the feature of
$p_0$ (with little dependence on frequency), 23 sets are ``unpolarized'', and 26 sets are ``complex'' (with complex polarization behavior).
In order to investigate the wavelength-dependent polarization features of blazars, we neglect all the data sets counted as $p_0$,
``unpolarized'', and ``complex''. Then, 20 blazars have the data with $dp/d\lambda<0$
and/or $dp/d\lambda>0$. We further identify these blazars. 17 blazars have the feature of $dp/d\lambda<0$, 6 blazars have the feature of $dp/d\lambda>0$, and 3
blazars present the features with both $dp/d\lambda<0$ and $dp/d\lambda>0$. In this blazar sample, 16 objects are BL Lacs and 4 objects are
flat-spectrum radio quasars (FSRQs). We list all the detailed counts in Table 1. Although the cases of $dp/d\lambda>0$ are shown, we note that
the depolarization of $dp/d\lambda<0$ is dominated in the sample. Hence, we can use the depolarization results obtained from this sample to further investigate
the depolarization properties proposed by \cite{2016ApJ...818..178L}.
Some special details of each blazar are included in Appendix. We also suggest some possibilities for the blazars with the case of $dp/d\lambda>0$ in Section 4.

We use the form of
\begin{equation}\label{fit_eq}
  p(\lambda)=a \lambda^{-b}
\end{equation}
to fit the depolarization feature in the sample, where $a$ is the fitting coefficient, and $b$ is the fitting parameter, which we need in order to determine the depolarization
properties.
The value of $b$ in each fitting is listed in Table 1. We plot the statistics of $b$ in Figure 1. For the cases of $dp/d\lambda<0$, we see that most
$b$-values are within the range of $0.0-1.0$. The significant distribution peak is shown at $0.1-0.2$. In particular, we note that the fitting results of
GC 0109+224 and IZw 186 have $b>1$. The fitting details of each multi-band depolarization are shown in Figures $5-24$.

We pay attention to some fitting results that are not reliable. The source GC 0109+224 observed on 1986 August 4 shows the polarization feature of $dp/d\lambda>0$,
but the fitting result of $b=-0.40\pm 0.46$ has a large error bar. The observation of the source 3C 279 on 1988 February 18 shows the polarization parameter of
$b=0.01\pm 0.01$. Although this observation was identified as an FDP by Mead et al. (1990), we cannot get the strong evidence of
wavelength-dependent polarization from the fitting. The observation of the source BL Lacertac on 1987 July 30 has the fitting result of
$b=-0.06\pm 0.08$, and the evidence of wavelength-dependent polarization from our fitting is very weak. Thus, we neglect the cases mentioned above in our statistics.
We present all fitting details of each object in the Appendix.

The blazars listed in Table 1 include both BL-Lac and FSRQ types. The number of observed wavelength-dependent polarization blazars is limited in the sample,
and we cannot see any difference on the depolarization property between BL Lac and FSRQ types. A large sample is expected for this investigation.

Although the polarization measurements at a particular frequency were carried out multiple times for each source in Mead et al. (1990), we only take the 
results for the cases of $dp/d\lambda<0$ and $dp/d\lambda>0$. Because the $b$-values of some sources have large difference as shown in Figure 2, we cannot 
average the $b$-value for each source. 
Since different sources have totally different $b$-values, we should consider the distribution of the $b$-value from all source measurements in Figure 1 as a
statistical result. By this statistical result, we can compare the modeling results of Lazarian \& Pogosyan (2016).

\section{Faraday Rotation with Turbulence} \label{result}
We take the wavelength-dependent polarization sample of blazars given by Mead et al. (1990) and fit the depolarization feature. Although it is widely accepted that
the blazar population has a power-law spectrum,  \citet{1990MNRAS.243..640B} examined the spectral property of the sample given by Mead et al. (1990). They
found that some spectra of those blazars present a curvature effect. This curvature effect means the lower flux at the shorter wavelength in a given spectrum,
and it was proposed as one reason for the wavelength-dependent polarization. However, there were no further explorations regrading the reason of the curvature
effect. In general, Faraday rotation is important in the study of the depolarization feature of the optical/infrared blazars. Shocks and turbulence
in relativistic jets are also considered. Here, we apply the theoretical analysis proposed by \cite{2016ApJ...818..178L} to explain the
wavelength-dependent polarization feature. We further distinguish the effects from Faraday rotation and turbulence to the blazar depolarization.

Lazarian \& Pogosyan (2016) comprehensively investigated the multi-wavelength polarization feature by the method of the statistic correlation function.
The Faraday rotation fluctuations come from anisotropic MHD turbulence.
A Faraday window described as $L_{\sigma_\phi,\bar \phi}\equiv {\rm min}[(\lambda^2 \bar \phi)^{-1},(\sqrt{2}\lambda^2 \sigma_\phi)^{-1}]$ was proposed, where
$\bar \phi=\kappa \bar n_e \bar B_z \nonumber$ can be treated as the average Faraday RM linear density, $\sigma^2_\phi = \kappa^2 \langle \Delta(n_eB_z)^2\rangle$ is the
fluctuation variance in the Faraday RM linear density, and $\lambda$ is the wavelength. The Faraday rotation window indicates the distance of one rad revolution by the Faraday rotation.
Lazarian \& Pogosyan (2016) proposed two complicated cases. (1) The case of
$L_{\sigma_\phi,\bar \phi}\ll r_i$, where $r_i$ is the correlation length of the polarization at the source.
In this case, the Faraday rotation window is small enough to resolve $r_i$, and it may be possible to measure the correlations of the underlying magnetic field.
When the mean Faraday rotation is dominated ($\bar \phi >\sigma_\phi$) and the condition is $\lambda^2 \bar{\phi} r_i\gg 1$, the wavelength-dependent polarization
is presented by $\langle p^2(\lambda^2)\rangle\varpropto \lambda^{-2-2m}$,
where $m$ is the correlation index for the polarization.
When the turbulent Faraday rotation is dominated ($\bar \phi <\sigma_\phi$) and the condition is $\lambda^2 \sigma_\phi r_i\gg 1$, the wavelength-dependent polarization
is presented by $\langle p^2(\lambda^2)\rangle\varpropto \lambda^{-2}$.
(2) The case of $L_{\sigma_\phi,\bar \phi}\gg r_i$. In this case, we cannot use the Faraday rotation window $L_{\sigma_\phi,\bar \phi}$ to resolve $r_i$. 
When the mean Faraday rotation is dominated ($\bar \phi >\sigma_\phi$) and the condition is $\lambda^2 \bar{\phi} r_i\ll 1$, the wavelength-dependent polarization
is presented by $\langle p^2(\lambda^2)\rangle\varpropto \lambda^{-2+2m}$. When the turbulent Faraday rotation is dominated ($\bar \phi <\sigma_\phi$) and the condition
is $\lambda^2 \sigma_\phi r_i\ll 1$, the wavelength-dependent polarization is presented by
$\langle p^2(\lambda^2)\rangle\varpropto \lambda^{\frac{-2+2m}{1-\tilde m_\phi/2}}$, where $\widetilde m_\phi={\rm min}(m_\phi,1)$, and $m_\phi$ is the correlation index of the Faraday RM density. Moreover, \cite{2016ApJ...825..154Z} used multi-dimensional simulations to further test the results of the analytical correlations given by
Lazarian \& Pogosyan (2016). The numerical results are in good agreement with the analytical study.

We examine whether the above cases can be applicable for the blazar sample of Mead et al. (1990). First, we propose one condition that the length scale of the emission region should be larger than the size of the Faraday rotation window. This gives $R>L_{\sigma_\phi,\bar \phi}$, where $R$ is the emission region scale. If 
$L_{\phi}=1/(\lambda^2\bar\phi)$, we get $\lambda^2>1/(R\bar\phi)$; If $L_{\sigma_\phi}=1/(\sqrt{2}\lambda^2\sigma_\phi)$, we get 
$\lambda^2=1/(\sqrt{2}R\sigma_\phi)$. Thus, we have $\lambda^2\bar\phi r_i=r_i/R$ and $\lambda^2\sigma_\phi r_i=r_i/(\sqrt{2}R)$. Here, we assume $r_i<R$. It indicates that the polarization correlation scale is smaller than the emission region scale. This assumption is valid for point sources in this paper. It is also reasonable for the case that the extent of the source in the plane of the sky is much smaller than that along the line of sight. Finally, we obtain  
$\lambda^2(\bar\phi, \sigma_\phi)r_i<1$. This means that the Faraday rotation window is not small enough to resolve the polarization correlation length scale.
If the mean Faraday rotation is dominated in the blazar jets, Lazarian \& Pogosyan (2016) provided the form of
$\langle p^2(\lambda^2)\rangle\varpropto \lambda^{-2+2m}$. If the turbulent Faraday rotation is dominated in the blazar jets, the form of 
$\langle p^2(\lambda^2)\rangle\varpropto \lambda^{\frac{-2+2m}{1-\widetilde m_\phi/2}}$ was given,
where $\widetilde m_\phi={\rm min}(m_\phi,1)$, and $m_\phi$ is the correlation index of the Faraday RM density.

In order to obtain the Faraday RM density and constrain the polarization correlation length scale, we further perform some simple estimations. 
Ghisellini et al. (2010) provided the general properties of blazars. In our paper,
we also consider the work of Tavecchio \& Ghisellini (2016), in which the magnetically dominated flows are dominated in blazar jets. 
Thus, we get the reasonable parameters to estimate the Faraday RM density. 
When the mean Faraday rotation is dominated in blazar jets, we have the case of $\bar \phi>\sigma_\phi$. 
We suggest that the regular magnetic field in the line of sight and the electron density have the values of $\bar B_z=10.0$ G and $\bar n_e=1.0\times 10^3~\rm{cm^{-3}}$, respectively.
The Faraday RM density is $\bar \phi =\kappa \bar n_e \bar B_z =8.1\times 10^9~\rm{m^{-2}~pc^{-1}}$,
where $\kappa=8.1\times 10^5~\rm{m^{-2}}\rm{pc^{-1}}\rm{cm^3}\rm{G^{-1}}$ is the constant. Within the wavelength range from U band to K band, we obtain that the emission region has the range from 26.0 to 952.6 pc. This is roughly consistent with the length scale of blazar jets. The polarization correlation scale should be smaller than the value in this range.  

When the turbulent Faraday rotation is dominated in blazar jets, 
we have the case of $\sigma_\phi>\bar \phi$.
The fluctuation variance of the Faraday RM linear density is 
$\sigma^2_\phi = \kappa^2 \langle \Delta(n_e  B_z)^2\rangle=\kappa^2\left({\bar n_e}^2 \langle(\Delta B_z)^2\rangle+{\bar B_z}^2 \langle(\Delta n_e)^2\rangle+\langle(\Delta n_e)^2\rangle \langle(\Delta B_z)^2\rangle\right),$
where $\bar n_e$ is the average electron density, $\Delta n_e$ is the fluctuation of the electron density, $\bar B_z$ is the mean magnetic field in the line of sight,
and $\Delta B_z$ is the fluctuation of the magnetic field (Lazarian \& Pogosyan 2016). 
We take the average electron density $\bar n_e= 1.0\times10^3~\rm{cm^{-3}}$ and the mean magnetic field
$\bar B_z=10.0~\rm{G}$. 
Because usually strong fluctuation values are much larger than the mean values, we assume that the fluctuations are strong enough such that the electron density fluctuation $\Delta n_e =10\bar n_e$ and the magnetic field fluctuation $\Delta B_z=10\bar B_z$.
Then, we estimate $\sigma_\phi=8.1\times 10^{11}~\rm{m^{-2}~pc^{-1}}$.
Within the wavelength range from U band to K band, we obtain that the emission region has the range from 0.3 to 9.5 pc. 
The polarization correlation scale should be smaller than the value in this range. 

We caution that the emission region sizes estimated above are the lower limits. Moreover, the estimated values are strongly dependent on the density and 
magnetic field in the blazar jet. The fluctuation values of the density and magnetic field are also quite uncertain.

It is important to note that the energy spectrum of the MHD turbulence ($E(k)\propto k^{-3-m}$) was involved in the physical description of Lazarian \& Pogosyan (2016)
and \cite{2016ApJ...825..154Z}. In particular, $m=2/3$ corresponds to the three-dimensional anisotropic Kolmogorov scaling. Therefore, we can link the fitting results of the blazar sample to the MHD turbulent properties. When the mean Fraday rotation is dominated in blazar jets, the variation of polarization with wavelength
is $p(\lambda)\propto \lambda^{(-1+m)/2}$ as we identified by the condition of Equation (3). In the case of the three-dimensional anisotropic Kolmogorov turbulence,
we obtain $p(\lambda)\propto \lambda^{-1/6}$.
When the turbulent Faraday rotation is dominated in blazar jets, the variation of polarization with wavelength is described as
$p(\lambda)\propto \lambda^{\frac{-1+m}{2-\widetilde m_\phi}}$.
The range of $\widetilde m_\phi$ are from 0 to 1. If we assume $\widetilde m_\phi=0$, the variation of polarization with wavelength is $p(\lambda)\propto \lambda^{(-1+m)/2}$, which is
the same as the result of the mean Faraday rotation dominated case.
If we assume $\widetilde m_\phi=1$, the variation of polarization with wavelength is $p(\lambda)\propto \lambda^{-1+m}$. In the case of the three-dimensional anisotropic
Kolmogorov turbulence, we obtain $p(\lambda)\propto \lambda^{-1/3}$.

We list all the results for each polarization observation in Table 1. We suggest that the depolarization property ($dp/d\lambda<0$) in the blazar sample is
originated from the turbulence. We also plot the $m$-value distributions derived from both the mean Faraday rotation dominated
case and the turbulent Faraday rotation dominated case in Figure 3 and Figure 4, respectively. It looks like that the fitting results from the observational sample
are roughly consistent with either the mean Faraday rotation dominated case or the turbulent Faraday rotation dominated case.

\section{Discussion} \label{discussion}

We examine the depolarization feature of the optical/infrared blazars given by Mead et al. (1990) and apply the correlation description of the polarization
given by Lazarian \& Pogosyan (2016) to identify that the turbulence may have a dominated role on the blazar depolarization. However, we caution contributing this direct link
between the observational results
to the theoretical analysis. First, the wavelength-dependent interstellar polarization makes contributions to the observational results \citep{serk75,martin90}.
Second, we should consider the radiation transfer process of the synchrotron radiation
in blazar jets when we perform optical/infrared observations \citep{jones77}. We can assume that the absorption used to calculate the Stokes
parameters is the same in each optical/infrared band, such that the polarization degree is not affected by the absorption through the radiation transfer. However, due to
the rotativity and
convertibility from the radiation transfer, the intrinsic polarization degree and the observed polarization degree have differences in each optical/infrared band.
Third, because the turbulent effect is believed to be one possible reason for the depolarization of the optical/infrared blazars, the polarized synchrotron radiation
transportation in the turbulent media is important \citep{spangler82}. It was recently proposed that the Faraday conversion in the turbulent blazar jets makes the
circular polarization \citep{macdonald17}. Because this result is related to the radiation transfer, it can cause the change of the intrinsic linear polarization.
In our paper, we take statistic results from a sample of optical/infrared blazars, and we simply neglect the effects mentioned above.

There is only one dataset containing one polarization degree value in each band for each blazar observational night in Mead et al. (1990). This indicates that a polarization value is a mean value in one night. Thus, in order to compare the time-averaged scaling given by Lararian \& Pogosyan (2016), we assume that the blazar polarization is stable during the observing night. We do not consider the case of strong polarization variation. 
Therefore, the depolarization feature presented in this paper is applicable for the quiescent blazars\footnote{Even for the sources with a rapid variability, when the polarization measurements can be carried out within sufficiently small time-scale intervals, we can also apply the modeling results of Lazarian \& Pogosyan (2016) to analyze the polarization features.}. However, polarization of some blazars may have strong evolution in short timescales. This can be due to some short-timescale magnetic activities and plasma instabilities, which are common in the flaring stage of blazar jets. 
Moreover, the quasi-periodic oscillation of the optical polarization during an enhanced high-energy brightness phase of blazar PKS 2155-304 was found, and the time period was determined as 15-30 minutes (Pekeur et al. 2016). 
These short-timescale polarization variations are difficult to explain using the time-averaged MHD turbulent model scaling given by Lazarian \& Pogosyan (2016).

The calculation of Lazarian \& Pogosyan (2016) takes the electron energy spectrum as a simple power law. Thus, the intrinsic polarization degree is not related to the wavelength. While the observed polarization degree is wavelength dependent, which comes from the Faraday rotation only. The power-law distribution of the electron energy distribution is originated from shock acceleration (the first-order Fermi acceleration). When we consider turbulent acceleration (the second-order Fermi acceleration), the electron energy spectrum has a mixture shape of a Maxwellian distribution with a power-law tail (e.g., Stawarz \& Petrosian 2008; Giannios \& Spitkovsky 2009). 
A kinetic study of the particle acceleration in magnetized plasma also indicated that the electron energy spectrum is not a simple power law (Yuan et al. 2016). 
Moreover, magnetic reconnection can make particle acceleration as well. The acceleration processes are more complicated, and the electron energy distribution does not seem to be a simple power law (Zenitani \& Nagai 2016). 
With the complicated electron energy spectra, the wavelength dependence of the intrinsic polarization cannot be scaled out. The Faraday rotation is not the only reason for the wavelength-dependent polarization. We may further investigate this issue in the future.

We focus on the depolarization feature of $dp/d\lambda<0$ in the blazar sample. However, we also note a few cases of $dp/d\lambda>0$ listed in Table 1. It is hard
to apply the simple physical scenario of the MHD turbulence or the Faraday rotation to explain the observational cases of $dp/d\lambda>0$.
\citet{sokoloff1998} already mentioned the anomalous depolarization feature, which is the polarization increasing with the wavelength. They proposed that a twisted
magnetic field can reduce an unusual Faraday rotation. Thus, this anomalous depolarization can happen. Recently,
both depolarization ($dp/d\lambda<0$) and anomalous depolarization ($dp/d\lambda>0$) features were found in some AGN jets \citep{krav2017}. The observations suggest a
spine-sheath jet structure, which is also possible for the polarized blazars in our sample.

Observing blazars in the high-energy band can also provide a possible explanation for the depolarization features. In Table 1, we particularly note some blazars
that have short-timescale flares in the high-energy band. These high-energy observational results are mostly obtained from the {\it Fermi}-Catalog of the Flaring Sources
\citep{2016arXiv161203165A}. The optical polarization feature has the possibility to link with the short-timescale flaring in the high-energy band. For example,
\cite{2015ApJ...809..130C} noticed the rapid variation of the optical polarization during the TeV-flare phase of the blazar S5 0716+714. They indicate an event of
the shock-initiated magnetic reconnection. The collision-induced magnetic reconnection was proposed to be a unified interpretation for the blazar
polarization \citep{deng16}. With some blazar samples, Blinov et al. (2016b) found the possible association of the optical polarization swing event with the
gamma-ray activity, and Itoh et al. (2016) identified a correlation between the maximum degree of the optical polarization and the
gamma-ray luminosity. We hope that this kind of observation can provide more physical details in the future.

\section{Summary}
We have comprehensively investigated the optical depolarization feature with the blazar sample of Mead et al. (1990). We confirm that the form of
$p(\lambda)\propto \lambda^{-b}$ can be applied to study the optical/infrared blazar depolarization.
The theoretical analysis of Lazarian \& Pogosyan (2016) presented that the Faraday rotation fluctuations come from anisotropic MHD turbulence. Our fitting statistical results in the blazar sample show that the optical/infrared depolarization roughly obeys the universal Kolmogorov scaling. We find that the effective Faraday rotation window length scale is not small enough to resolve the polarization correlation length scale in the blazar sample. 
The depolarization and the related turbulent features show diversities in different blazar sources.

\acknowledgments
We thank the referee for his/her helpful suggestions. J. M. is supported by the National Natural Science Foundation of China 11673062, the Hundred Talent Program of the Chinese Academy of Sciences, the Major Program 
of the Chinese Academy of Sciences (KJZD-EW-M06),  and the Oversea Talent Program of Yunnan Province. J. W. is supported by the Strategic Priority Research 
Program ``The Emergence of Cosmological Structures" of the Chinese Academy of Sciences (XDB09000000) and the National Natural Science Foundation of 
China (11573060 and 11661161010).

\section*{APPENDIX} \label{appendix}

We carefully collect and describe some special points for each polarization observation given by Mead et al. (1990) and put these descriptions in the Appendix, in case
some useful details are necessary for further research. The fitting plots are shown in Figure 5-24.

The source 0048$-$097 (OB$-$081) has an inverted spectrum with a turnover at 10 GHz in the radio band \citep{2001AJ....121.1306T}.
It was observed 11 times presented by \cite{1990A&AS...83..183M}. The polarization results include 5 times of $dp/d\lambda<0$ and 6 times of $p_0$.
We fit the data of $dp/d\lambda<0$.

The blazar GC 0109 + 224 was discovered in the radio survey \citep{1971AJ.....76..980D}. The strong flux and polarization variability was reported both
in the radio and optical bands \citep{2000A&AS..143..357K, 2003A&A...400..487C, 2004MNRAS.348.1379C, 2009AJ....137..337S}.
This source was observed 10 times presented by \cite{1990A&AS...83..183M}. The polarization results include 3 times of $dp/d\lambda<0$, 1 time of $dp/d\lambda>0$, and 6
times of $p_0$.
There is a set of $dp/d\lambda<0$ data that has only two data points, and we neglect this dataset.
We note the result of $dp/d\lambda>0$ because of the fitting of $b=-0.40\pm0.46$ with a large error bar.

PKS 0118$-$272 was observed 7 times presented by \cite{1990A&AS...83..183M}. The polarization results include 3 times of $dp/d\lambda<0$, 3 times of $p_0$, and 1 time of
``complex''. We fit the observed data of $dp/d\lambda<0$. 

The source 0138$-$097 has a smooth IR-optical spectrum \citep{1983A&A...117...60F}. \cite{1997ApJ...489L..17S} identified the redshift of $z=0.733$ by
the weak emission lines of Mg II and O II through the 2.1 m telescope at the Kitt Peak National Observatory and the Multiple Mirror Telescope Observatory.
This source was also observed by the Hubble Space Telescope in 1996 September 28 \citep{1999ApJ...521..134S}.
It was observed 8 times presented by \cite{1990A&AS...83..183M}. The polarization results include 2 times of $dp/d\lambda<0$, 1 time of $dp/d\lambda>0$, and 5 times of $p_0$.
We fit all the the observed data with the FDP feature.

3C 66A (0219+428) is a bright source in the high-energy ($E\geqslant 100 GeV$) band \citep{2009ApJ...693L.104A,2009ApJ...692L..29A}.
The maximum value of the polarization degree in the optical band is 33\%. From the presentation of \cite{1990A&AS...83..183M},
this source was observed 10 times. The polarization results include 1 time of $dp/d\lambda<0$, 8 times of $p_0$, and 1 time of ``complex''. We fit
the observed data of $dp/d\lambda<0$.

0754+100 with the feature of the near-Infrared flare was detected \citep{2010ATel.2516....1C}.
This source was observed 4 times, presented by \cite{1990A&AS...83..183M}. The polarization results include 2 times of $dp/d\lambda<0$ and 2 times of $p_0$.
We the observed data of $dp/d\lambda<0$.

0818$-$128 (OJ-131) was observed 4 times, presented by \cite{1990A&AS...83..183M}. The polarization results include 2 times of $dp/d\lambda<0$ and
2 times of $p_0$. We fit the observed data of $dp/d\lambda<0$.

OJ 287 (0851+202) hosts a supermassive binary black hole system at its center \citep{1988ApJ...325..628S}.
This blazar is bright in the high-energy $\gamma$-ray band \citep{2009ApJ...693L.104A}.
Some dedicated studies of the optical polarization were given \citep{2009ApJ...697..985D, 2010MNRAS.402.2087V, 2010PASJ...62...69U}.
 this source was observed 4 times presented by \cite{1990A&AS...83..183M}. The polarization results include 2 times of $dp/d\lambda<0$, 1 time of $p_0$, and 1 time of
``unpolarized''. We fit the observed data of $dp/d\lambda<0$.

The source 1147+245 was only observed 1 time presented by \cite{1990A&AS...83..183M}. The observed result has the feature of $dp/d\lambda<0$. We fit the
observed data.

3C 279 (1253$-$055) is a superluminous source. The twin $\gamma$-ray flares with similar intensity were detected in 2013 December and 2014 April
\citep{2016ApJ...817...61P}. An exceptional $\gamma$-ray outburst was detected by the {\it Fermi}/Large Area Telescope (LAT) in 2015 June \citep{2015ApJ...809..130C}.
This source was observed 10 times, presented by \cite{1990A&AS...83..183M}. In particular, a polarization flare was shown during the observation time of August 1986,
and the polarization degree measured in the night of 1986 August 5 is about 46\% in the U band.
In general, the polarization results presented by \cite{1990A&AS...83..183M} include 6 times of $dp/d\lambda<0$, 3 times of $p_0$, and 1 time of ``unpolarized''.
We fit the observed data of $dp/d\lambda<0$.

The source 1418+546 (OQ 530) was detected 9 times, presented by \cite{1990A&AS...83..183M}. The polarization results include 5 times of $dp/d\lambda<0$, 3 times of
$p_0$, and 1 time of ``unpolarized''. We fit the observed data of $dp/d\lambda<0$.

3C 345 (1641 + 399) is a powerful high polarized quasar (HPQ), the redshift of $z=0.593$ is given by \citet{2012ApJ...748...81K}. This source
was found variable in X-ray, optical, and radio bands on the timescales ranging from minutes to months. The source was highly polarized ($p \gtrsim 10\% $)
at optical and radio wavelengths \citep{1968ApJ...152..357K}. It was observed 9 times presented by \cite{1990A&AS...83..183M}. The polarization results
include 1 time of $dp/d\lambda>0$, 3 times of ``unpolarized'' and 5 times of ``complex''.
We fit the observed data of $dp/d\lambda>0$. The value of $b$ is $-0.47\pm0.09$.

The source 1717+178 (OT 129) was observed 3 times presented by \cite{1990A&AS...83..183M}. The polarization results include 1 time of $dp/d\lambda>0$, 1 time of
$p_0$, and 1 time of ``unpolarized''.
We fit the observed data of $dp/d\lambda>0$. The fitted values of $b$ with $dp/d\lambda>0$ is $-0.47\pm0.09$.

I Zw 186 (1727+502) was simultaneously observed at radio, NIR, optical frequencies \citep{1982ApJ...253...19B}.
It was discovered as a $\gamma$-ray source by \textit{Fermi} \citep{2010ApJS..188..405A} and a very high energy (VHE) source by \textit{MAGIC}
\citep{2014A&A...563A..90A}. Its $\gamma$-ray flare was detected by VERITAS \citep{2015ApJ...808..110A}.
It was observed 3 times, presented by \cite{1990A&AS...83..183M}. The polarization results include 2 times of $dp/d\lambda<0$ and 1 time of $p_0$.
There is a set of $dp/d\lambda<0$ data, which only contains two data points, and we neglect this dataset. We fit the observed data of $dp/d\lambda<0$.
The value of $b$ is $3.22\pm0.03$. It is difficult to explain this fitting result.

The source 1749+096 (OT 081) was observed 5 times, presented by \cite{1990A&AS...83..183M}. The polarization results include 2 times of $dp/d\lambda<0$ and 5 times of
$p_0$. We fit the observed data of $dp/d\lambda<0$.

The source 1921$-$293 (OV 236) was clearly detected by WMAP \citep{2009A&A...508..107G}. This source is also one of the brightest extragalactic objects at millimeter
frequencies. It was observed 4 times presented by \cite{1990A&AS...83..183M}. The polarization results include 1 time of $dp/d\lambda>0$, 2 times of $p_0$, and 1
time of ``unpolarized''. 
We fit the observed data of $dp/d\lambda>0$. The value of $b$ is $-0.77\pm0.17$.

PKS 2155$-$304 is known for its short variability time scales at optical to X-ray wavelengths \citep{1993ApJS...85..265J}.
It was observed 6 times presented by \cite{1990A&AS...83..183M}. The polarization results include 2 times of $dp/d\lambda<0$, 3 times of $p_0$, and 1 time of
``complex''. We fit the observed data of $dp/d\lambda<0$.

The source 2200+ 420 is the prototype of BL Lacertae, which has been monitored intensively in multi-wavelengths \citep{2007A&A...464..175B, 1997ApJ...490L.145B,
1999ApJ...515..140S, 1999ApJ...521..145M, 2002A&A...390..407V,2003ApJ...596..847B, 2004A&A...421..103V}.
\cite{2006A&A...456..105B} suggested that the optical variations and the radio variations have a common origin in the inner portion of the jet.
\cite{1999ApJ...521..145M} reported the X-ray outburst during 1997 July.
\cite{2013ApJ...762...92A} reported the detection of a very rapid TeV gamma-ray flare that occurred on 2011 June 28 with the Very Energetic Radiation Imaging Telescope
Array System.
\cite{1997ApJ...490L.145B} reported the EGRET detection of a $\gamma$-ray flare that occurred during 1997 July 15-22.
It was observed 13 times presented by \cite{1990A&AS...83..183M}. The polarization results include 1 time of $dp/d\lambda<0$, 2 times of $dp/d\lambda>0$, 1 time of $p_0$, and 9 times of ``complex''.
We fit the observed data of FDP. The values of $b$ are $0.13\pm0.07$, $-0.12\pm0.04$, and $-0.06\pm0.08$.
We neglect the result of $b=-0.06\pm0.08$.

3C 446 (2223$-$052) exhibited a rapid variability in the optical, X-ray, and radio bands \citep{1997ApJ...487..536S, 1998A&AS..132..305T}.
It was observed 6 times, presented by \cite{1990A&AS...83..183M}. The polarization results include 1 time of $dp/d\lambda<0$, 4 times of $p_0$, and 1 time of
``unpolarized''. We fit the observed data of $dp/d\lambda<0$.

The source 2254+074 (OY 091) was observed 11 times presented by \cite{1990A&AS...83..183M}.
The polarization results include 5 times of $dp/d\lambda<0$, 3 times of $p_0$, and 3 times of ``complex''.
We fit the observed data of $dp/d\lambda<0$.

\clearpage

\begin{figure*}
  \includegraphics[scale=0.6,angle=-90]{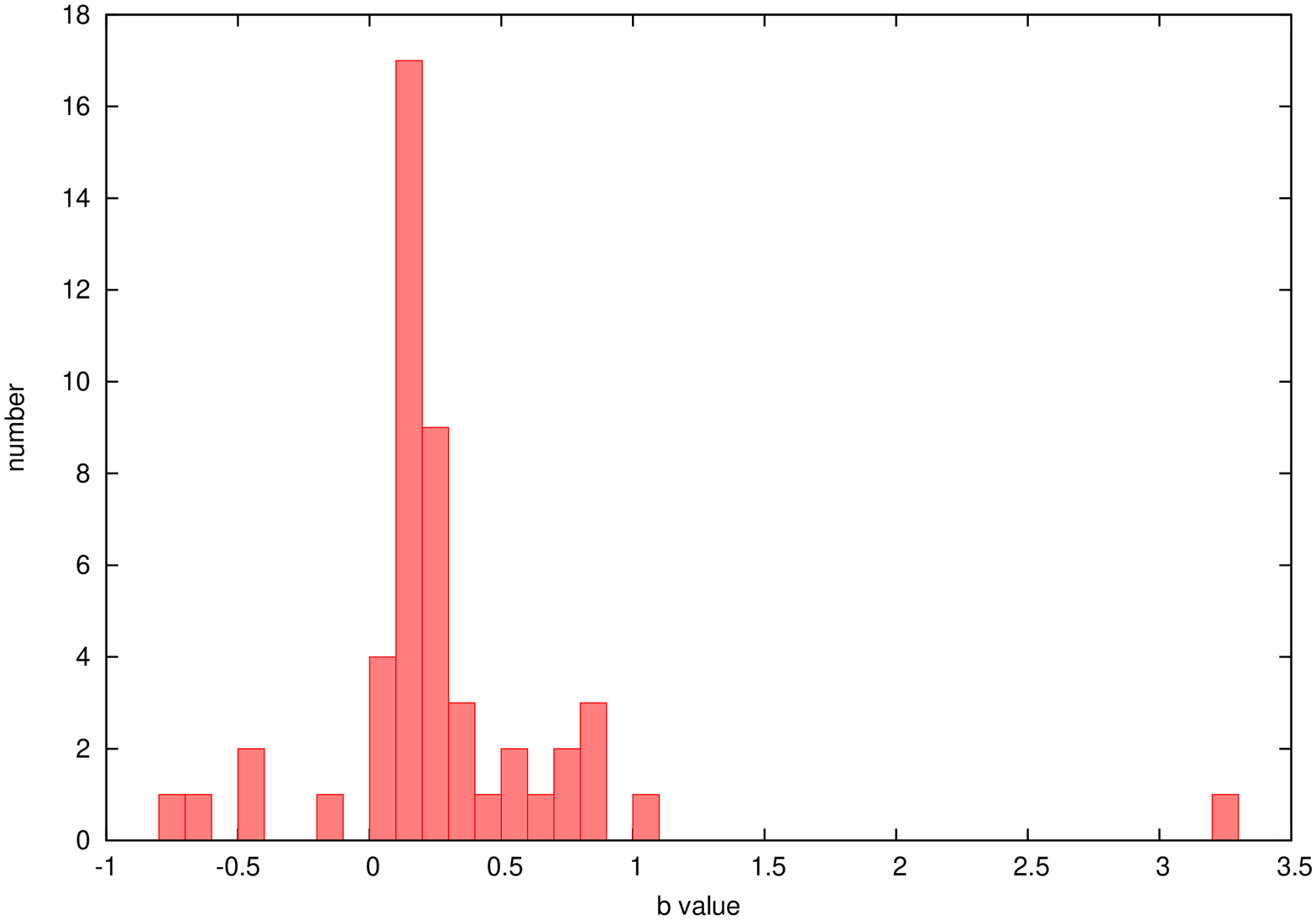}
  \caption{Distribution of the $b$-value. $b=1.03\pm 0.11$ is the fitting result of the source GC 0109+224, and $b=3.22\pm 0.03$ is the fitting result of the
source I Zw 186.}\label{blazar of b}
\end{figure*}

\begin{figure*}
  \includegraphics[scale=0.6,angle=-90]{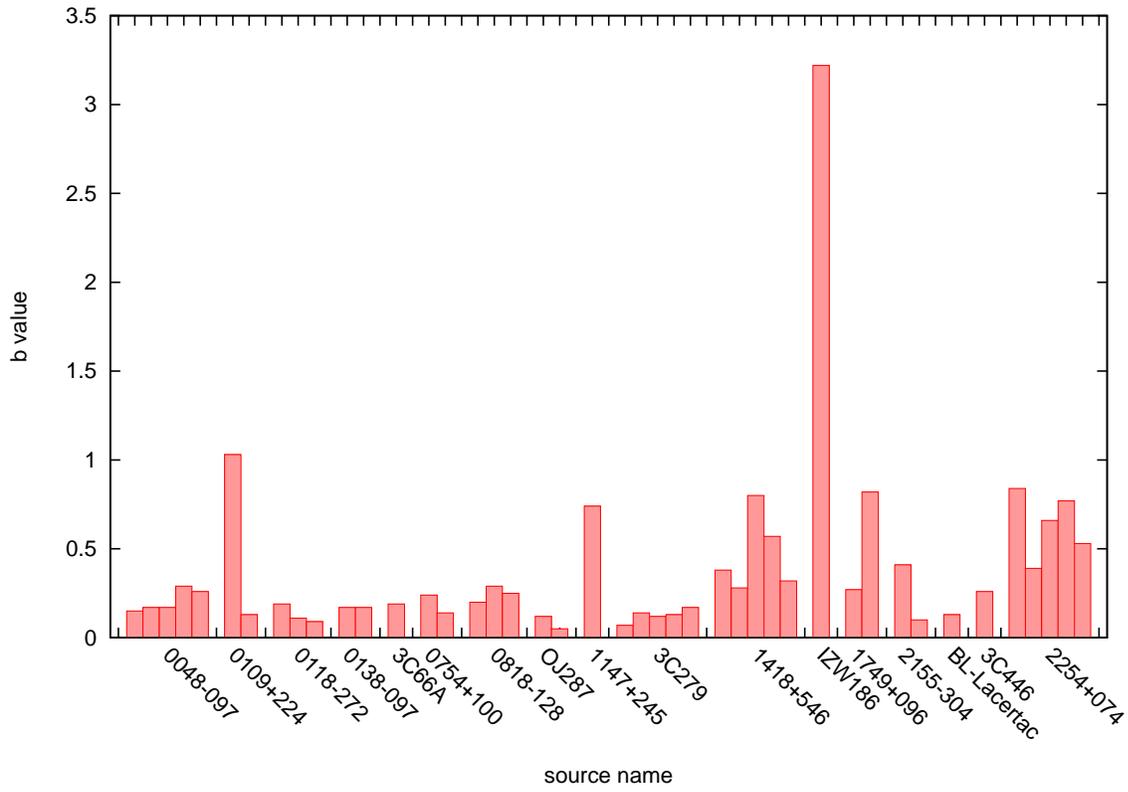}
  \caption{$b$-value in the case of $dp/d\lambda<0$ for each source.
}
  \label{b value}
\end{figure*}

\begin{figure*}
  \includegraphics[scale=0.6,angle=-90]{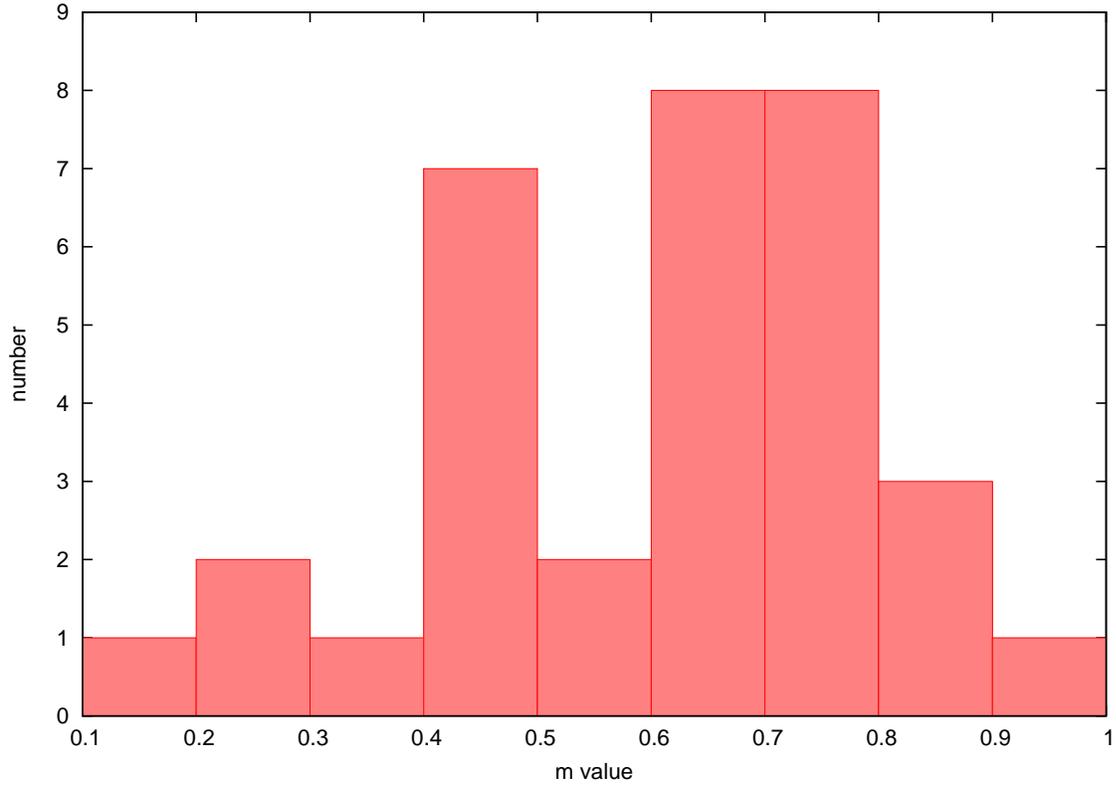}
  \caption{Distribution of the MHD turbulent index $m$ in the regular magnetic field dominated case. Each value of $m$ is derived from the fitting value of $b$.
We note that $m=2/3$ corresponds to the three-dimensional anisotropic Kolmogorov scaling.}
  \label{p1-m}
\end{figure*}
\begin{figure*}
  \includegraphics[scale=0.6,angle=-90]{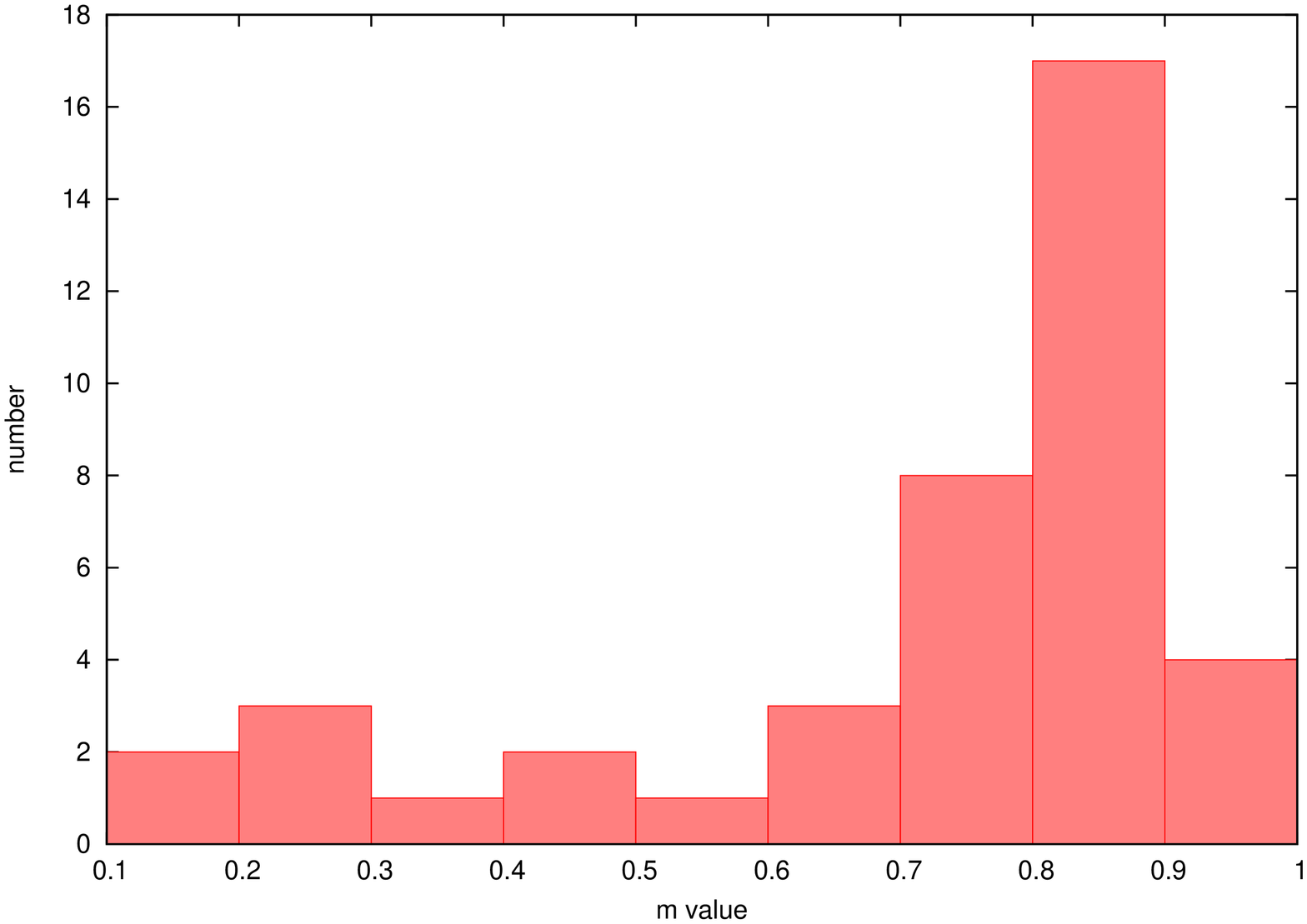}
  \caption{Distribution of the MHD turbulent index $m$ in the turbulent magnetic field dominated case when we assume $\widetilde m_\phi=1$.
Each value of $m$ is derived from the fitting value of $b$.
We note that $m=2/3$ corresponds to the three-dimensional anisotropic Kolmogorov scaling.
}
  \label{p1-m2}
\end{figure*}

\begin{figure*}
 \includegraphics[scale=0.2,angle=-90]{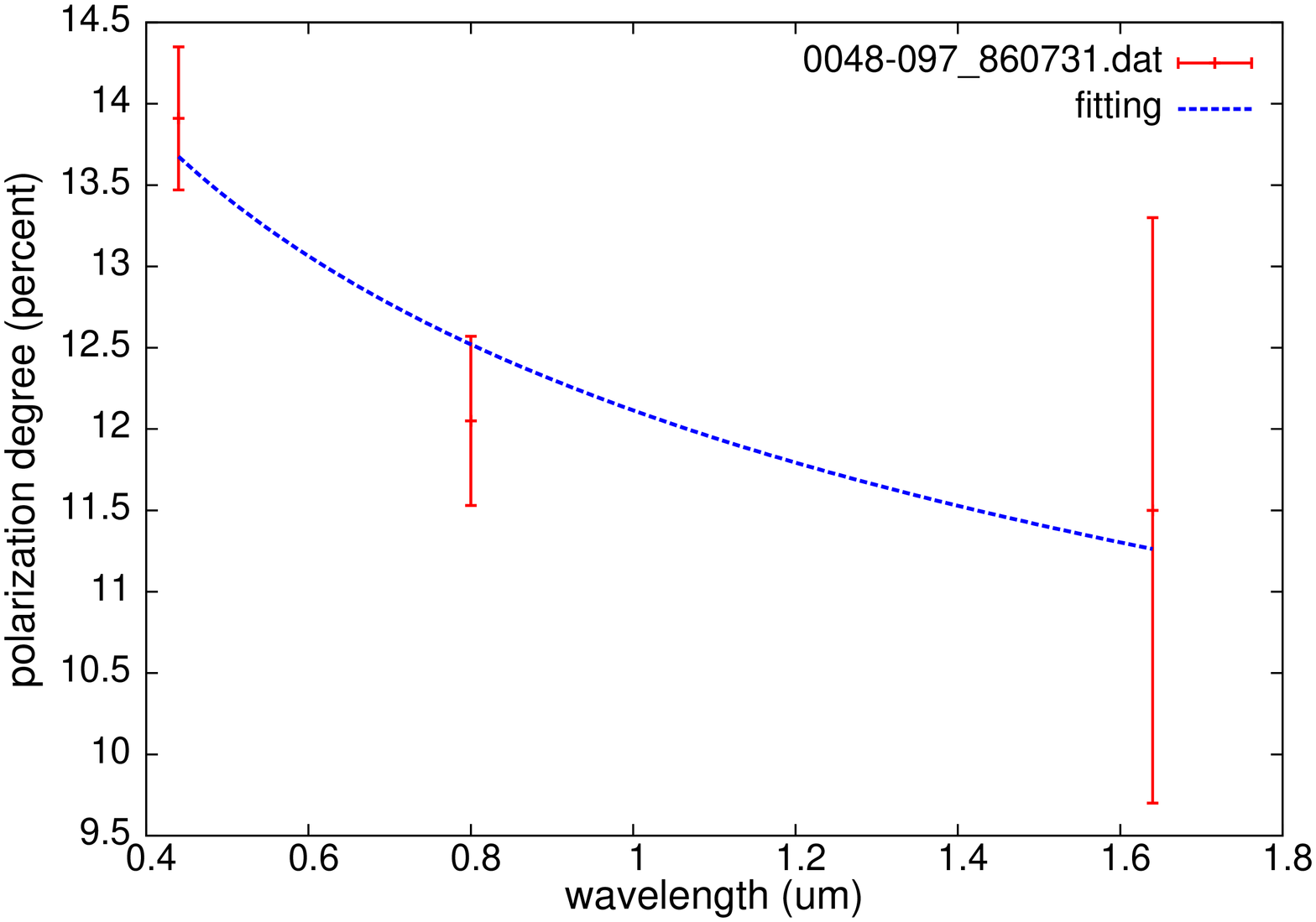}
 \includegraphics[scale=0.2,angle=-90]{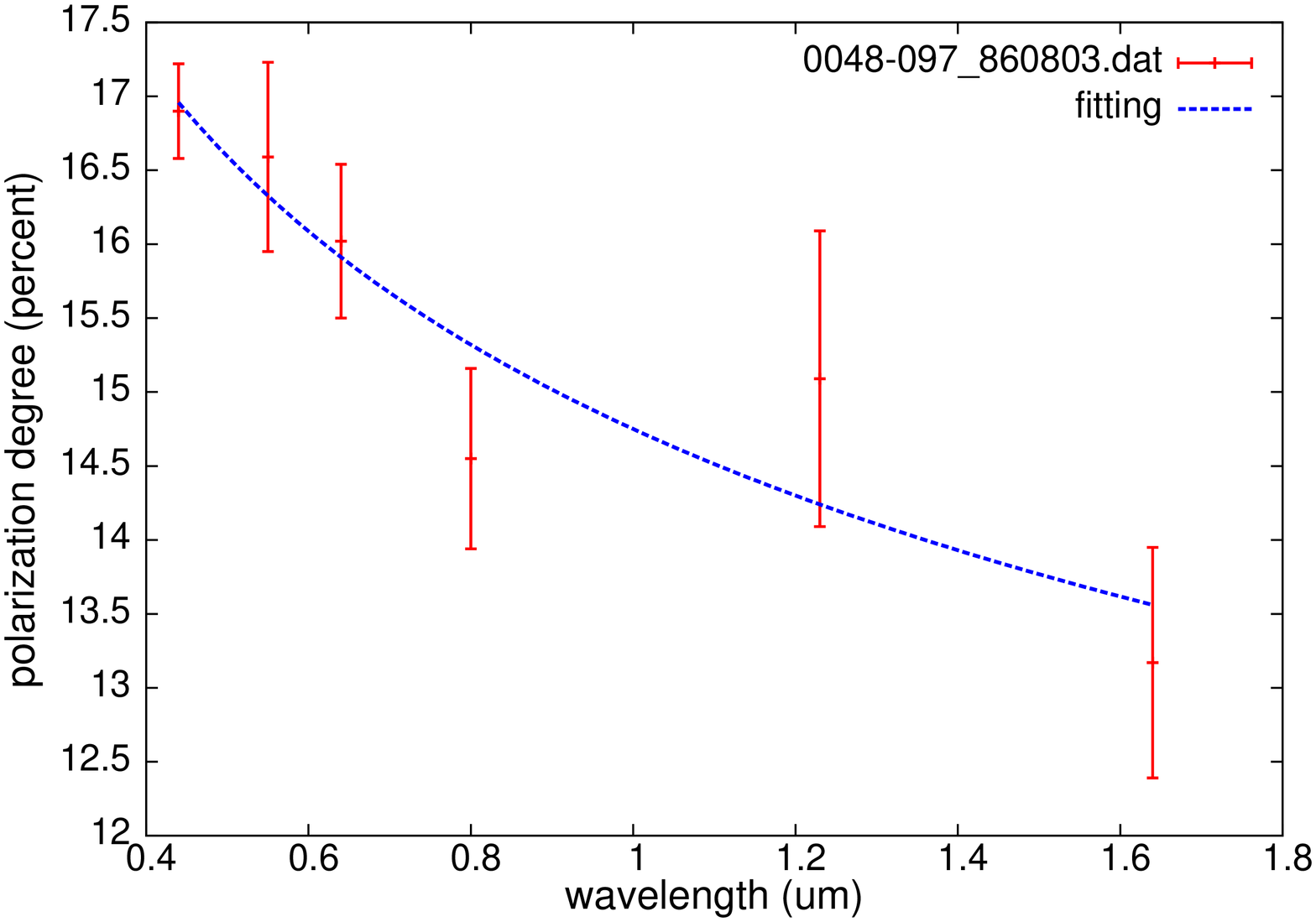}
 \includegraphics[scale=0.2,angle=-90]{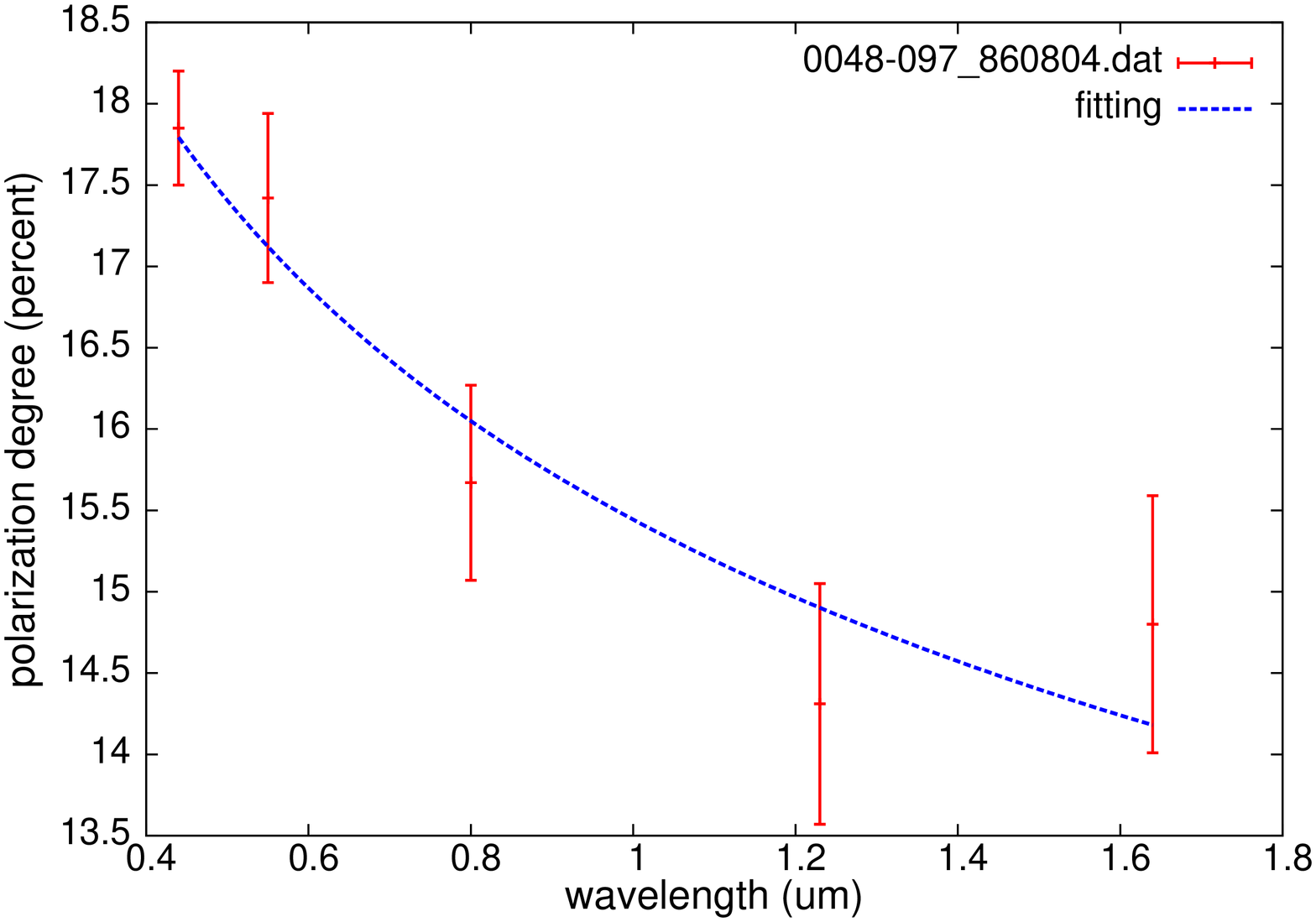} \\
 \includegraphics[scale=0.2,angle=-90]{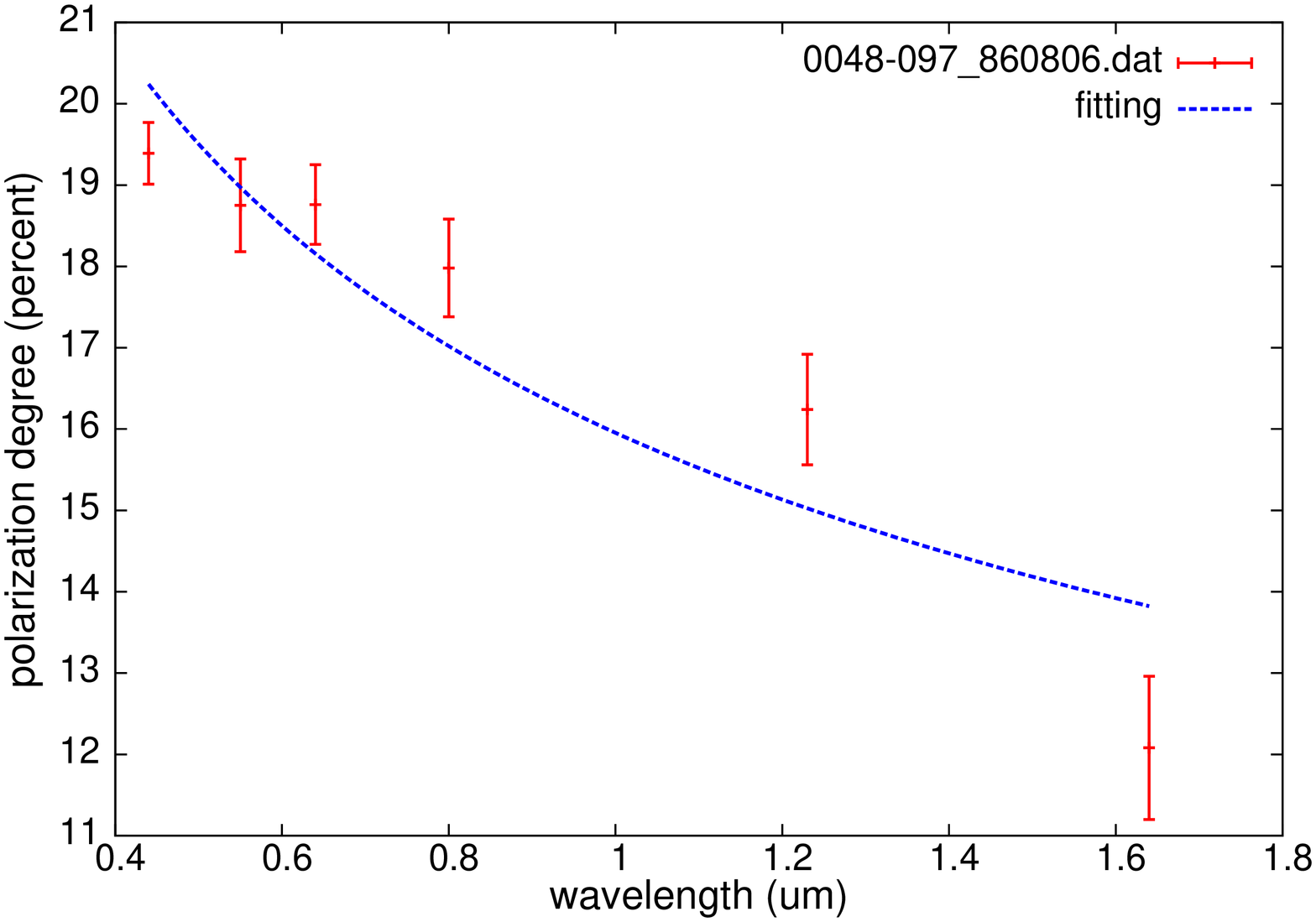}
 \includegraphics[scale=0.2,angle=-90]{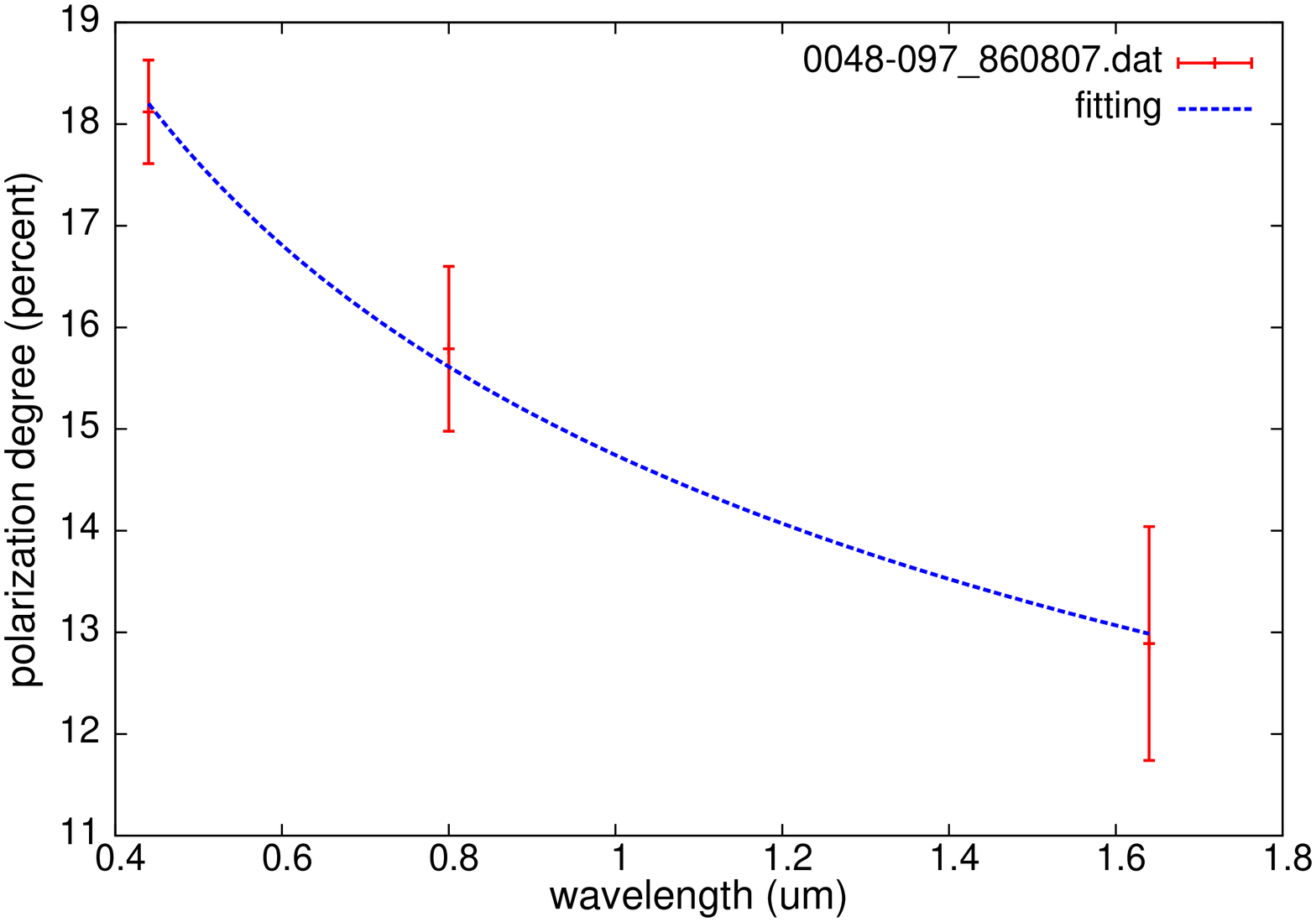}
  \caption{Depolarization fitting for 0048$-$097 (OB$-$081). The first panel with the data observed on 1986 July 31 presents the result of $b=0.15\pm0.05$ with
$\chi^2/{\rm d.o.f}=0.33$. The second panel with the data observed on 1986 Aug. 3 presents the result of $b=0.17\pm0.04$ with $\chi^2/{\rm d.o.f}=0.39$.
The third panel with the data observed on 1986 August. 4 presents the result of $b=0.17\pm0.03$ with $\chi^2/{\rm d.o.f}=0.32$.
The fourth panel with the data observed on 1986 August. 6 presents the result of $b=0.29\pm0.07$ with $\chi^2/{\rm d.o.f}=1.64$.
The fifth panel with the data observed on 1986 August. 7 presents the result of $b=0.26\pm0.02$ with $\chi^2/{\rm d.o.f}=0.05$.}
  \label{0048}
\end{figure*}

\begin{figure*}
 \includegraphics[scale=0.2,angle=-90]{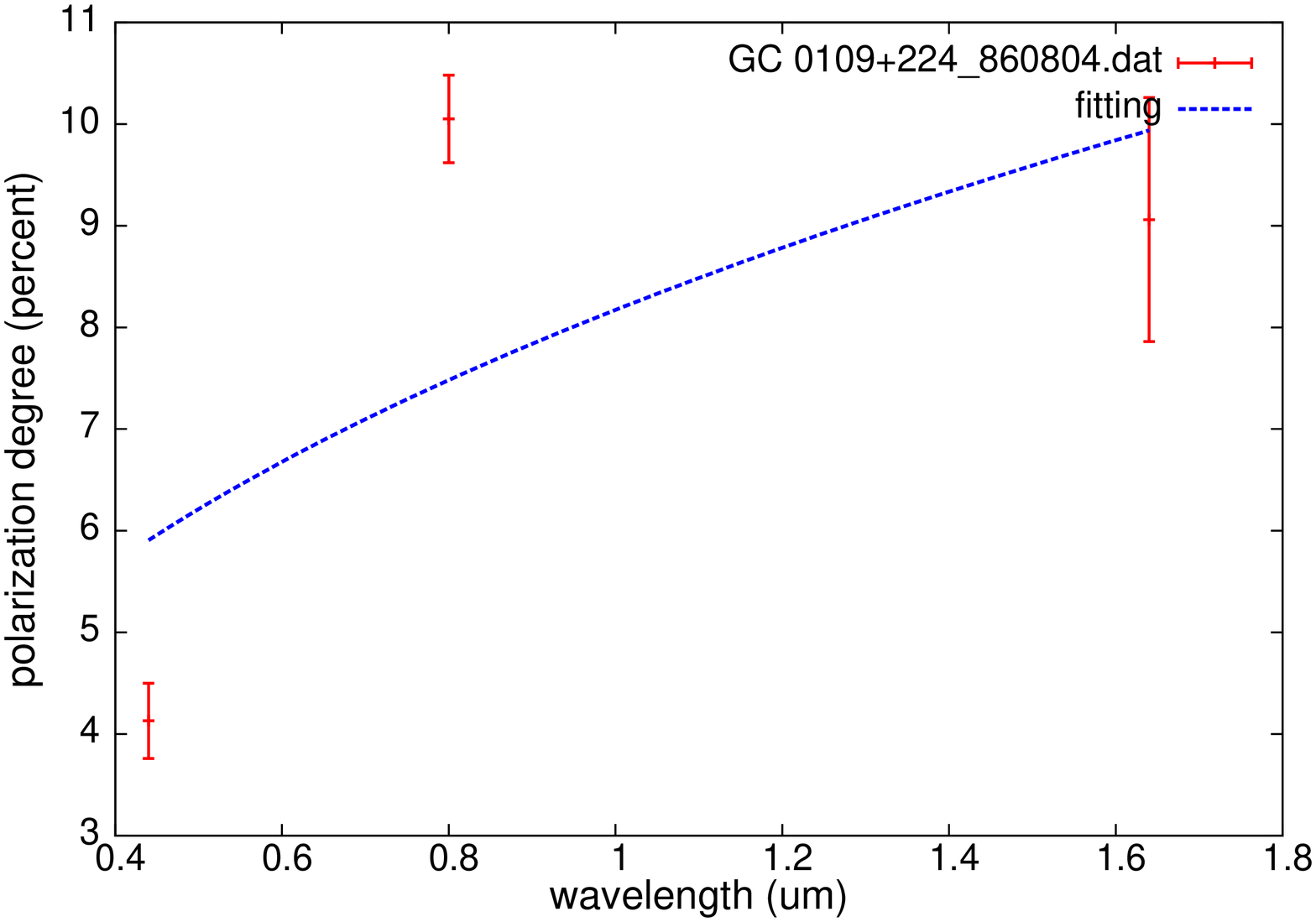}
 \includegraphics[scale=0.2,angle=-90]{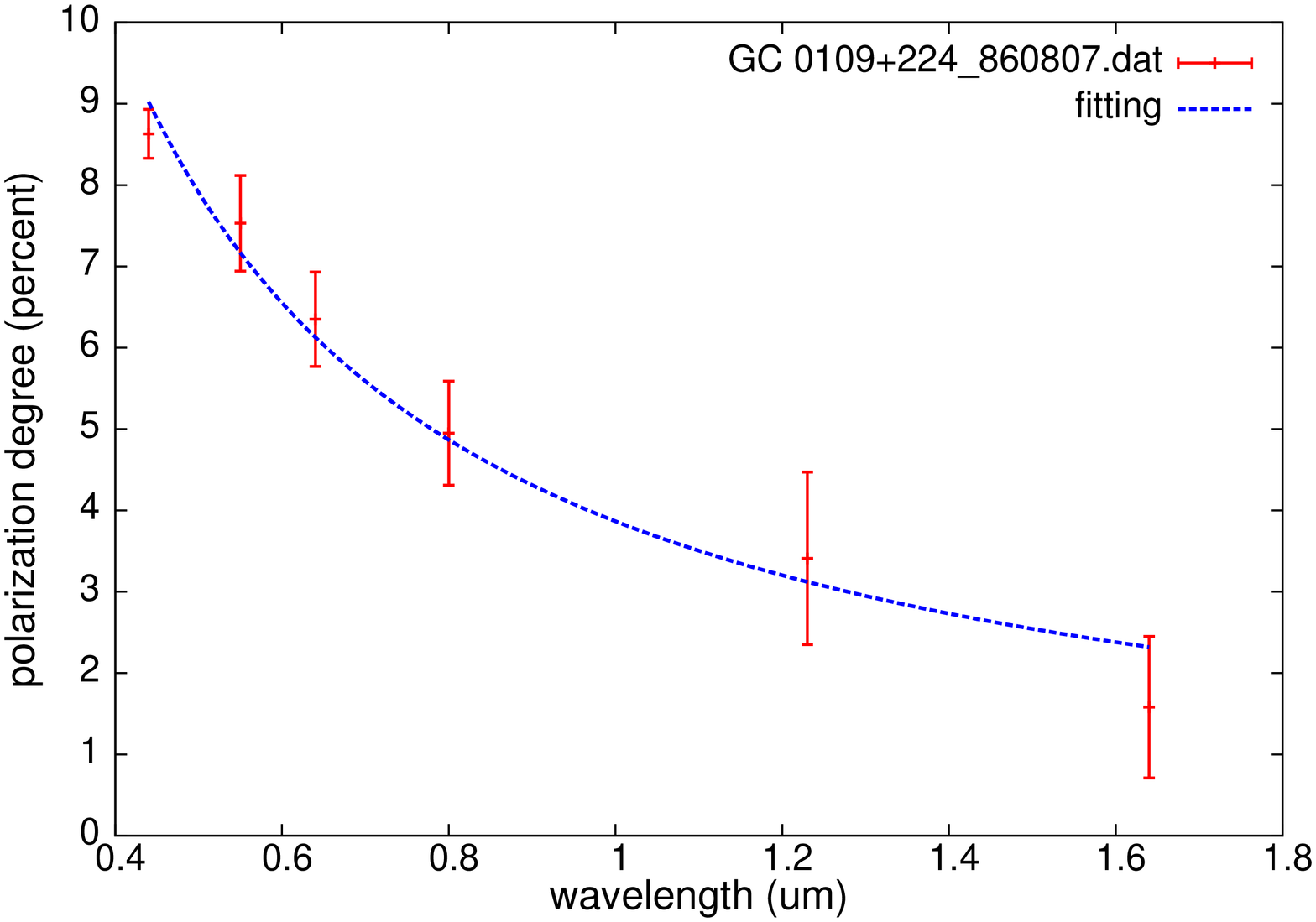}
 \includegraphics[scale=0.2,angle=-90]{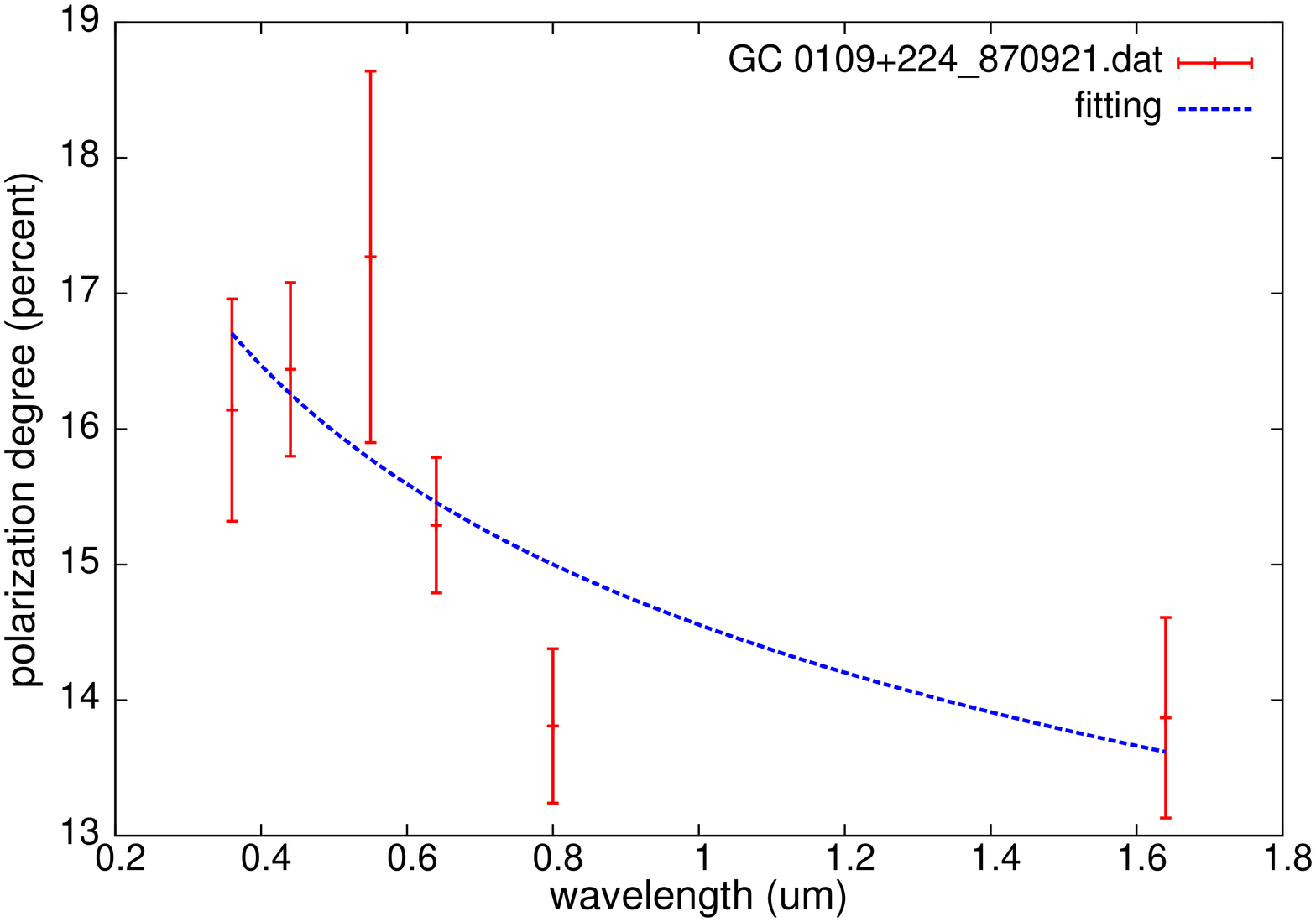}
  \caption{Depolarization fitting for GC 0109+224. The first panel with the data observed on 1986 August. 4 presents the result of $b=-0.39\pm0.46$ with
$\chi^2/{\rm d.o.f}=10.52$. The second panel with the data observed on 1986 August. 7 presents the result of $b=1.03\pm0.11$ with $\chi^2/{\rm d.o.f}=0.24$.
The third panel with the data observed on 1987 September 21 presents the result of $b=0.13\pm0.06$ with $\chi^2/{\rm d.o.f}=1.02$.}
  \label{0109}
\end{figure*}

\begin{figure*}
 \includegraphics[scale=0.2,angle=-90]{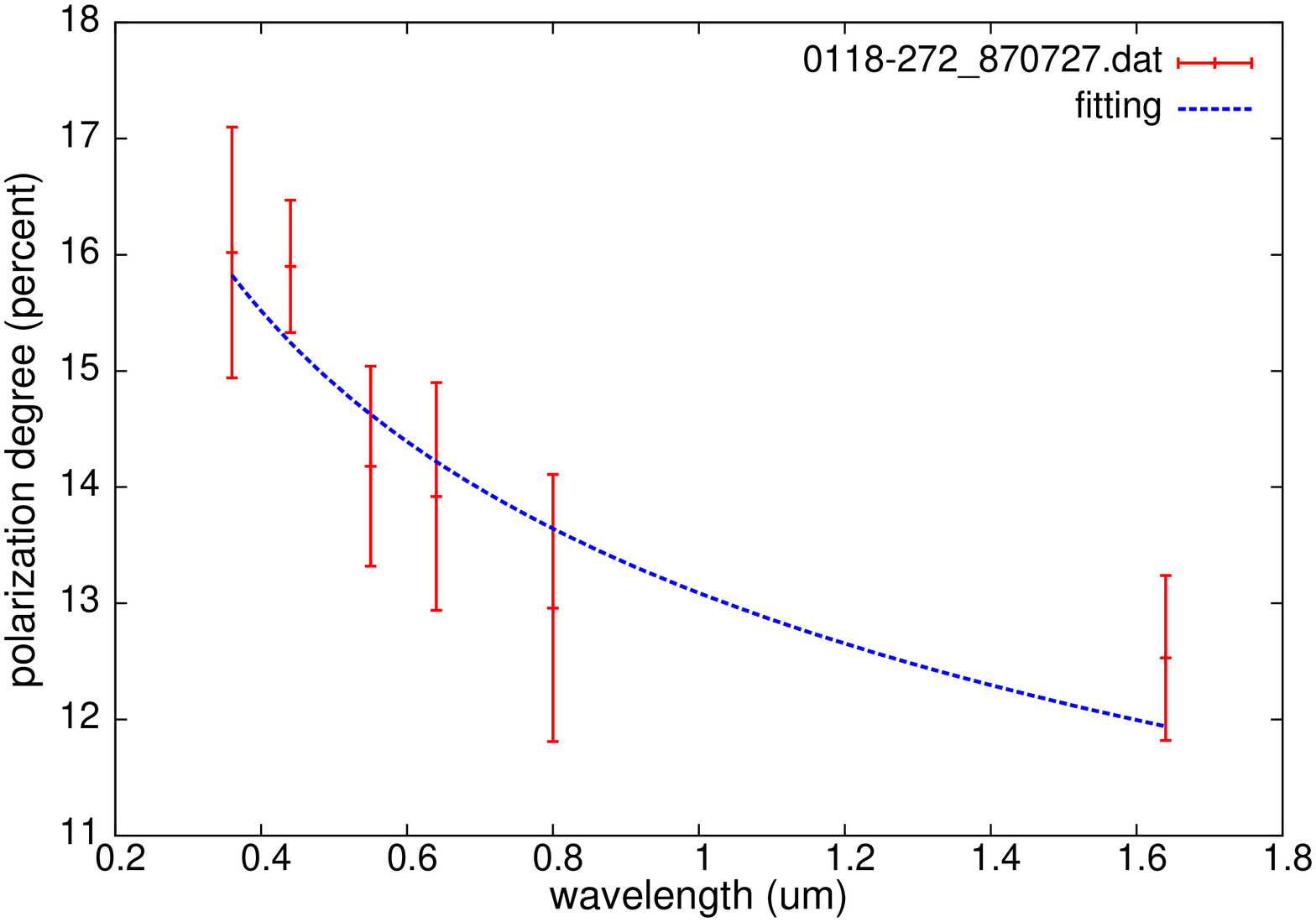}
 \includegraphics[scale=0.2,angle=-90]{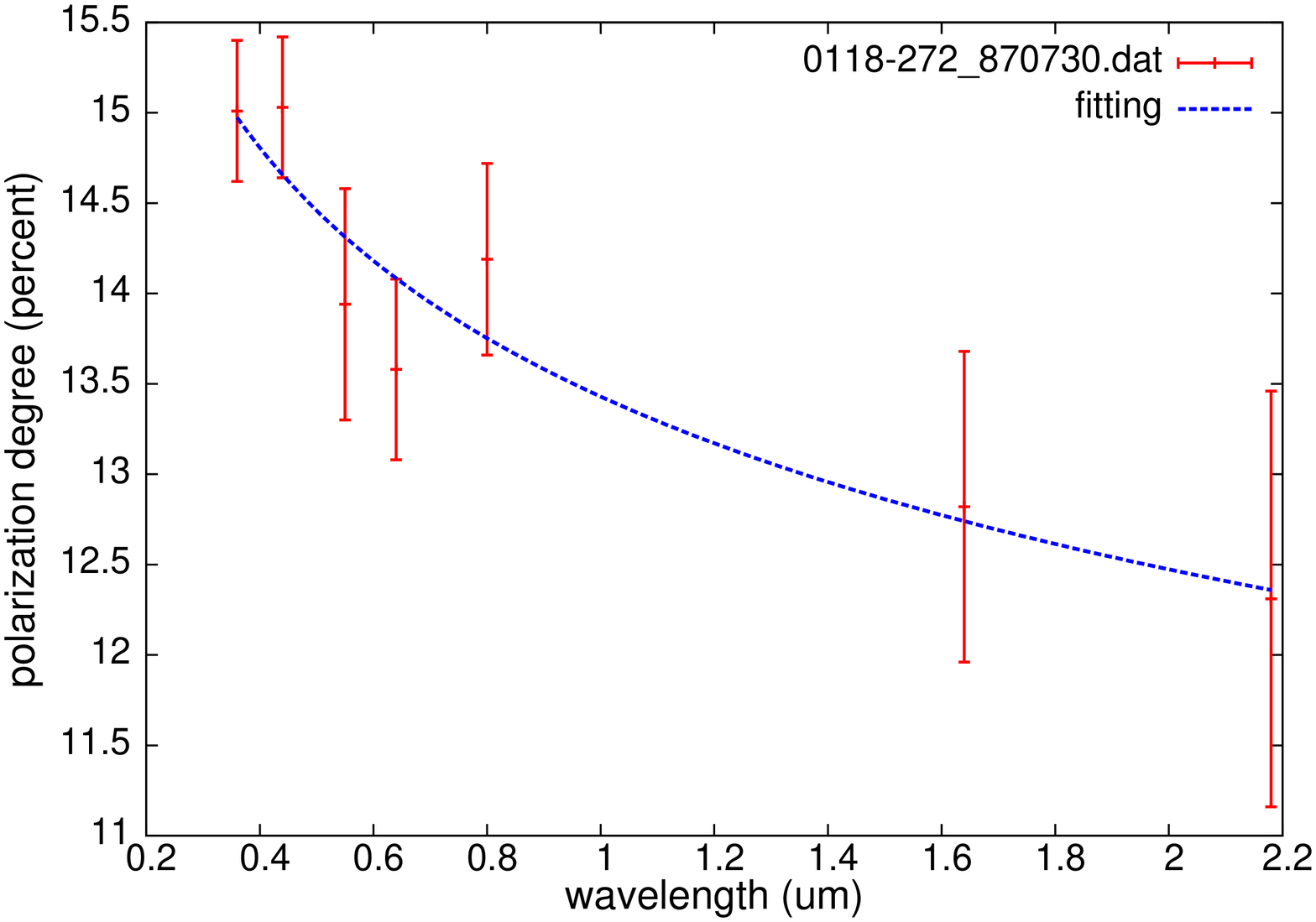}
 \includegraphics[scale=0.2,angle=-90]{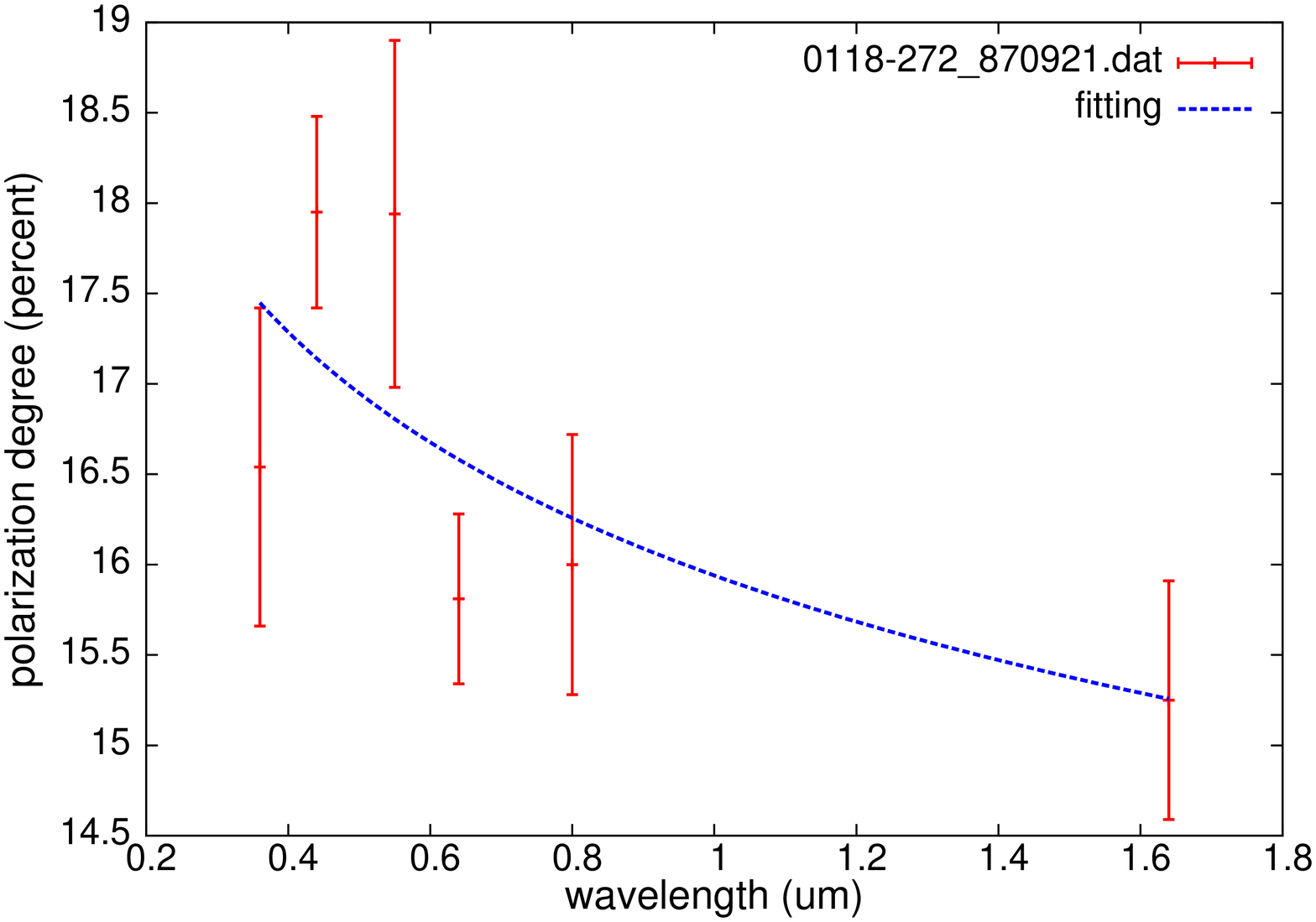}
  \caption{Depolarization fitting for PKS 0118$-$272. The first panel with the data observed on 1987 July 27 presents the result of $b=0.19\pm0.04$ with
$\chi^2/{\rm d.o.f}=0.39$. The second panel with the data observed on 1987 July 30 presents the result of $b=0.11\pm0.02$ with $\chi^2/{\rm d.o.f}=0.15$.
The third panel with the data observed on 1987 Sepember 21 presents the result of $b=0.09\pm0.05$ with $\chi^2/{\rm d.o.f}=0.86$.}
  \label{0118}
\end{figure*}

\begin{figure*}
 \includegraphics[scale=0.2,angle=-90]{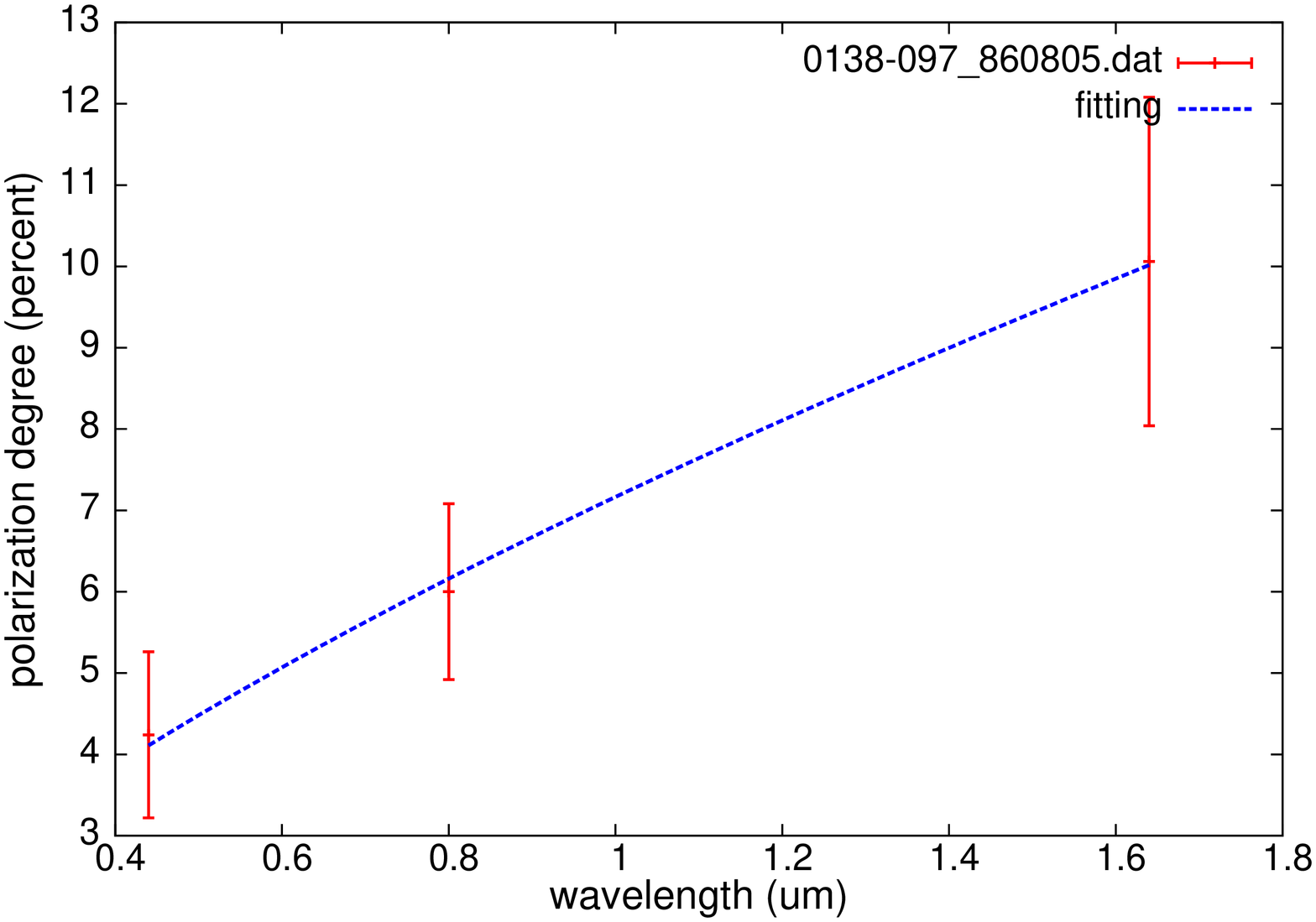}
 \includegraphics[scale=0.2,angle=-90]{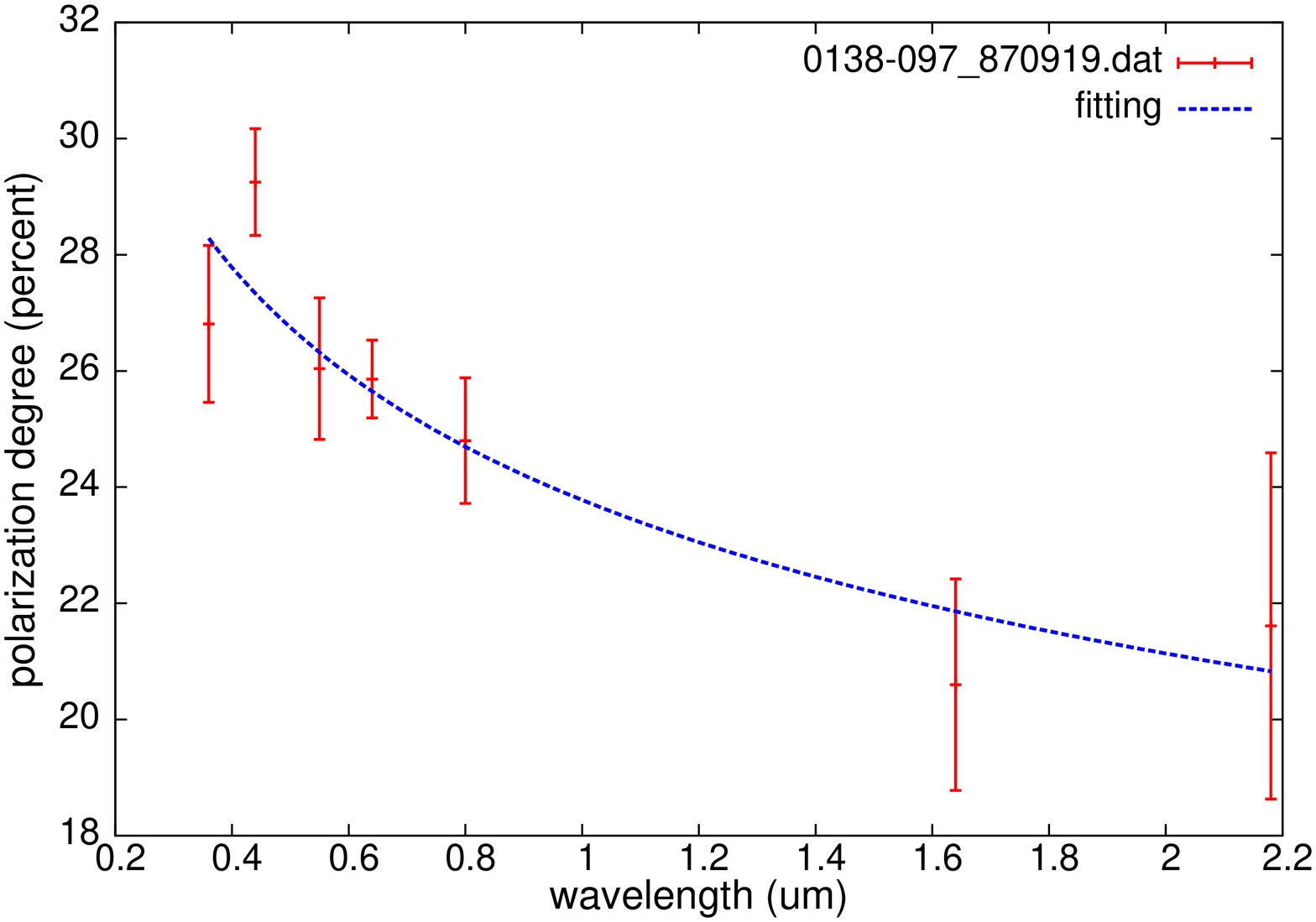}
 \includegraphics[scale=0.2,angle=-90]{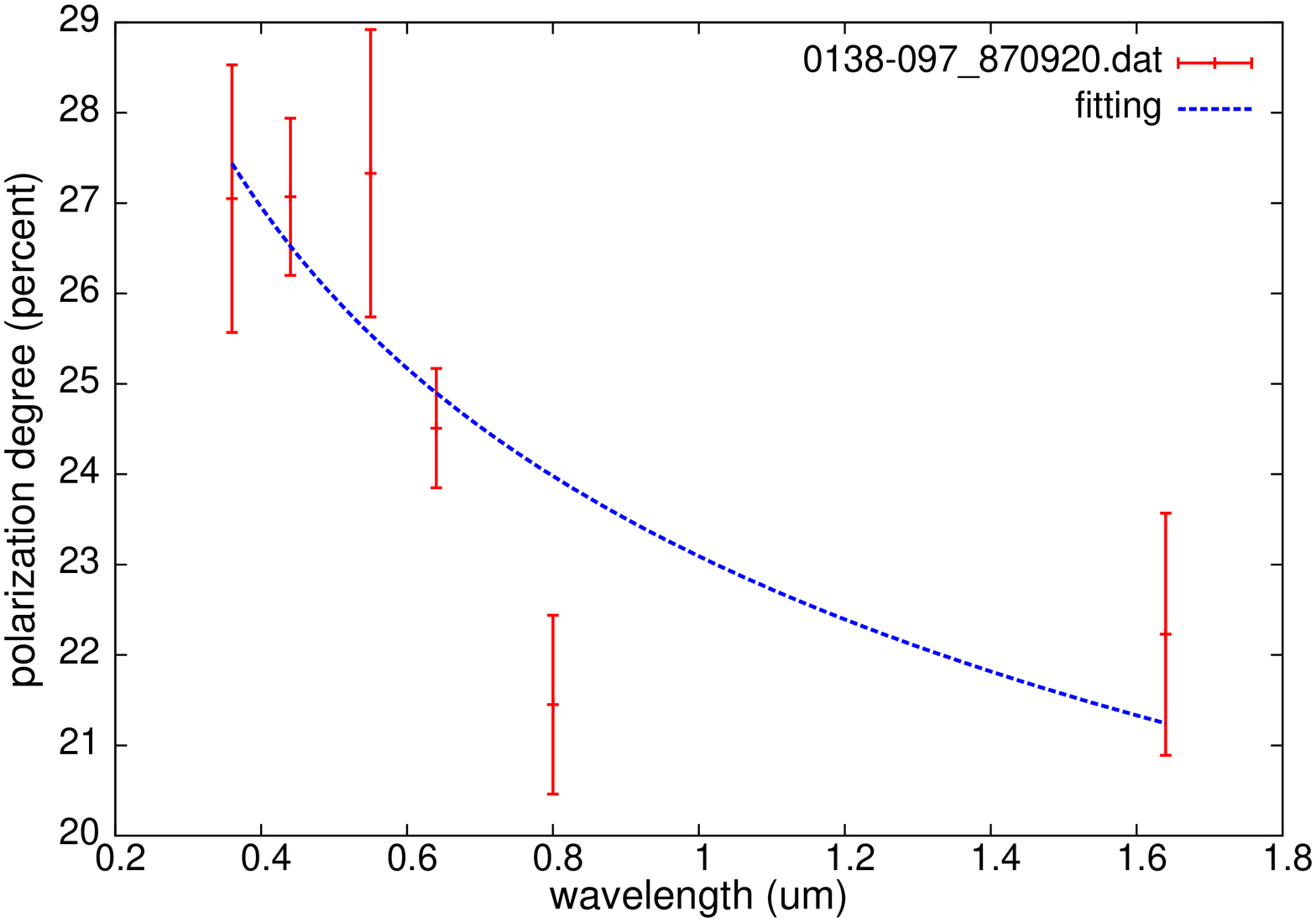}
  \caption{Depolarization fitting for 0138$-$097. The first panel with the data observed on 1986 August 5 presents the result of $b=-0.68\pm0.04$ with
$\chi^2/{\rm d.o.f}=0.04$. The second panel with the data observed on 1987 September 19 presents the result of $b=0.17\pm0.03$ with $\chi^2/{\rm d.o.f}=1.63$.
The third panel with the data observed on 1987 September 20 presents the result of $b=0.17\pm0.06$ with $\chi^2/{\rm d.o.f}=2.79$.}
  \label{0138}
\end{figure*}
\clearpage
\begin{figure}
  \includegraphics[scale=0.2,angle=-90]{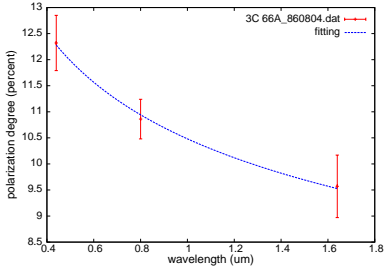}
  \caption{Depolarization fitting for 3C 66A (0219+428). The panel with the data observed on 1986 August 4 presents the result of $b=0.19\pm0.01$ with
$\chi^2/{\rm d.o.f}=0.01$.}
  \label{0219}
\end{figure}

\begin{figure*}
  \includegraphics[scale=0.2,angle=-90]{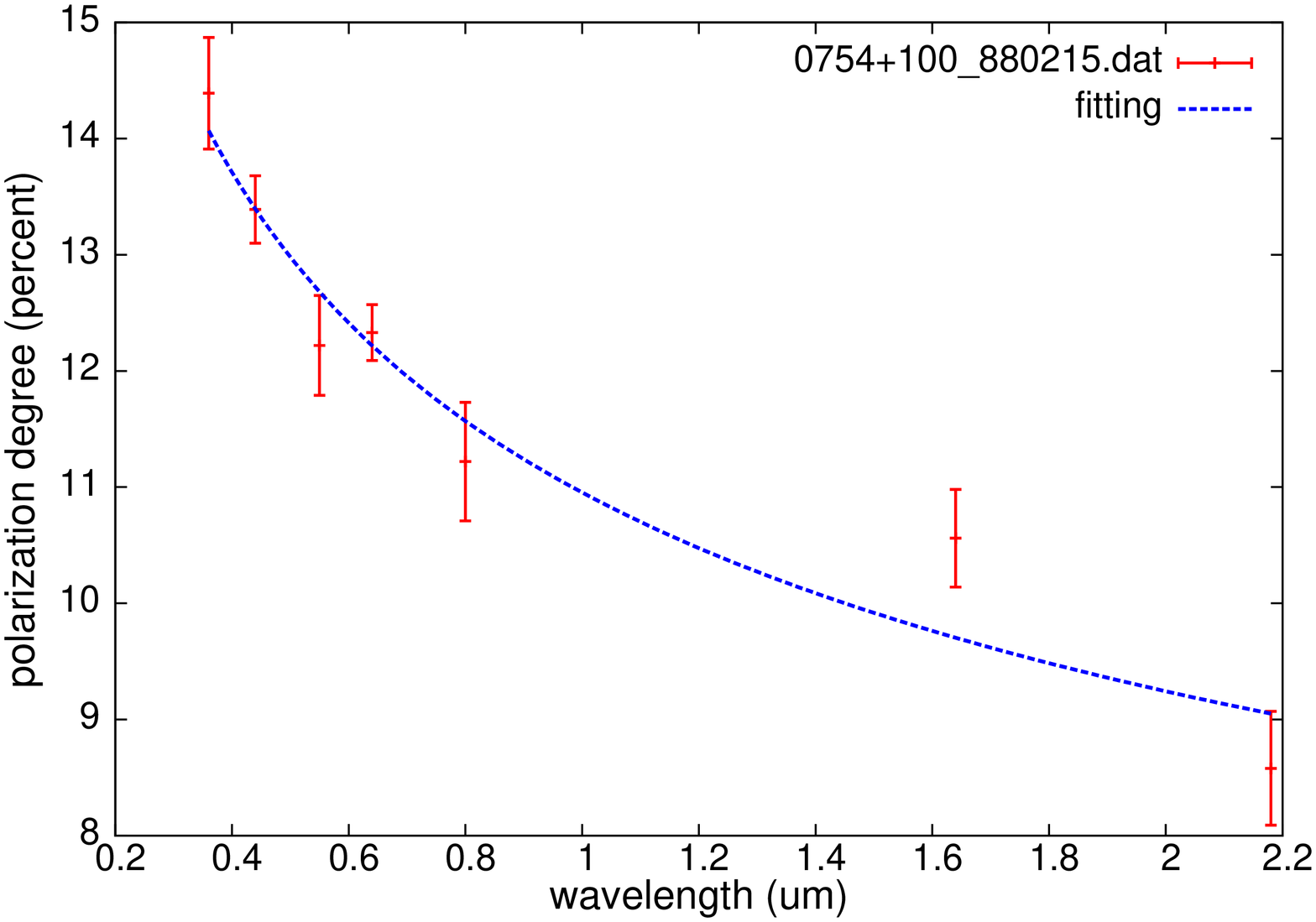}
  \includegraphics[scale=0.2,angle=-90]{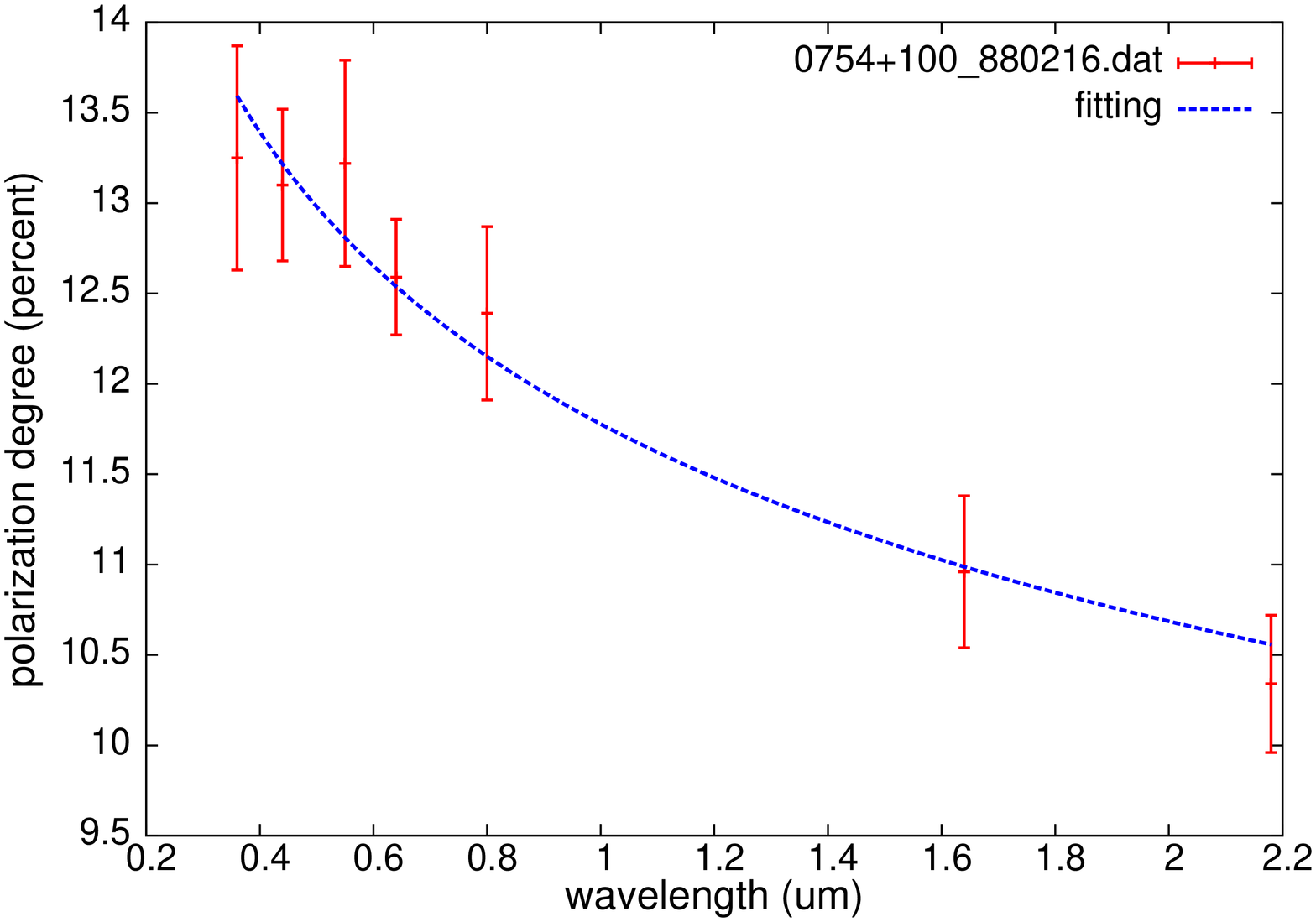}
  \caption{Depolarization fitting for 0754+100. The first panel with the data observed on 1988 February 15 presents the result of $b=0.24\pm0.03$ with
$\chi^2/{\rm d.o.f}=0.28$. The second panel with the data observed on 1988 February 16 presents the result of $b=0.14\pm0.01$ with $\chi^2/{\rm d.o.f}=0.08$.}
  \label{0754}
\end{figure*}

\begin{figure*}
  \includegraphics[scale=0.2,angle=-90]{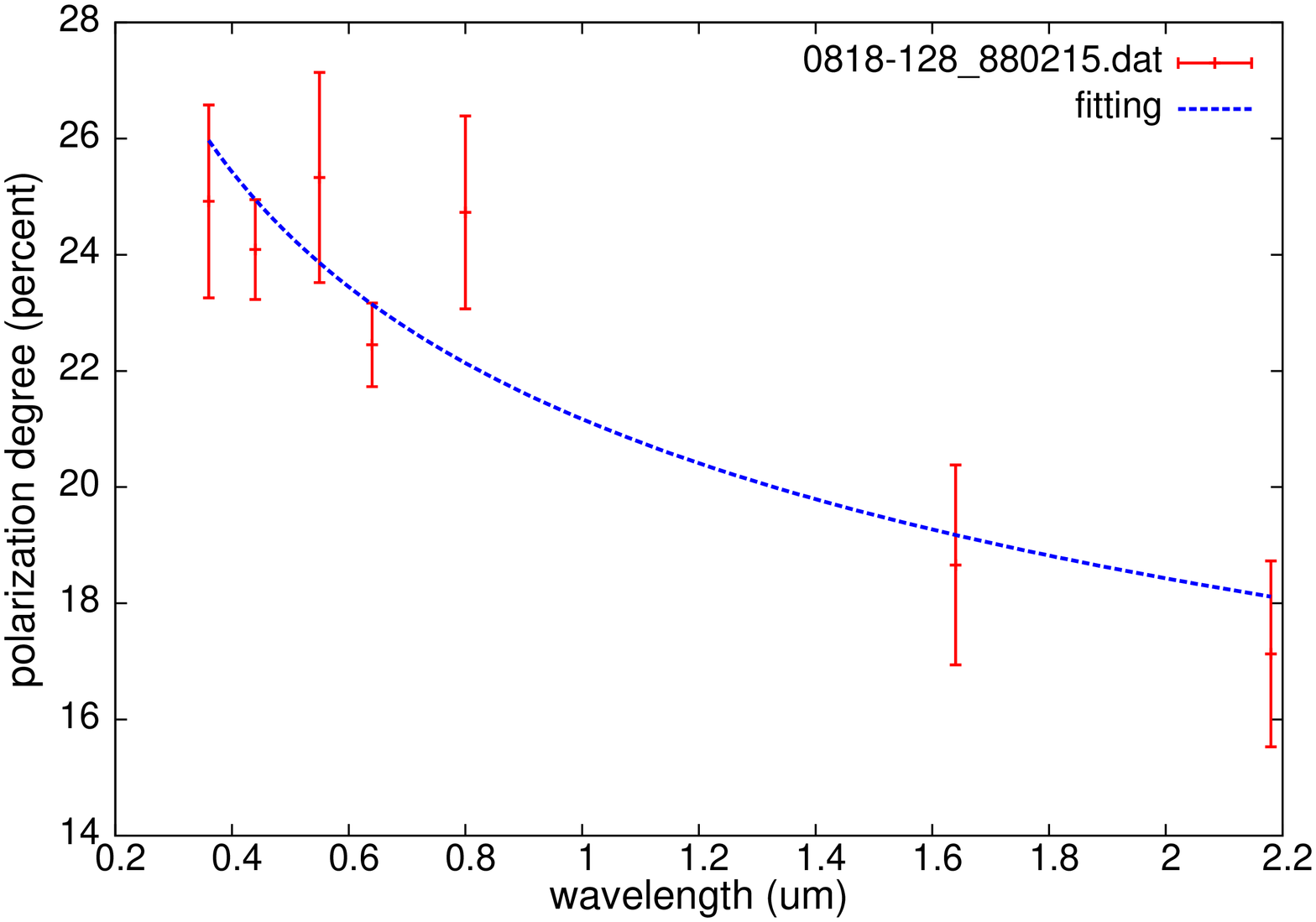}
  \includegraphics[scale=0.2,angle=-90]{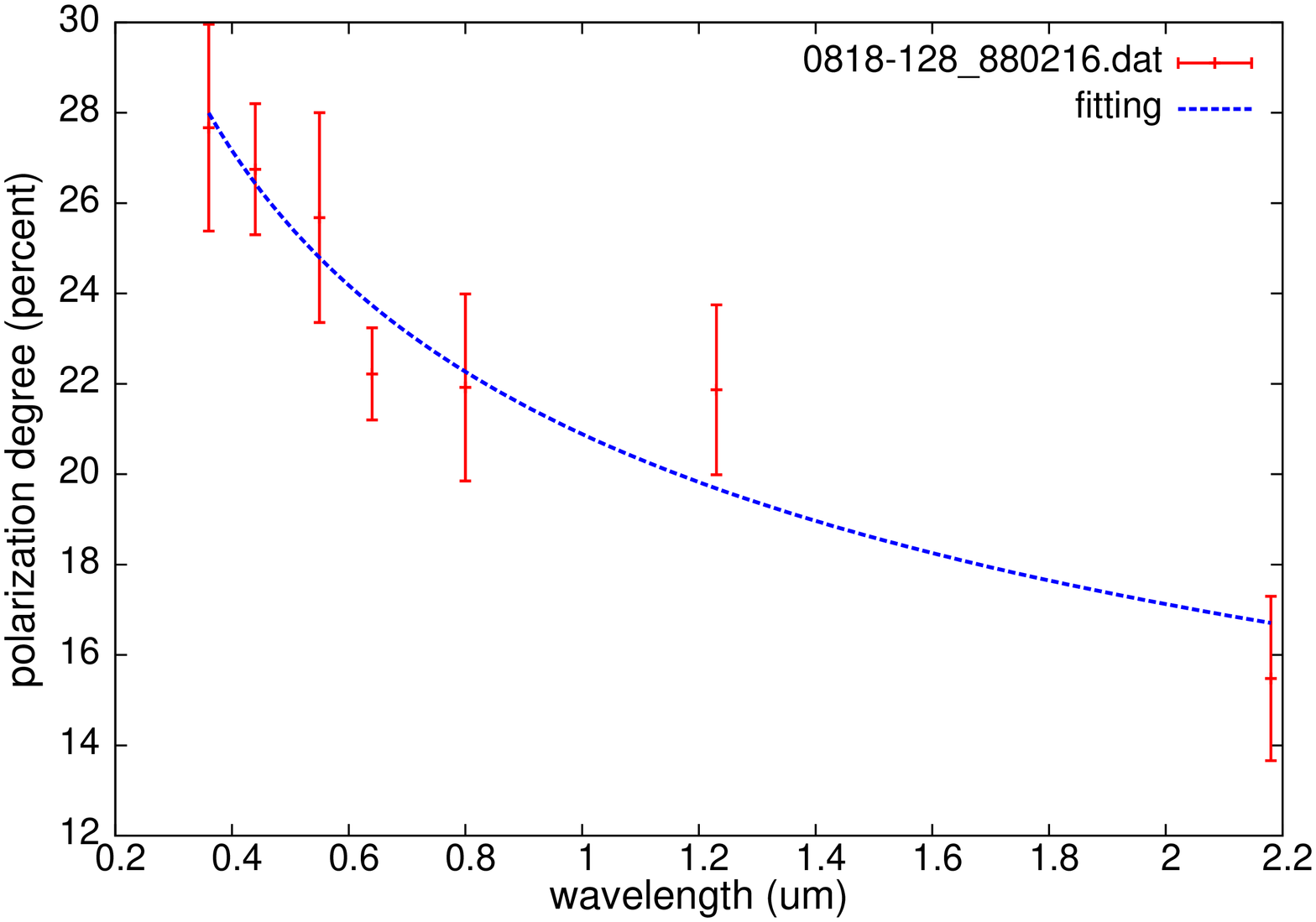}
  \includegraphics[scale=0.2,angle=-90]{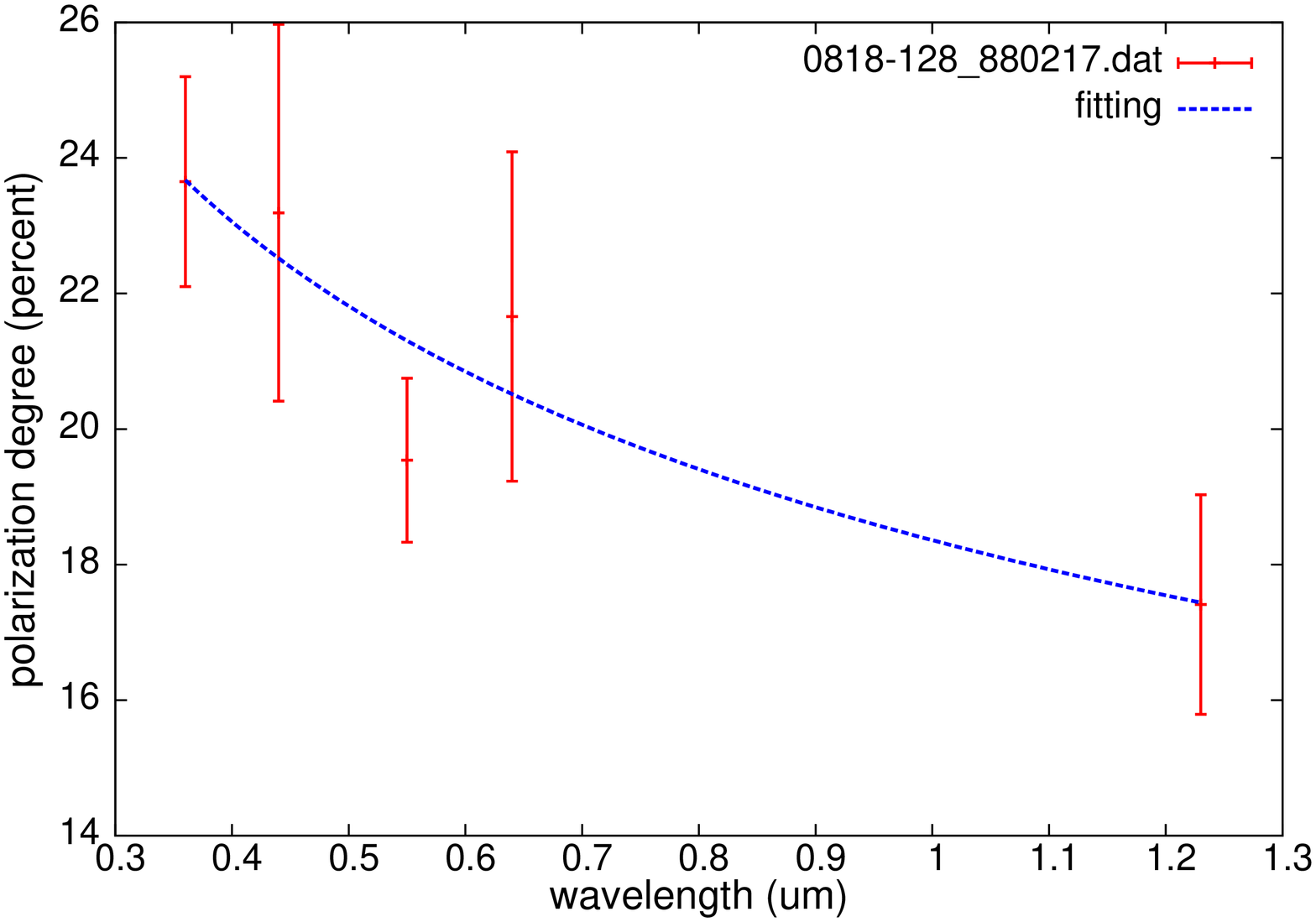}
  \caption{Depolarization fitting for 0818$-$128 (OJ-131). The first panel with the data observed on 1988 February 15 presents the result of $b=0.20\pm0.05$ with
$\chi^2/{\rm d.o.f}=2.49$. The second panel with the data observed on 1988 February 16 presents the result of $b=0.29\pm0.04$ with $\chi^2/{\rm d.o.f}=1.94$.
The third panel with the data observed on 1988 February 17 presents the result of $b=0.25\pm0.07$ with $\chi^2/{\rm d.o.f}=1.62$.}
  \label{0818}
\end{figure*}

\begin{figure*}
  \includegraphics[scale=0.2,angle=-90]{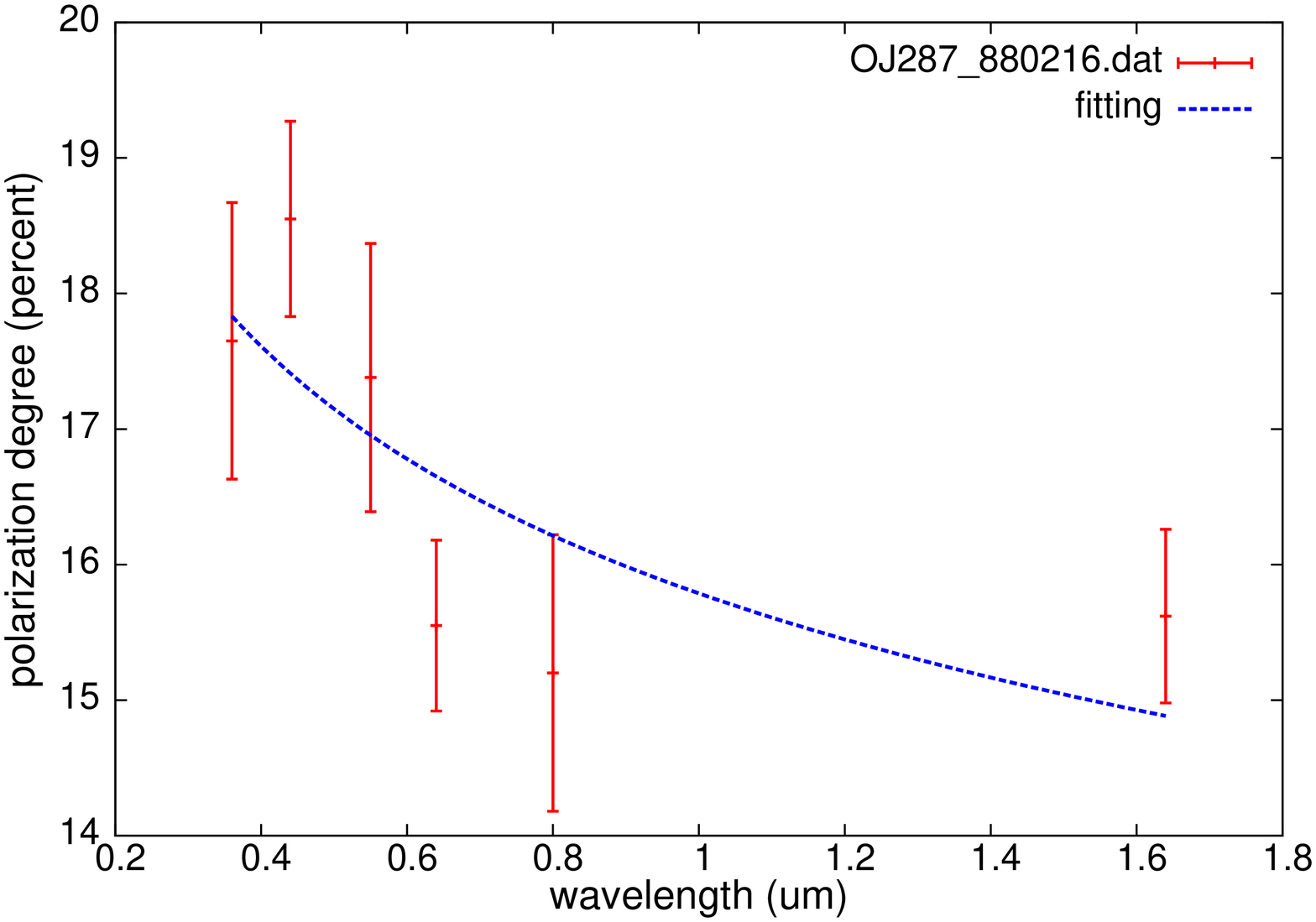}
  \includegraphics[scale=0.2,angle=-90]{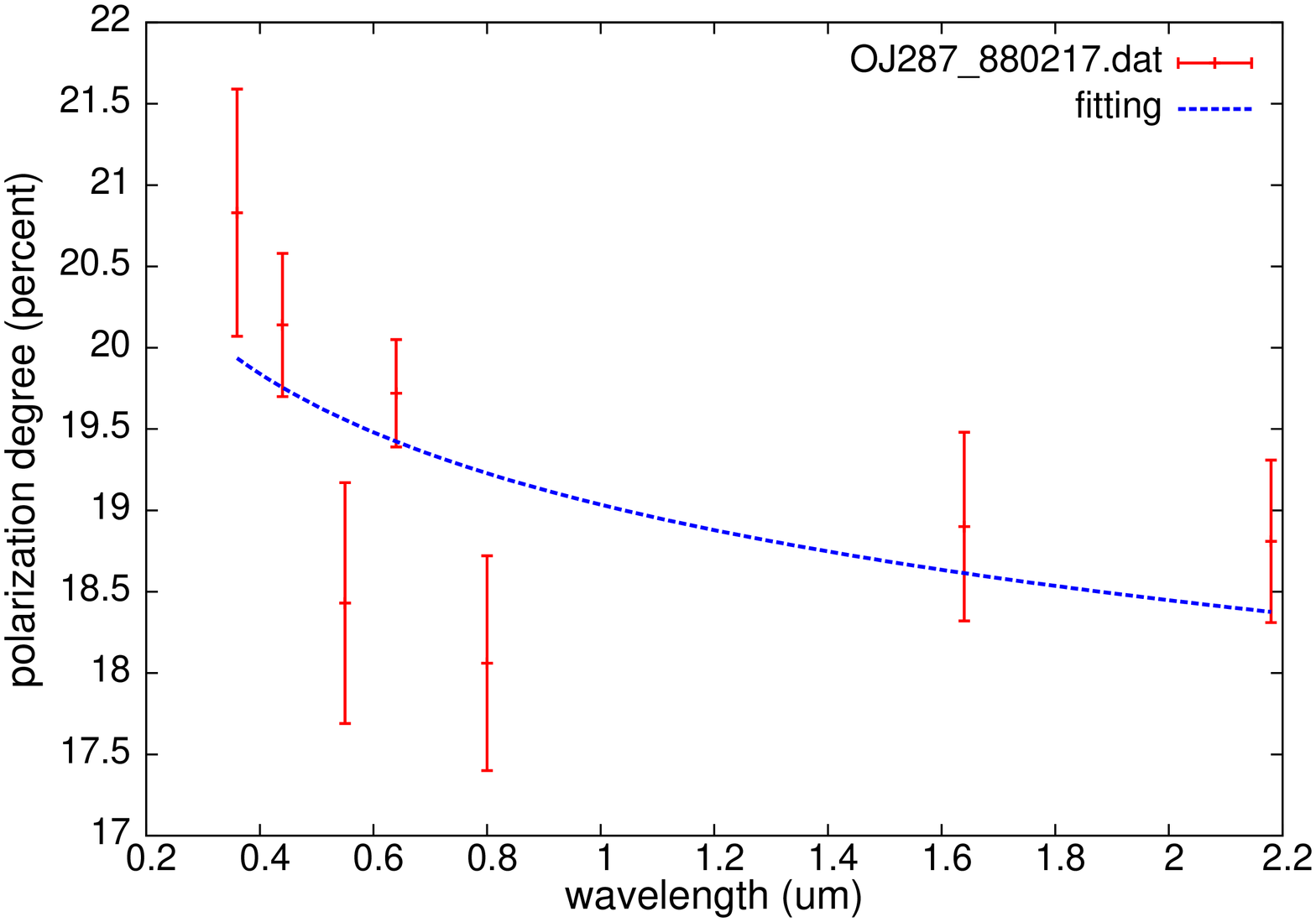}
  \caption{Depolarization fitting of OJ 287 (0851+202). The first panel with the data observed on 1988 February 16 presents the result of $b=0.12\pm0.05$ with
$\chi^2/{\rm d.o.f}=1.07$. The second panel with the data observed on 1988 February 17 presents the result of $b=0.05\pm0.03$ with $\chi^2/{\rm d.o.f}=0.79$.}
  \label{0851}
\end{figure*}

\begin{figure*}
  \includegraphics[scale=0.2,angle=-90]{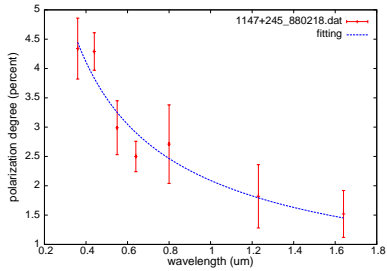}
  \caption{The depolarization fitting for 1147+245. The panel with the data observed on 1988 February 18 presents the result of $b=0.74\pm0.10$ with
$\chi^2/{\rm d.o.f}=0.10$.}
  \label{1147}
\end{figure*}

\begin{figure*}
  \includegraphics[scale=0.2,angle=-90]{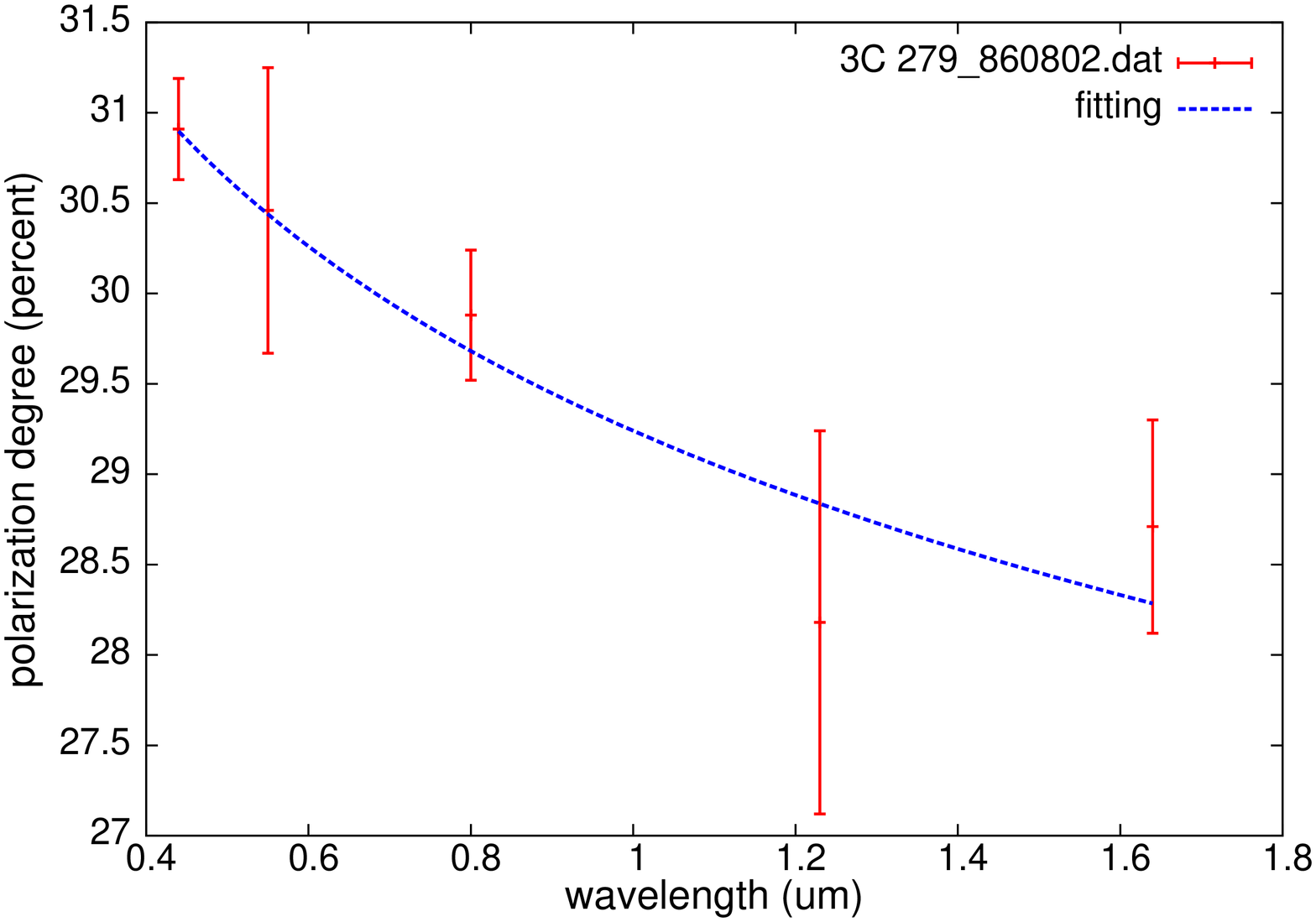}
  \includegraphics[scale=0.2,angle=-90]{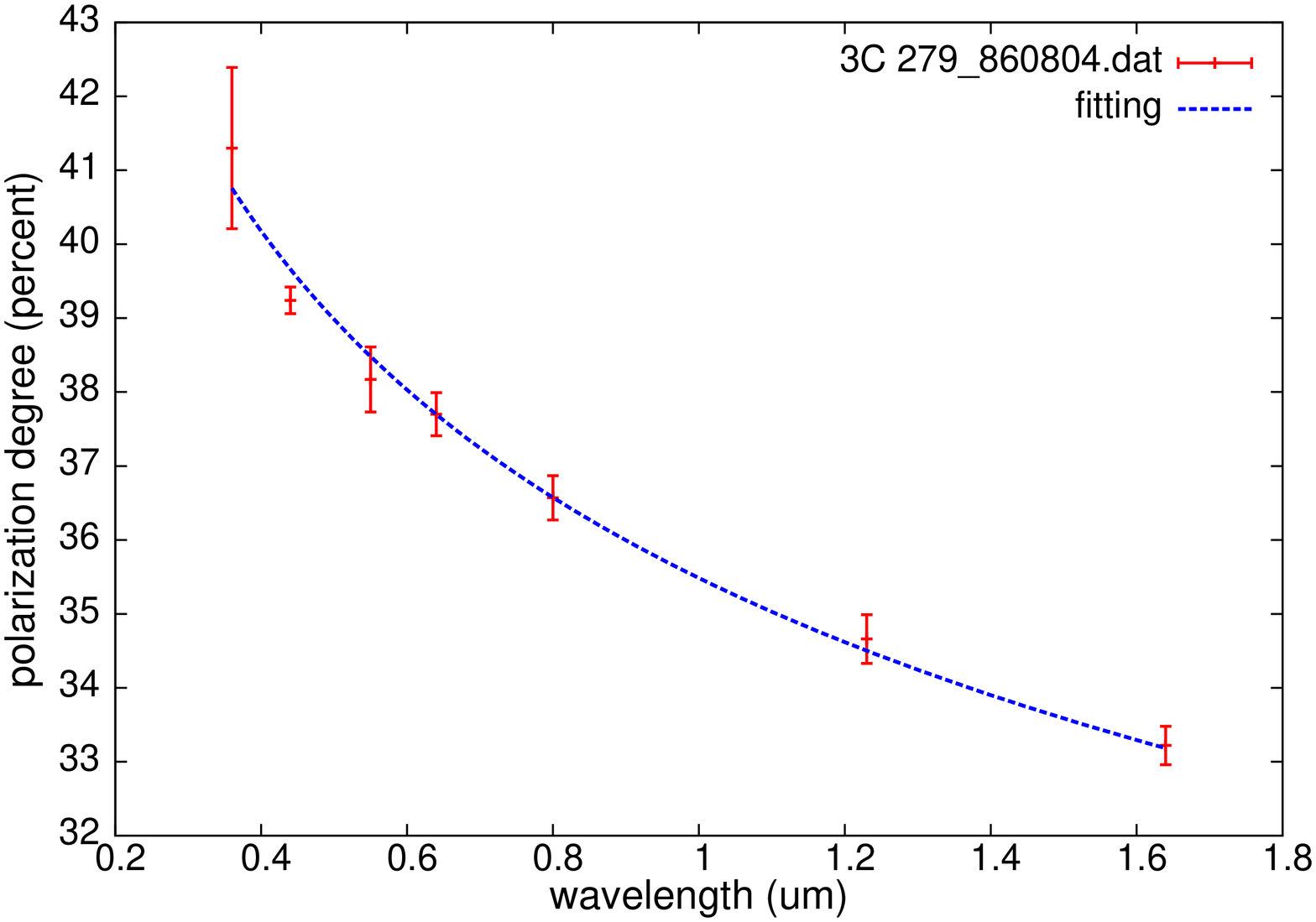}
  \includegraphics[scale=0.2,angle=-90]{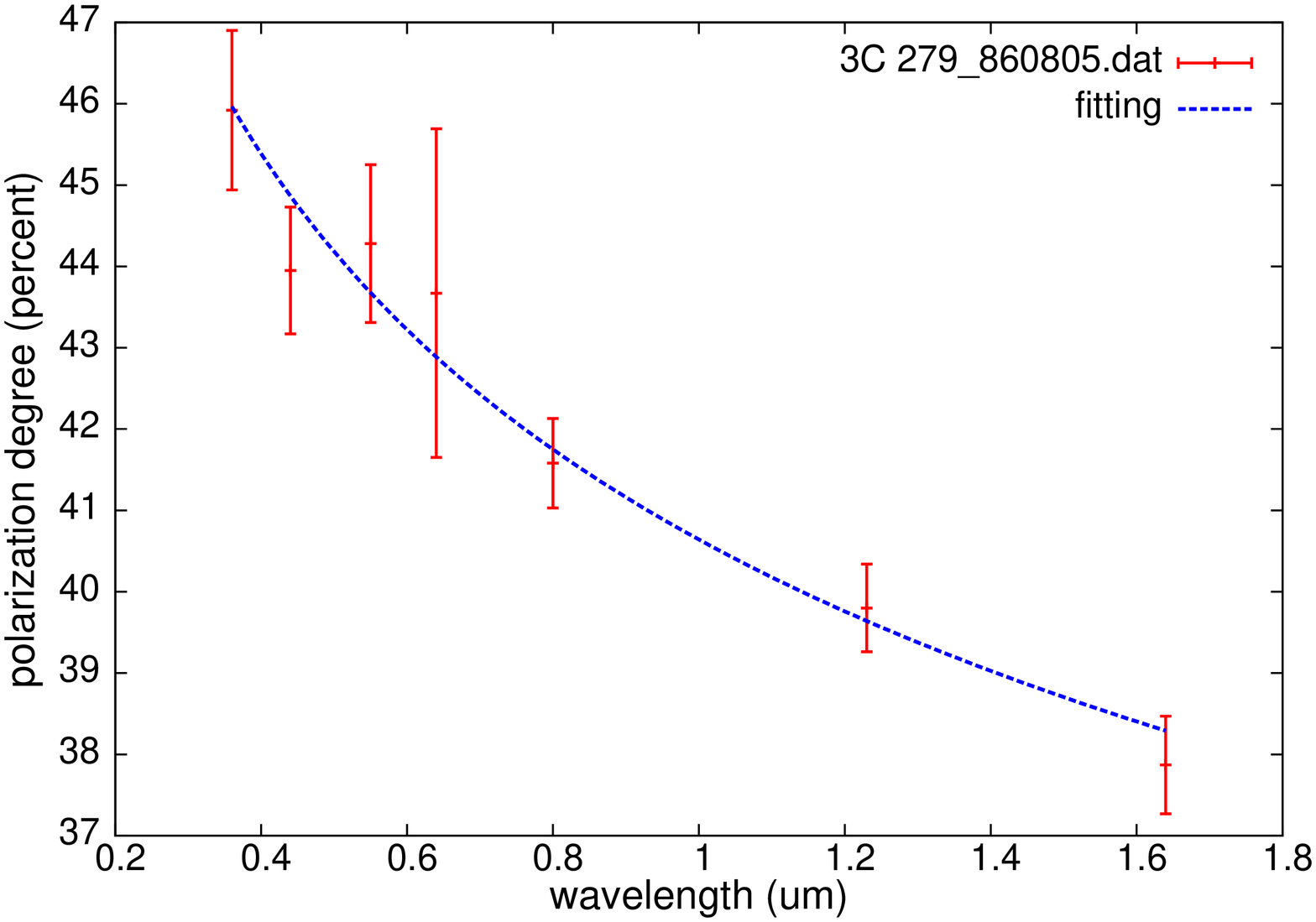} \\
  \includegraphics[scale=0.2,angle=-90]{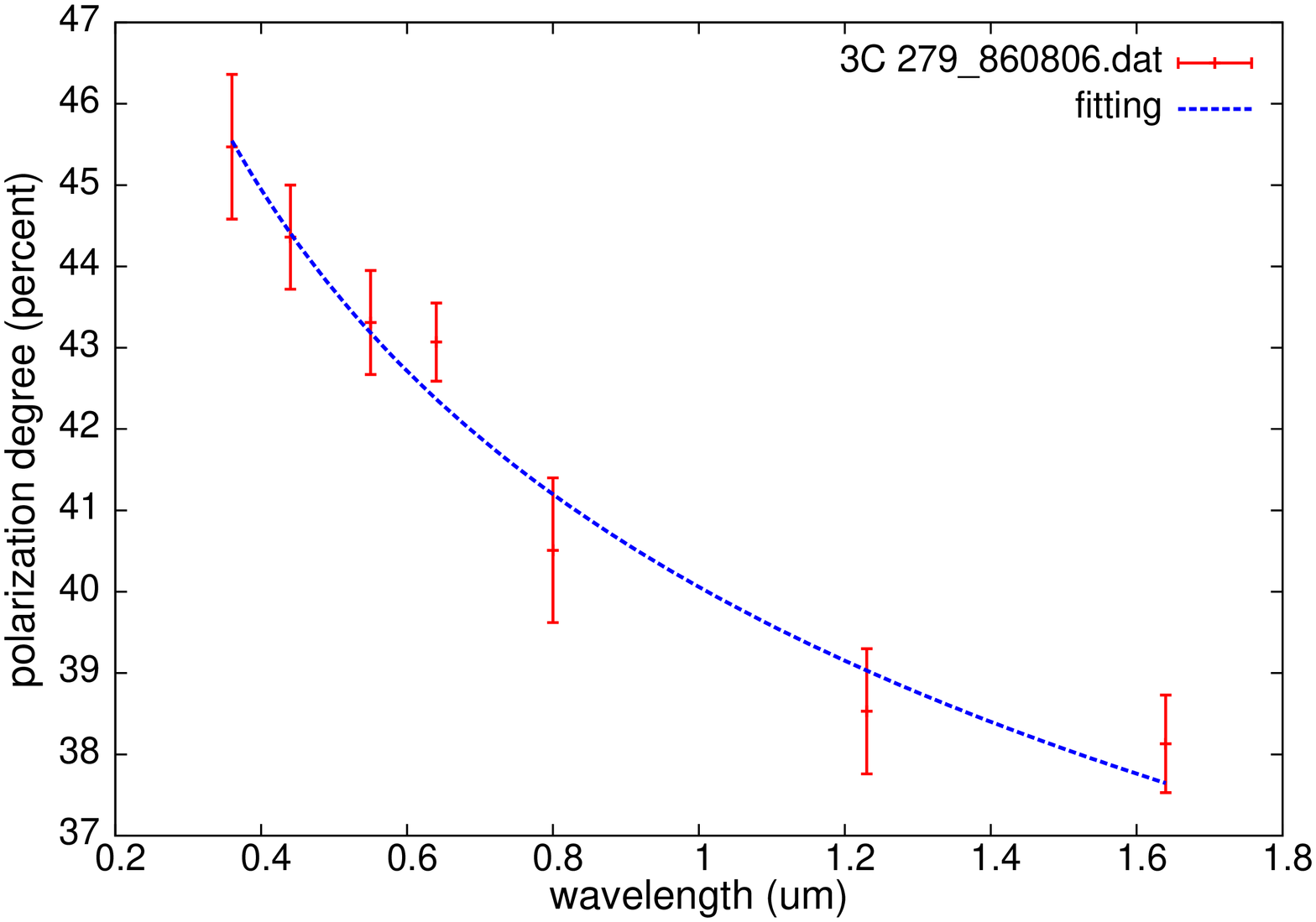}
  \includegraphics[scale=0.2,angle=-90]{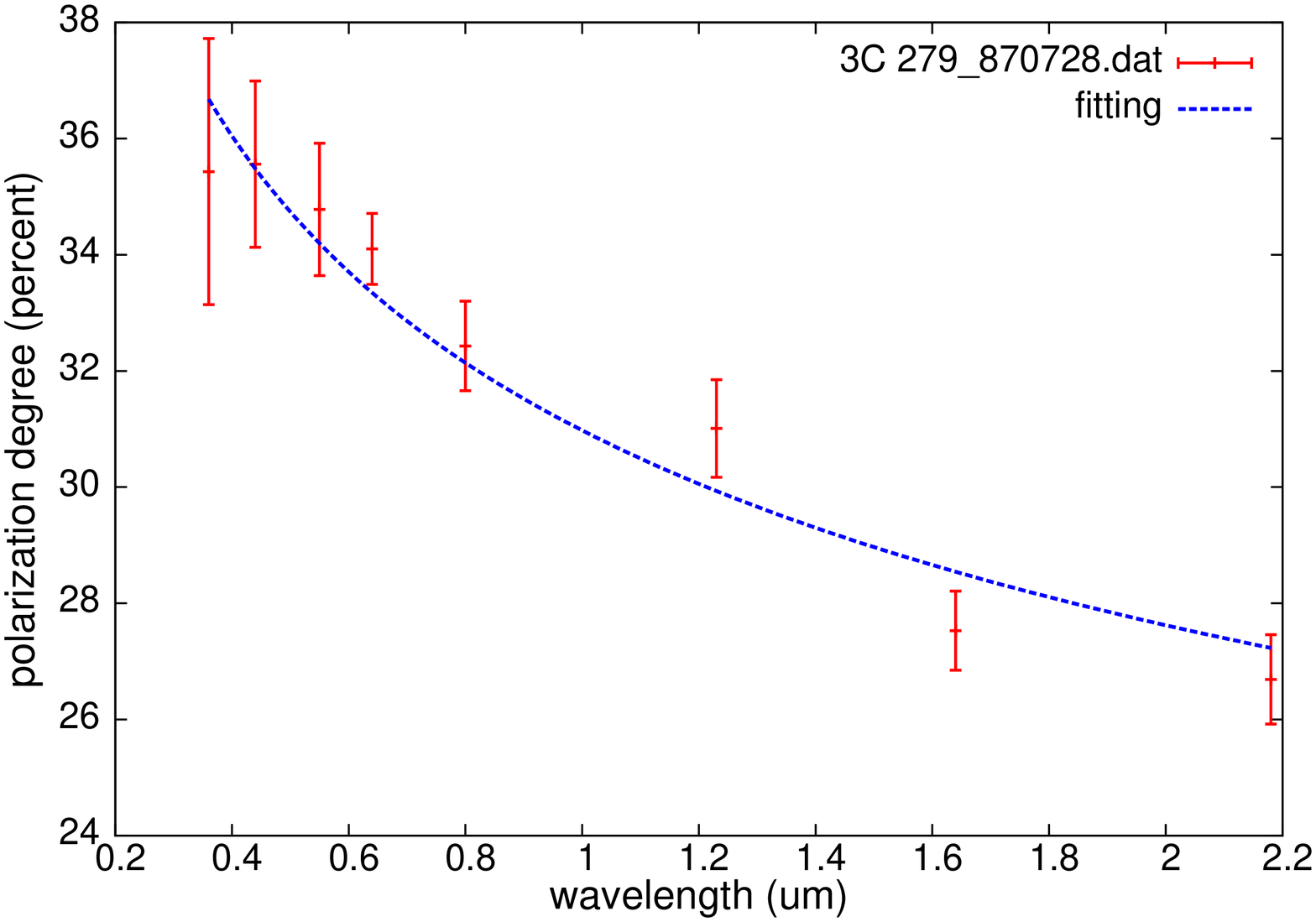}
  \includegraphics[scale=0.2,angle=-90]{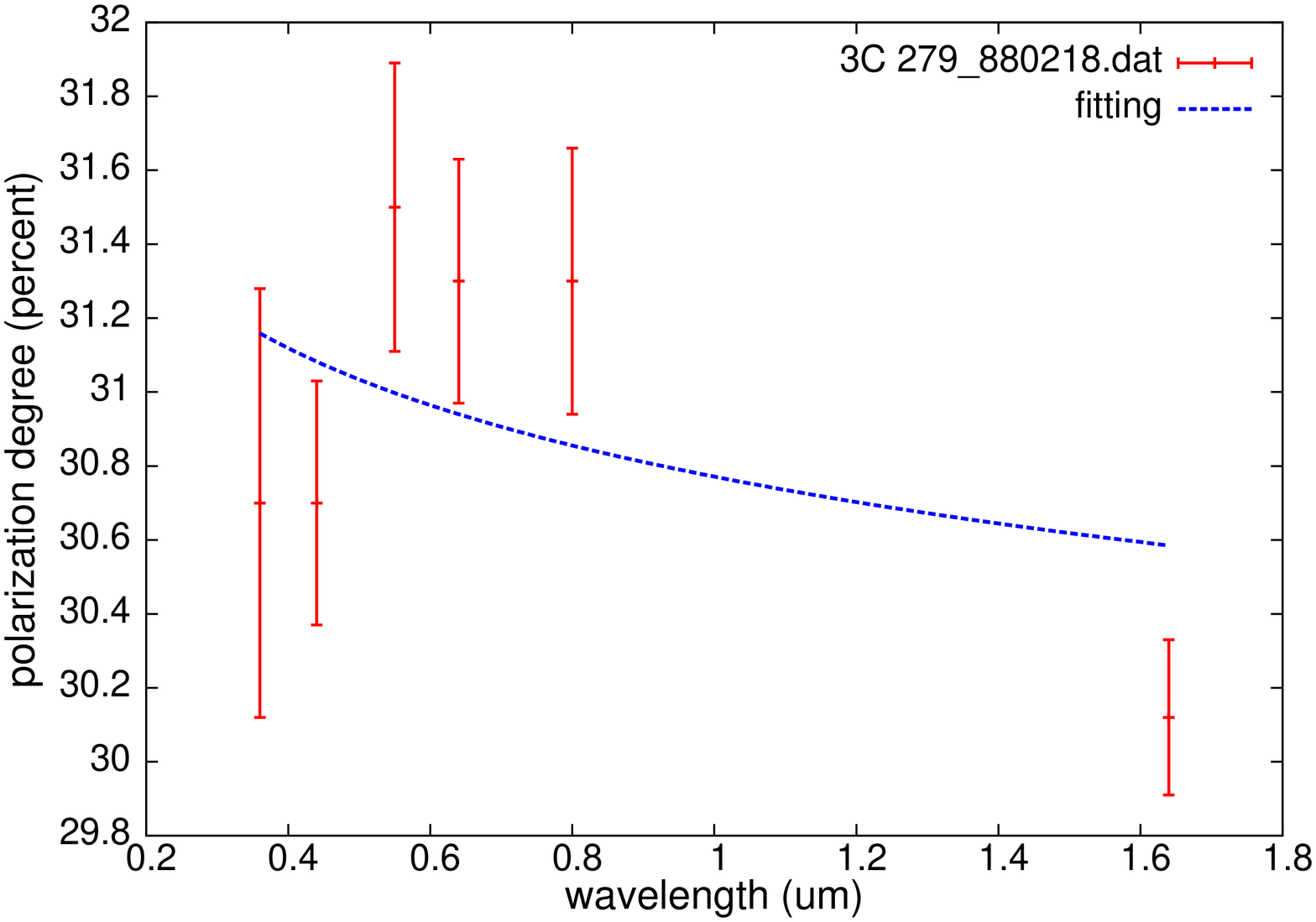}
  \caption{Depolarization fitting for 3C 279 (1253$-$055). The first panel with the data observed on 1986 August 2 presents the result of $b=0.07\pm0.01$ with
$\chi^2/{\rm d.o.f}=0.22$. The second panel with the data observed on 1986 August 4 presents the result of $b=0.14\pm0.01$ with $\chi^2/{\rm d.o.f}=0.12$.
The third panel with the data observed on 1986 August 5 presents the result of $b=0.12\pm0.01$ with $\chi^2/{\rm d.o.f}=0.41$.
The fourth panel with the data observed on 1986 August 6 presents the result of $b=0.13\pm0.01$ with $\chi^2/{\rm d.o.f}=0.29$.
The fifth panel with the data observed on 1987 July 28 presents the result of $b=0.17\pm0.02$ with $\chi^2/{\rm d.o.f}=0.84$.
The sixth panel with the data observed on 1988 February 18 presents the result of $b=0.01\pm0.01$ with $\chi^2/{\rm d.o.f}=0.29$.}
  \label{1253}
\end{figure*}

\begin{figure*}
 \includegraphics[scale=0.2,angle=-90]{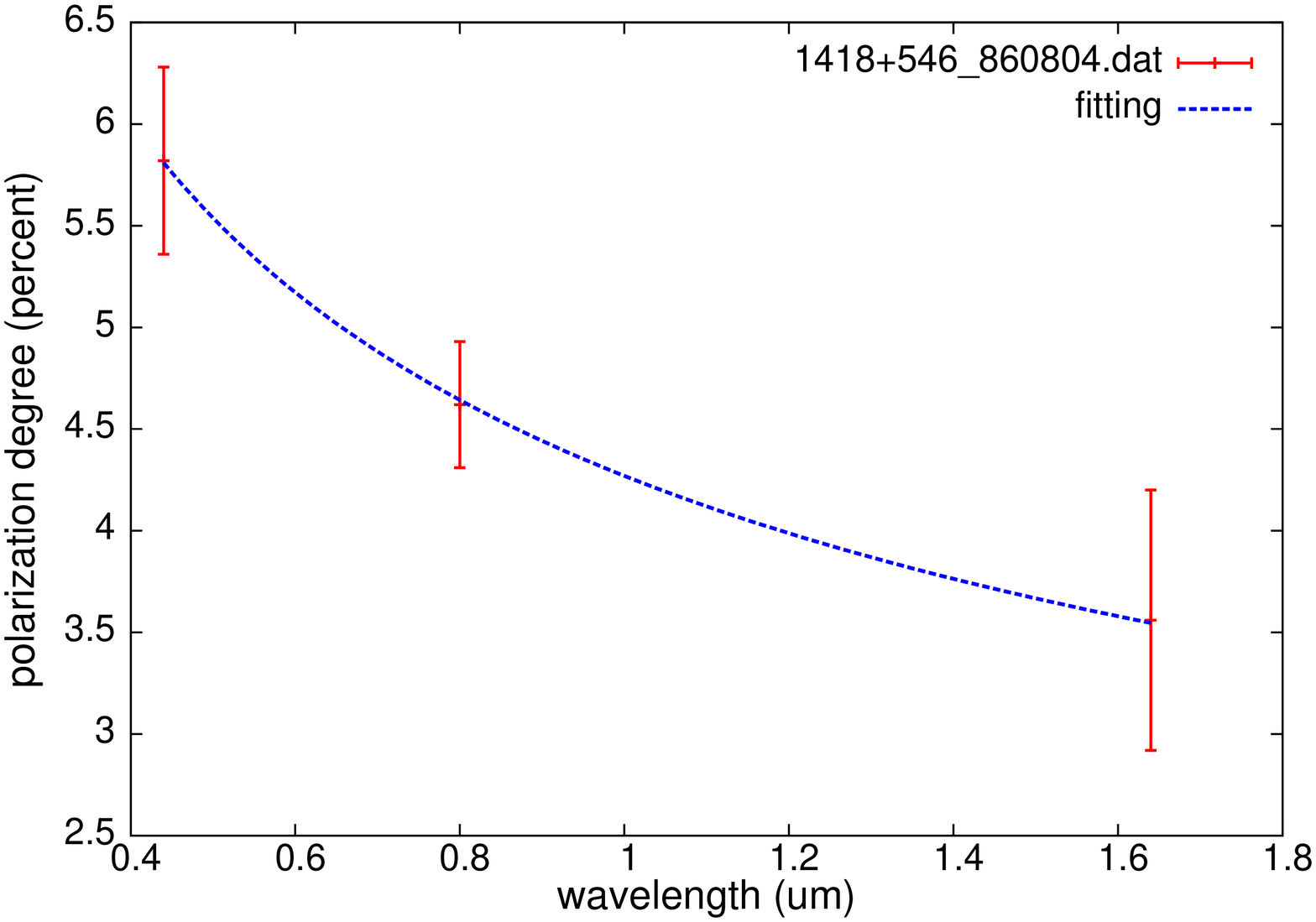}
 \includegraphics[scale=0.2,angle=-90]{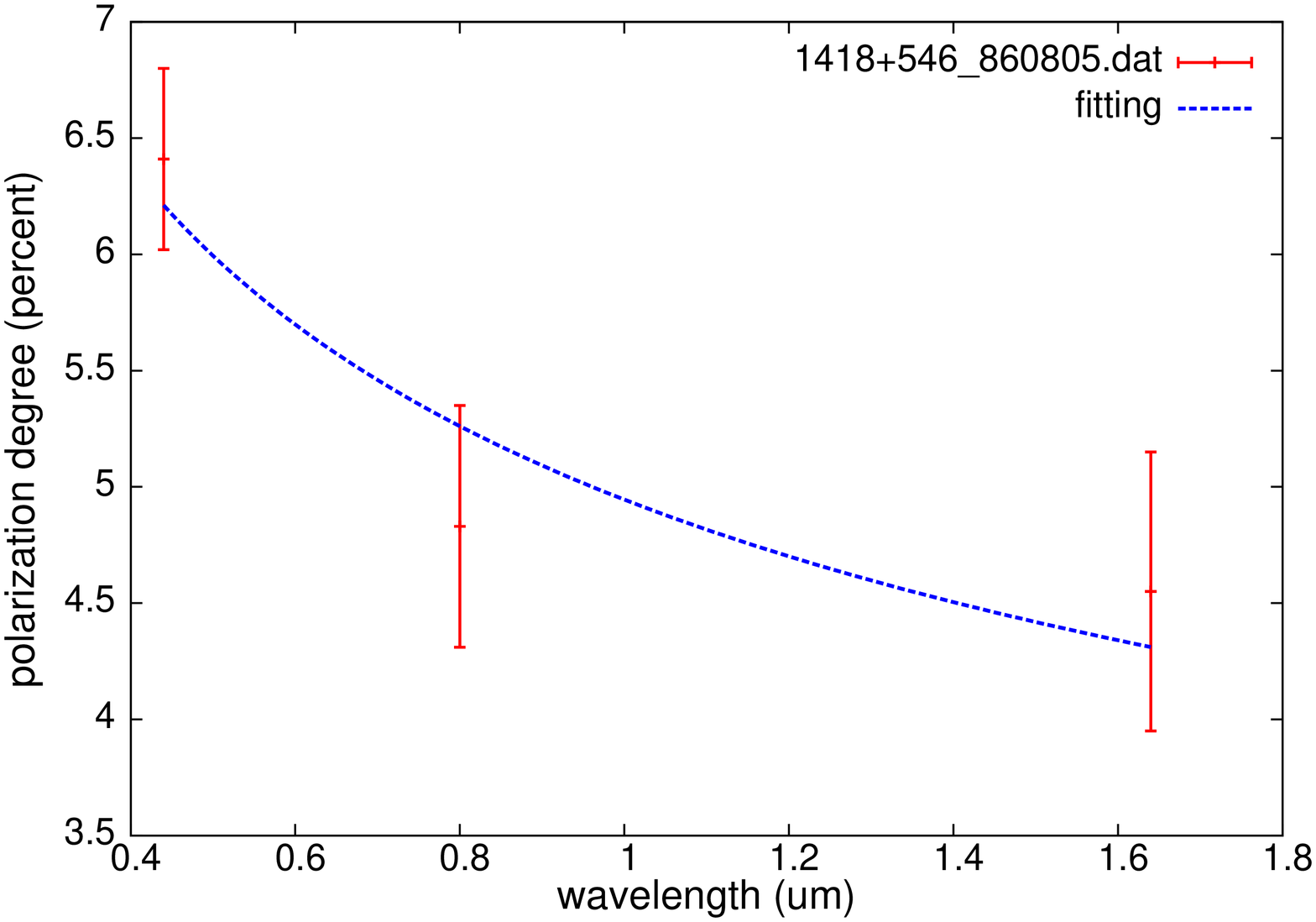}
 \includegraphics[scale=0.2,angle=-90]{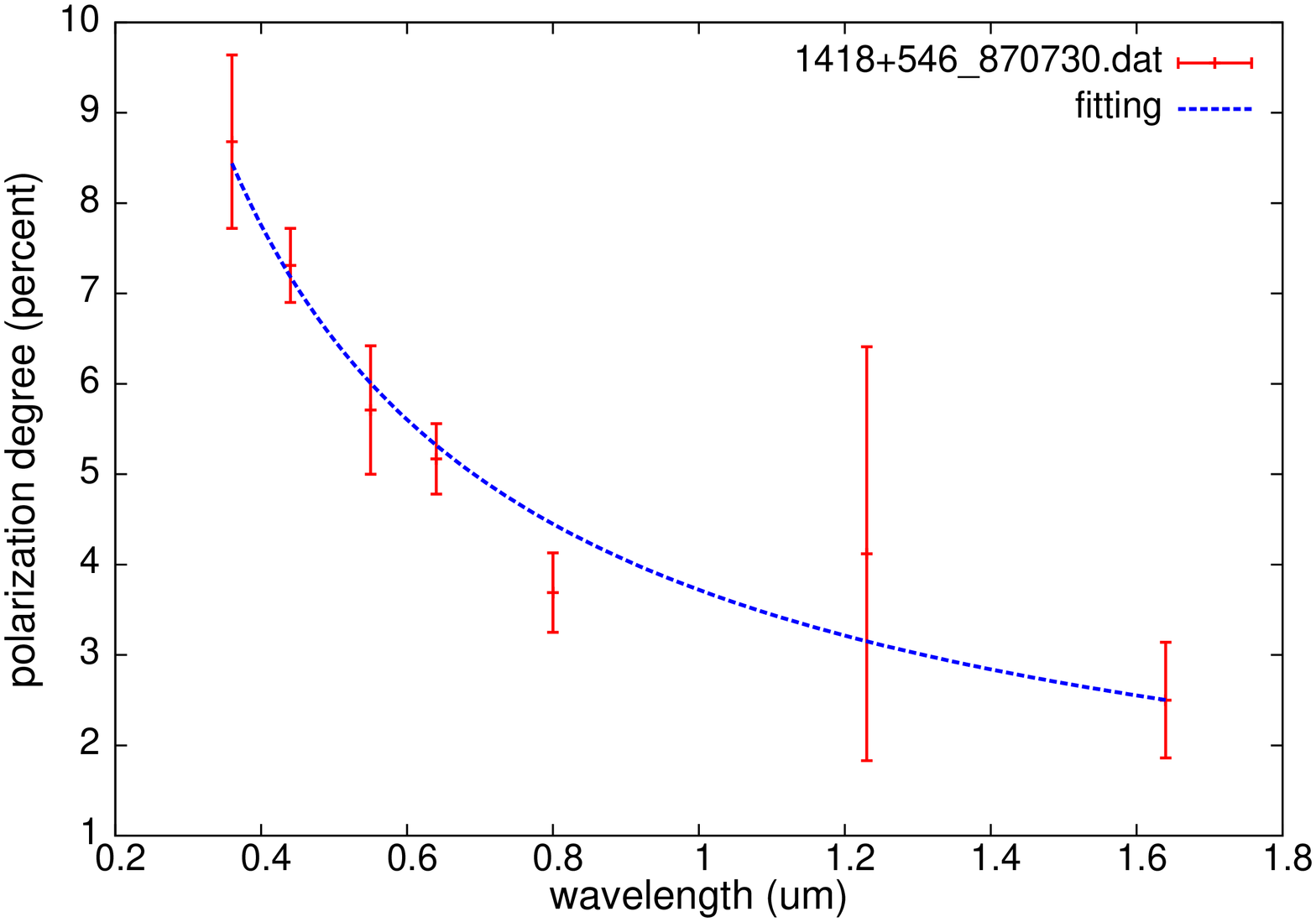} \\
 \includegraphics[scale=0.2,angle=-90]{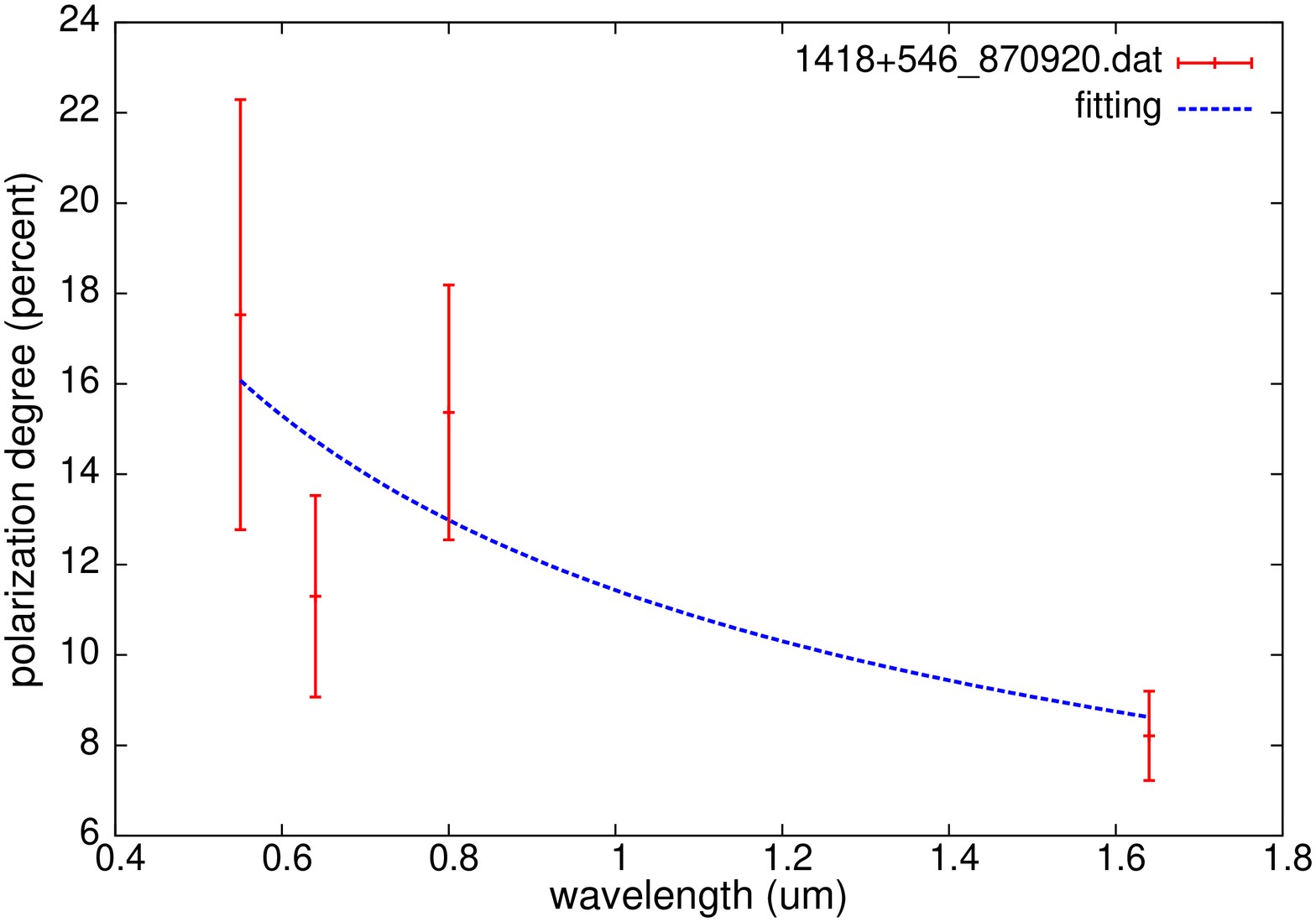}
 \includegraphics[scale=0.2,angle=-90]{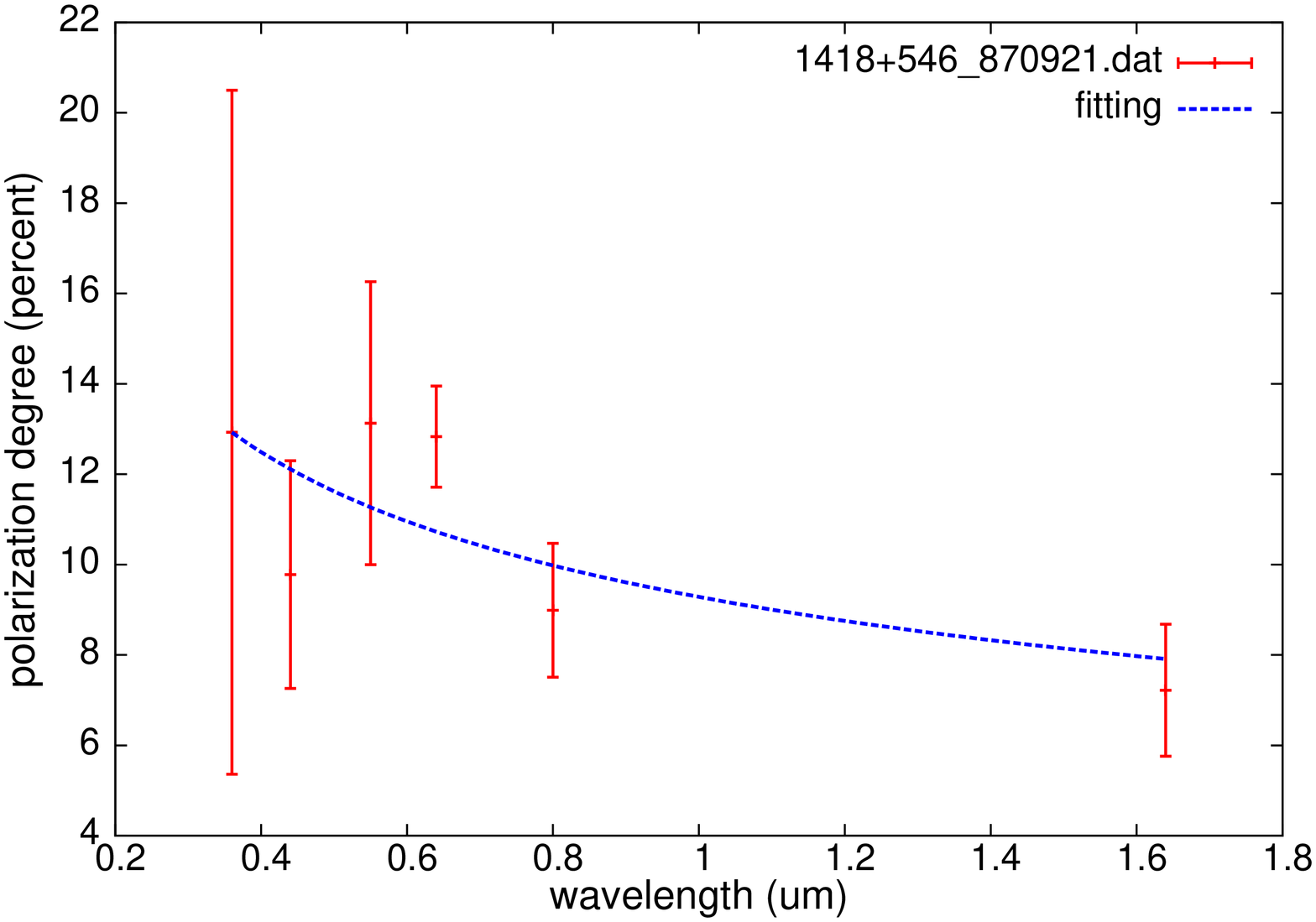}
  \caption{Depolarization fitting for 1418+546 (OQ 530). The first panel with the data observed on 1986 August 4 presents the result of $b=0.38\pm0.01$ with
$\chi^2/{\rm d.o.f}=8.0\times 10^{-4}$. The second panel with the data observed on 1986 August 5 presents the result of $b=0.28\pm0.11$ with $\chi^2/{d.o.f}=0.28$.
The third panel with the data observed on 1987 July 30 presents the result of $b=0.80\pm0.10$ with $\chi^2/{\rm d.o.f}=0.34$.
The fourth panel with the data observed on 1987 September 20 presents the result of $b=0.57\pm0.38$ with $\chi^2/{\rm d.o.f}=9.92$.
The fifth panel with the data observed on 1987 September 21 presents the result of the fitting $b=0.32\pm0.17$ with $\chi^2/{\rm d.o.f}=3.70$.}
  \label{1418}
\end{figure*}

\begin{figure*}
  \includegraphics[scale=0.2,angle=-90]{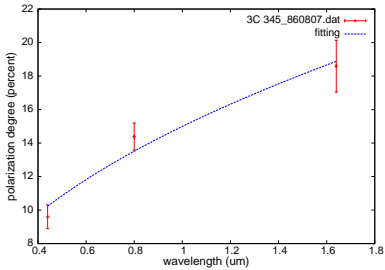}
  \caption{Depolarization fitting for 3C 345 (1641+399). The panel with the data observed on 1986 August 7 presents the result of $b=-0.47\pm0.09$ with
$\chi^2/{\rm d.o.f}=1.25$.}
  \label{1641}
\end{figure*}

\begin{figure*}
  \includegraphics[scale=0.2,angle=-90]{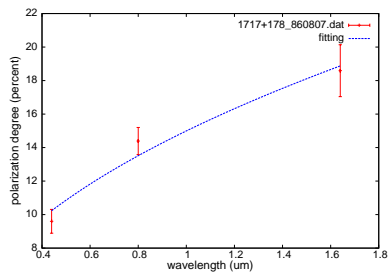}
  \caption{Depolarization fitting for 1717+178 (OT 129). The panel with the data data observed on 1986 August 7 presents the result of $b=-0.47\pm0.09$ with
$\chi^2/{\rm d.o.f}=1.25$.}
  \label{1717}
\end{figure*}

\begin{figure*}
 \includegraphics[scale=0.2,angle=-90]{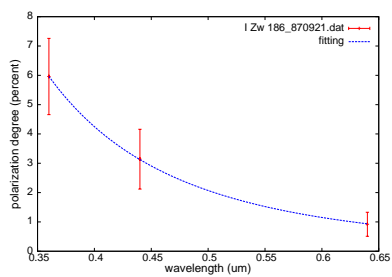}
  \caption{Depolarization fitting of I Zw 186 (1727+502). The panel with the data observed on 1987 September 21 presents the result of $b=3.22\pm0.03$ with
$\chi^2/{\rm d.o.f}=0.0005$.}
  \label{1727}
\end{figure*}

\begin{figure*}
 \includegraphics[scale=0.2,angle=-90]{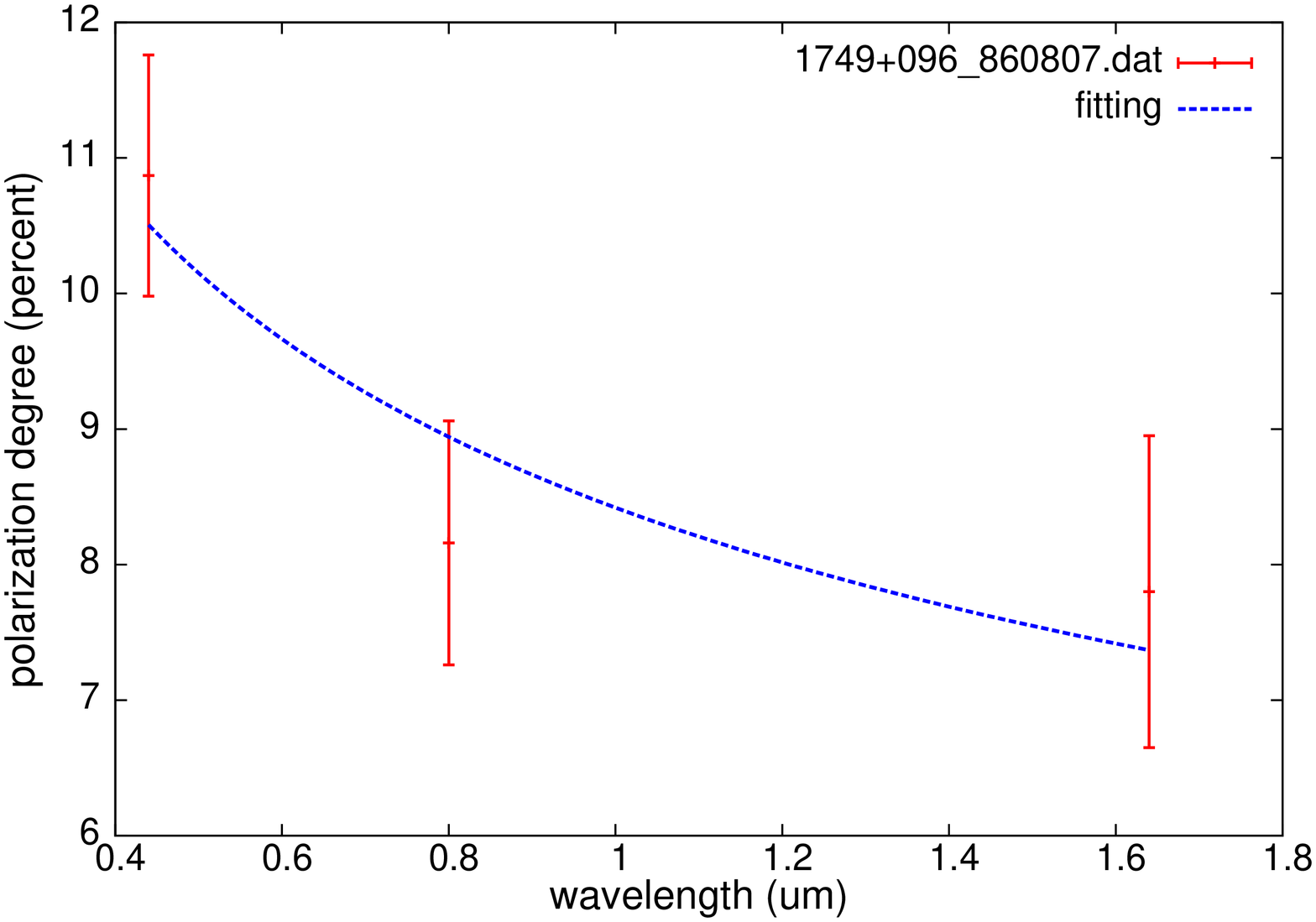}
 \includegraphics[scale=0.2,angle=-90]{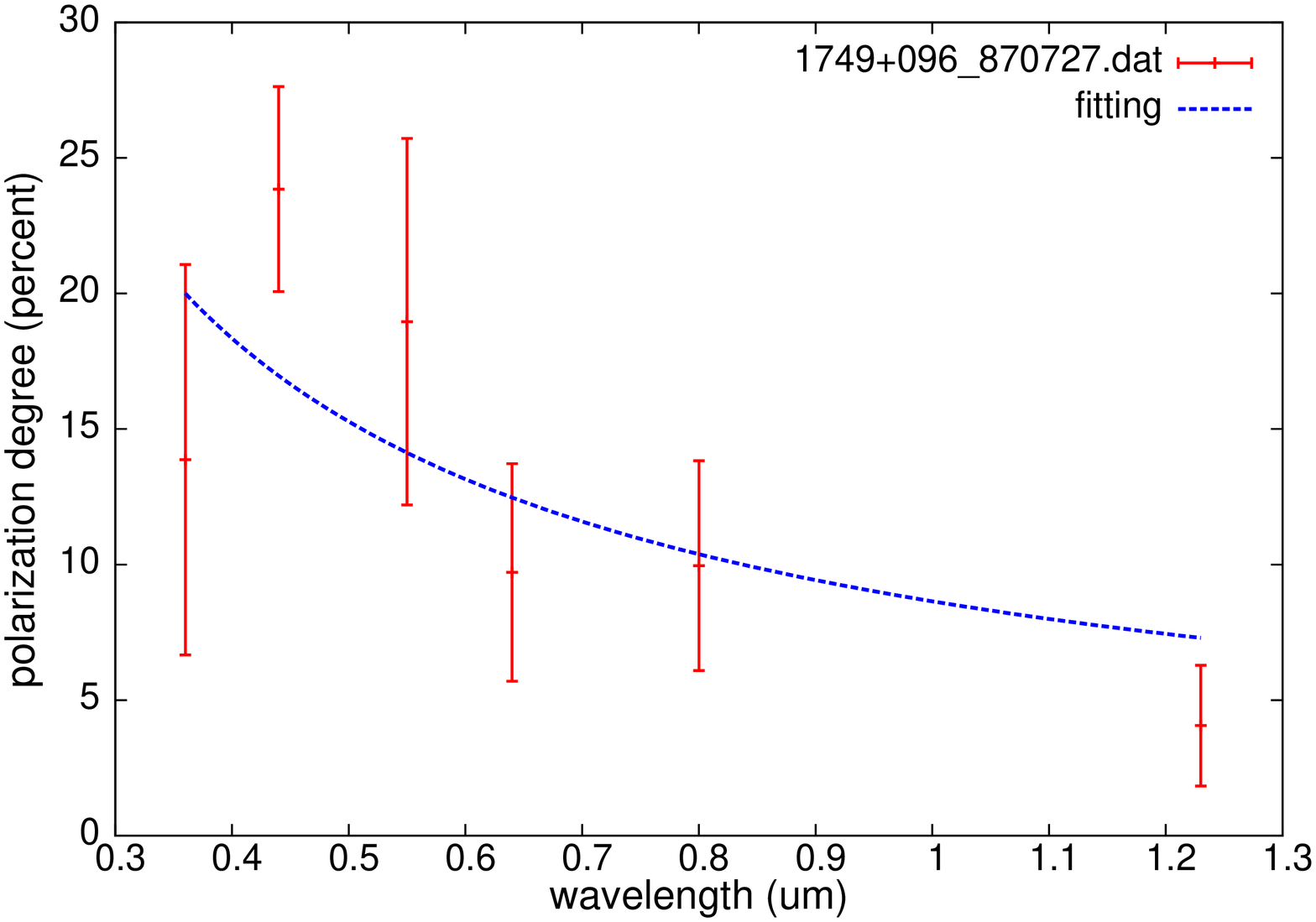}
  \caption{Depolarization fitting for 1749+096 (OT 081). The first panel with the data observed on 1986 August 7 presents the result of $b=0.27\pm0.12$ with
$\chi^2/{\rm d.o.f}=0.93$. The second panel with the data observed on 1987 July 27 presents the result of $b=0.82\pm0.50$ with $\chi^2/{\rm d.o.f}=31.67$.}
  \label{1749}
\end{figure*}

\begin{figure*}
  \includegraphics[scale=0.2,angle=-90]{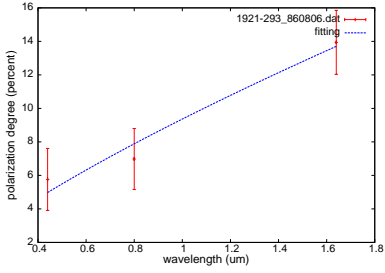}
  \caption{Depolarization fitting of 1921$-$293 (OV 236). The panel with the data observed on 1986 August 6 presents the result of $b=-0.77\pm0.17$ with
$\chi^2/{\rm d.o.f}=1.49$.}
  \label{1921}
\end{figure*}

\begin{figure*}
  \includegraphics[scale=0.2,angle=-90]{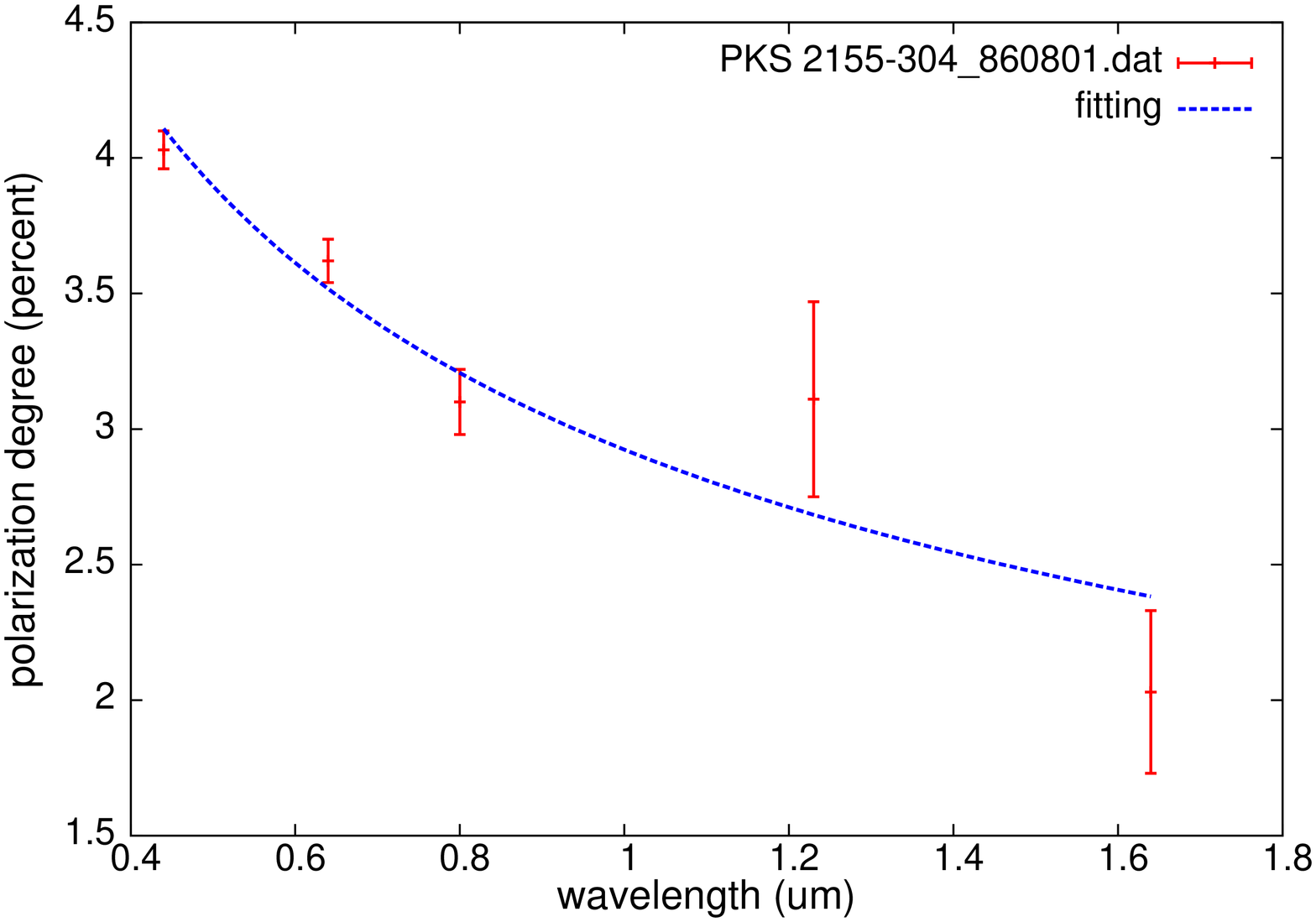}
  \includegraphics[scale=0.2,angle=-90]{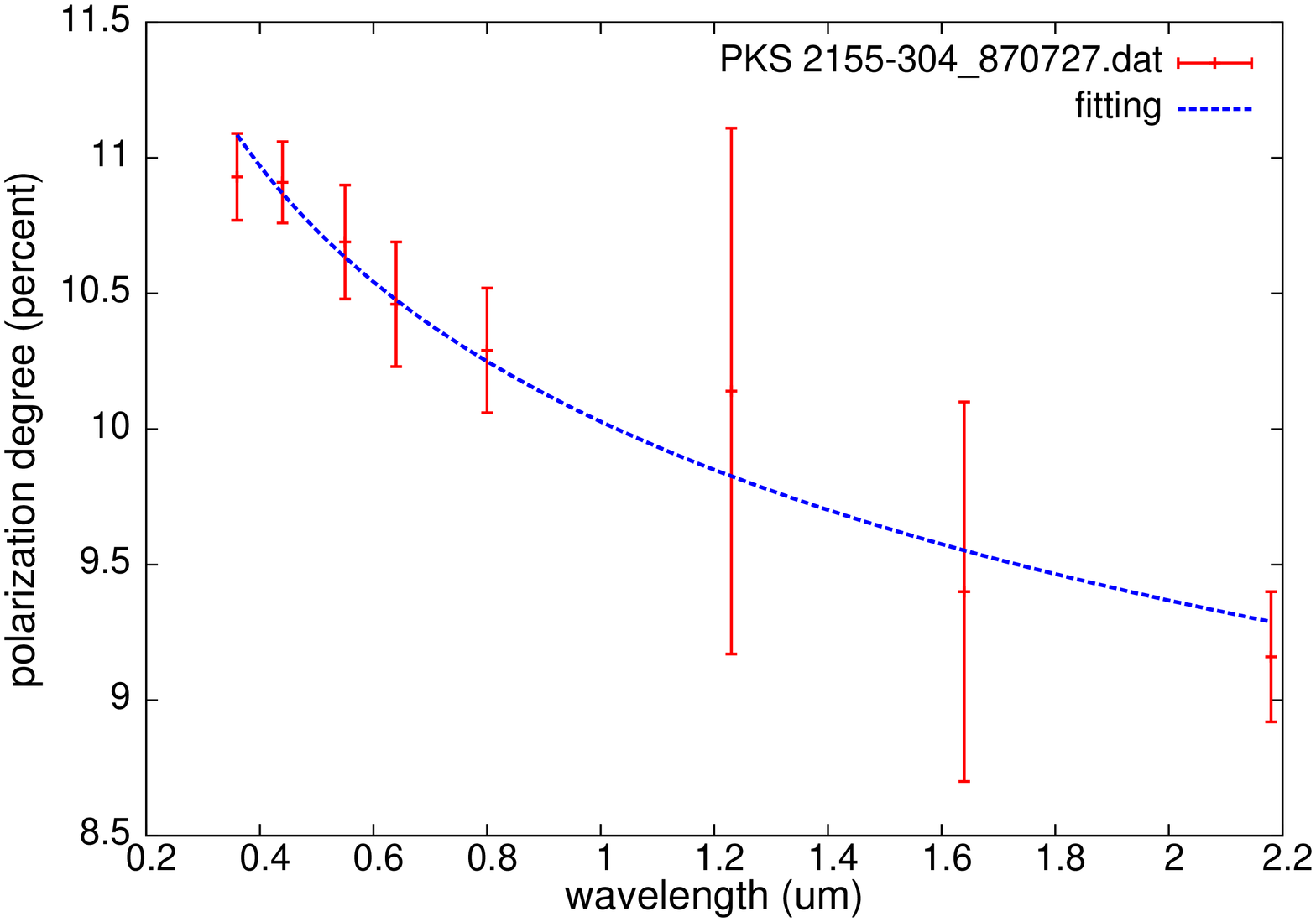}
  \caption{Depolarization fitting for PKS 2155$-$304.The first panel with the data observed on 1986 August 7 presents the result of $b=0.41\pm0.10$ with
$\chi^2/{\rm d.o.f}=0.11$. The second panel with the data observed on 1987 July 27 presents the result of $b=0.10\pm0.01$ with $\chi^2/{\rm d.o.f}=0.03$.}
  \label{2155}
\end{figure*}

\begin{figure*}
 \includegraphics[scale=0.2,angle=-90]{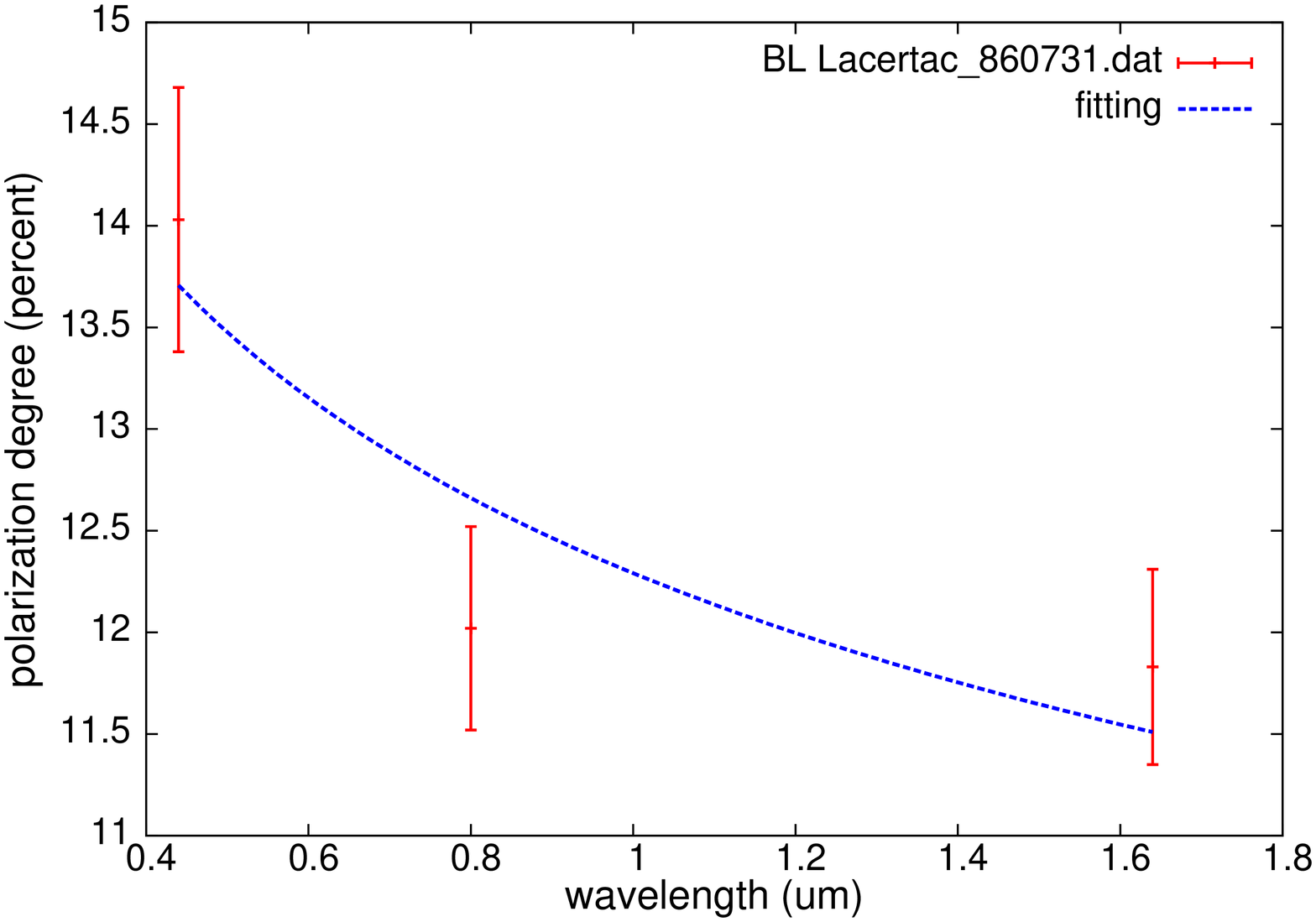}
 \includegraphics[scale=0.2,angle=-90]{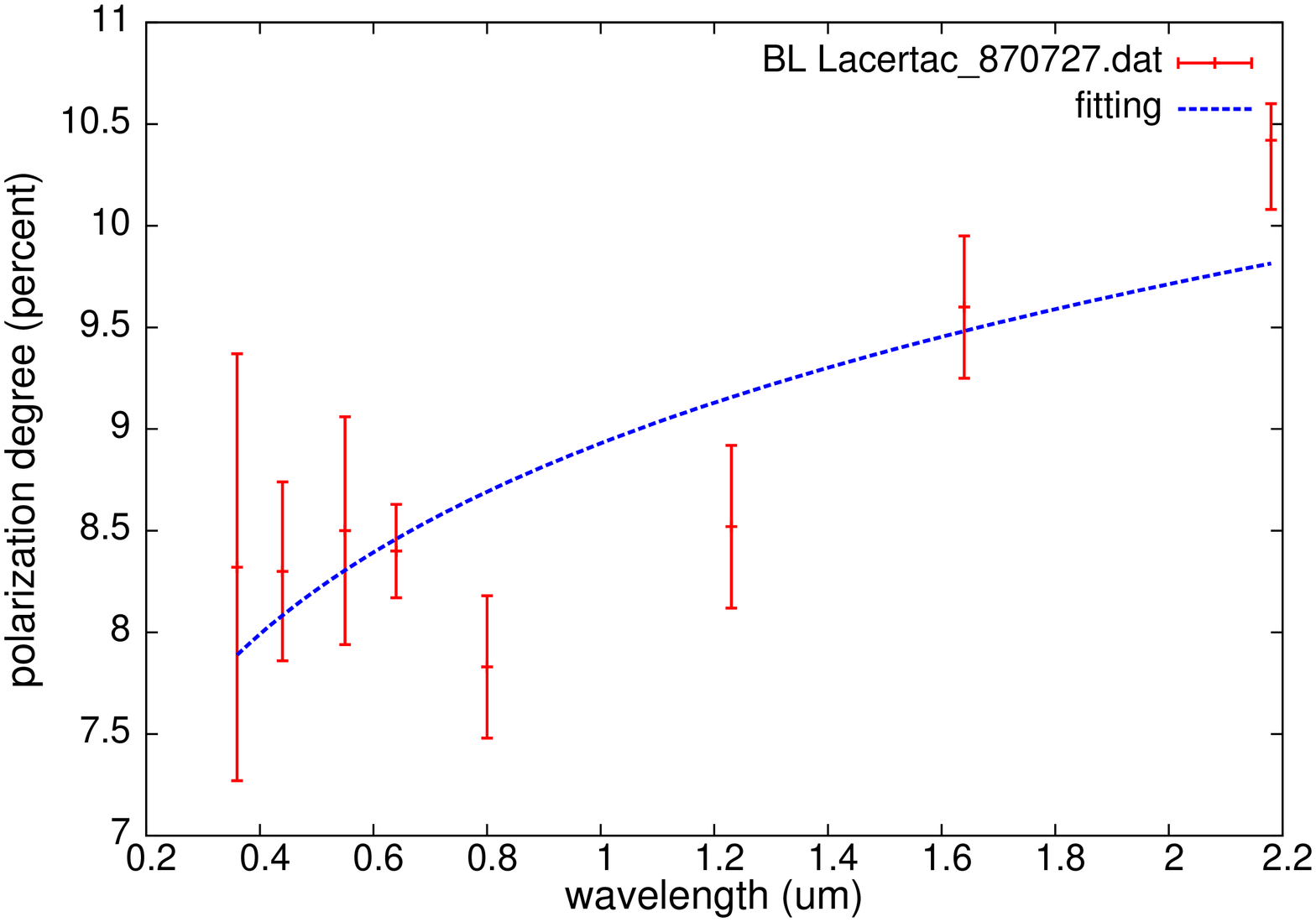}
 \includegraphics[scale=0.2,angle=-90]{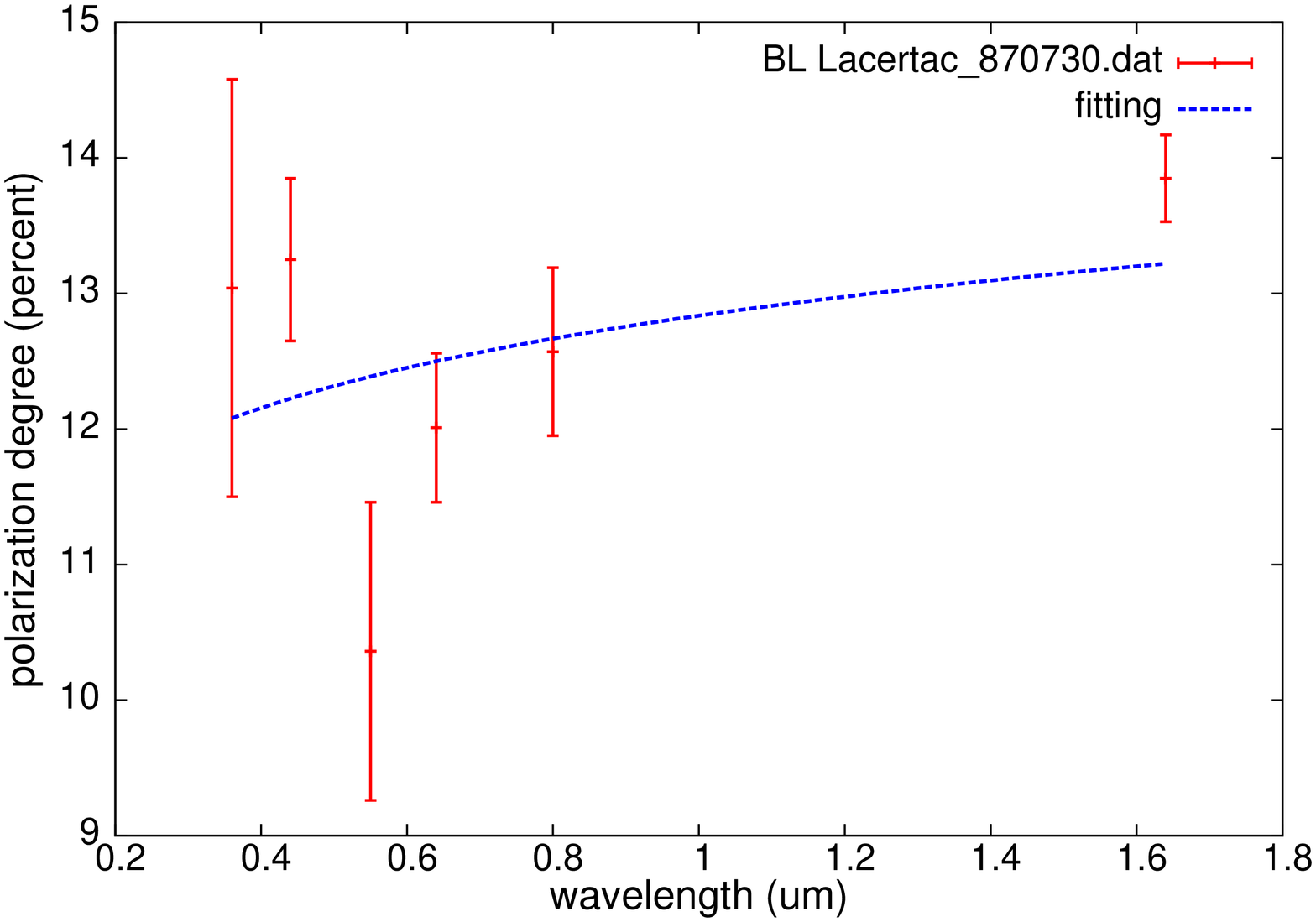}
  \caption{Depolarization fitting for BL Lacertae. The first panel with the data observed on 1986 July 31 presents the result of $b=0.13\pm0.07$ with
$\chi^2/{d.o.f}=0.62$. The second panel with the data observed on 1987 July 27 presents the result of $b=-0.12\pm0.04$, $\chi^2/{\rm d.o.f}=0.30$.
The third panel with the data observed on 1987 July 30 presents the result of $b=-0.06\pm0.08$ with $\chi^2/{\rm d.o.f}=1.68$.}
  \label{2200}
\end{figure*}

\begin{figure*}
  \includegraphics[scale=0.2,angle=-90]{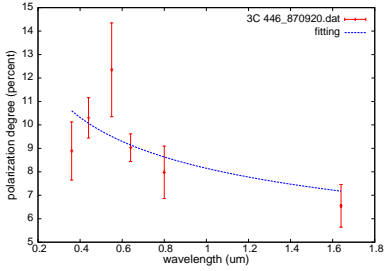}
  \caption{Depolarization fitting of 3C 446 (2223$-$052). The panel with the data observed on 1987 September 20 presents the result of $b=0.26\pm0.17$ with
$\chi^2/{\rm d.o.f}=2.98$.}
  \label{2223}
\end{figure*}

\begin{figure*}
 \includegraphics[scale=0.2,angle=-90]{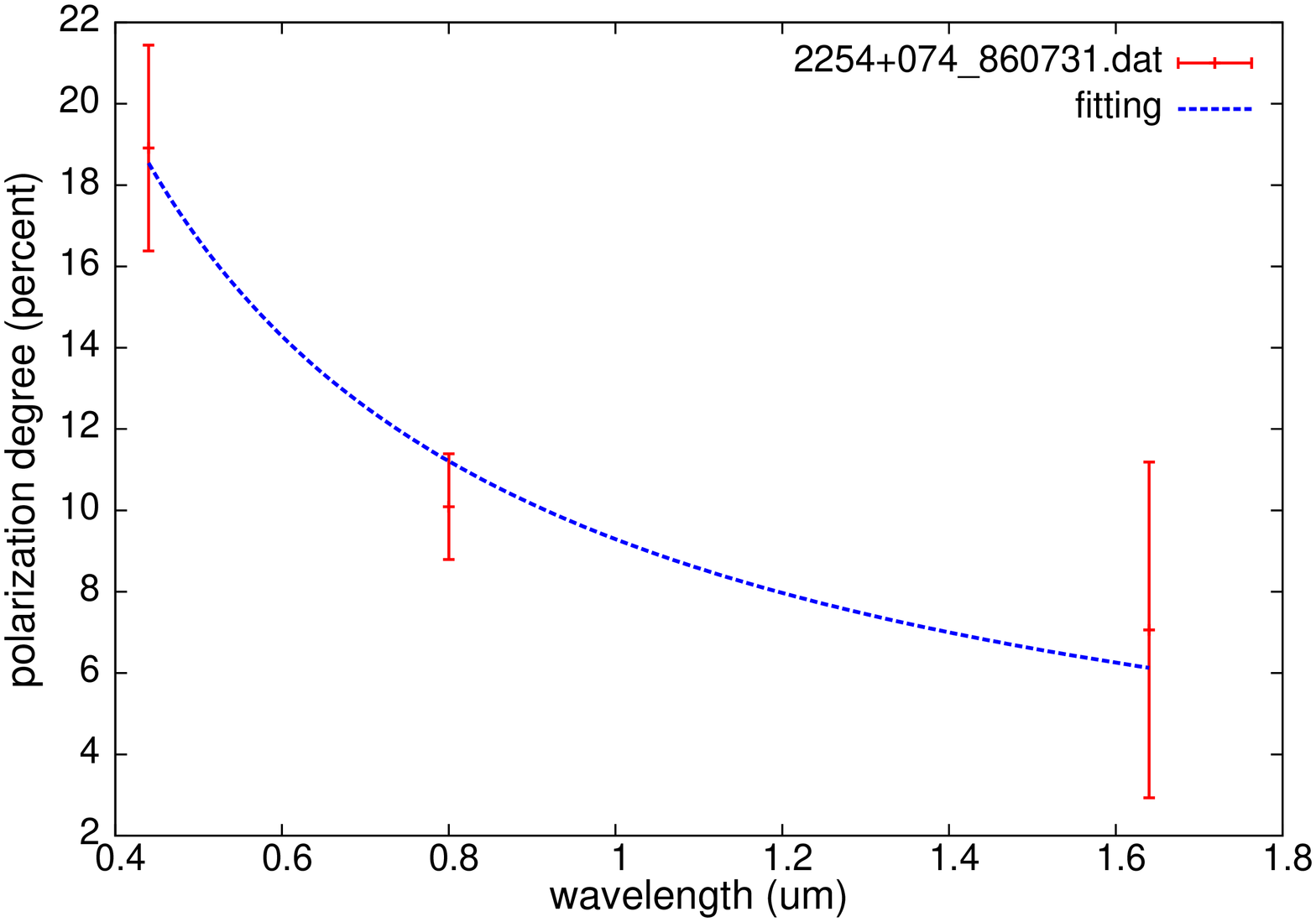}
 \includegraphics[scale=0.2,angle=-90]{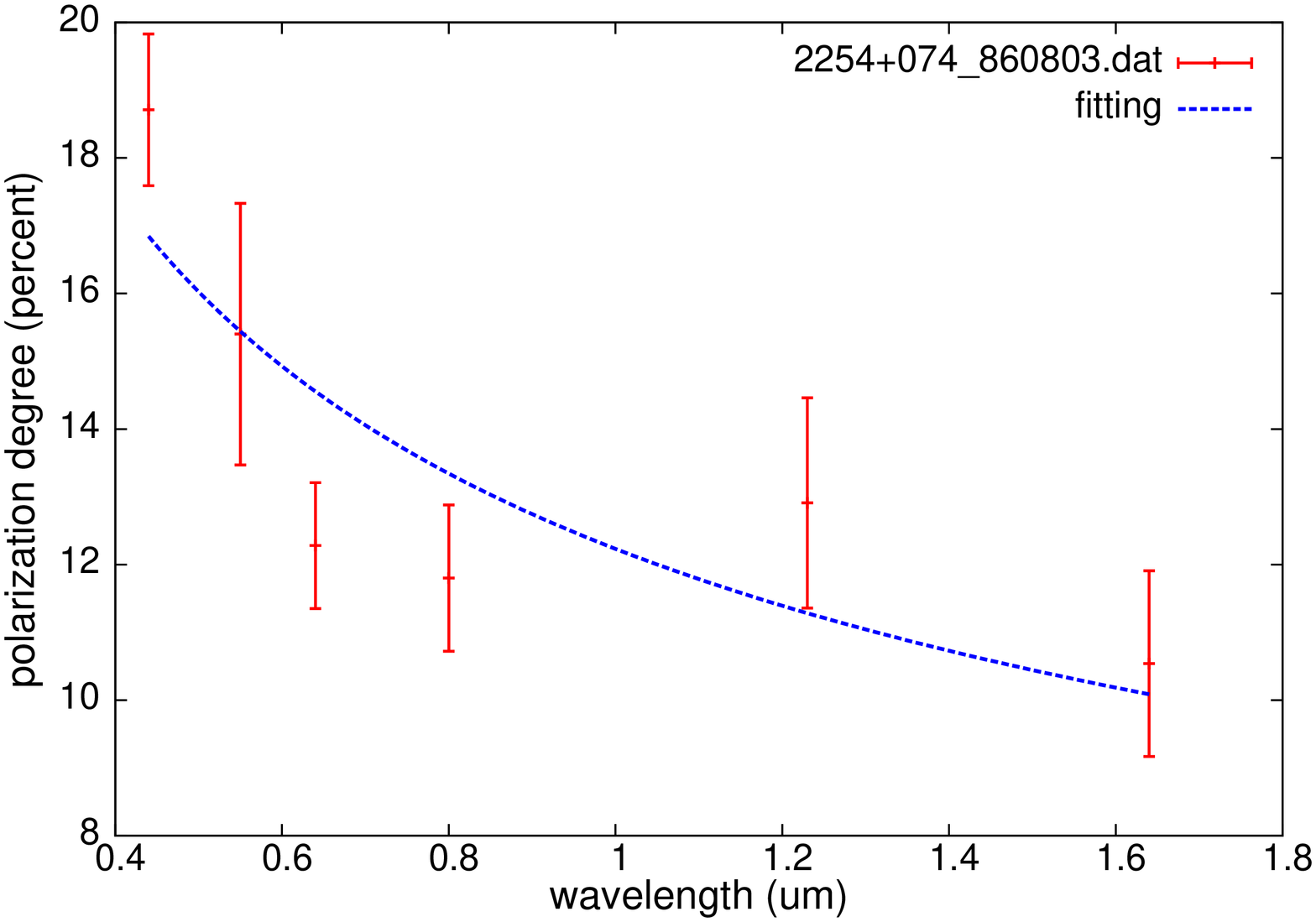}
 \includegraphics[scale=0.2,angle=-90]{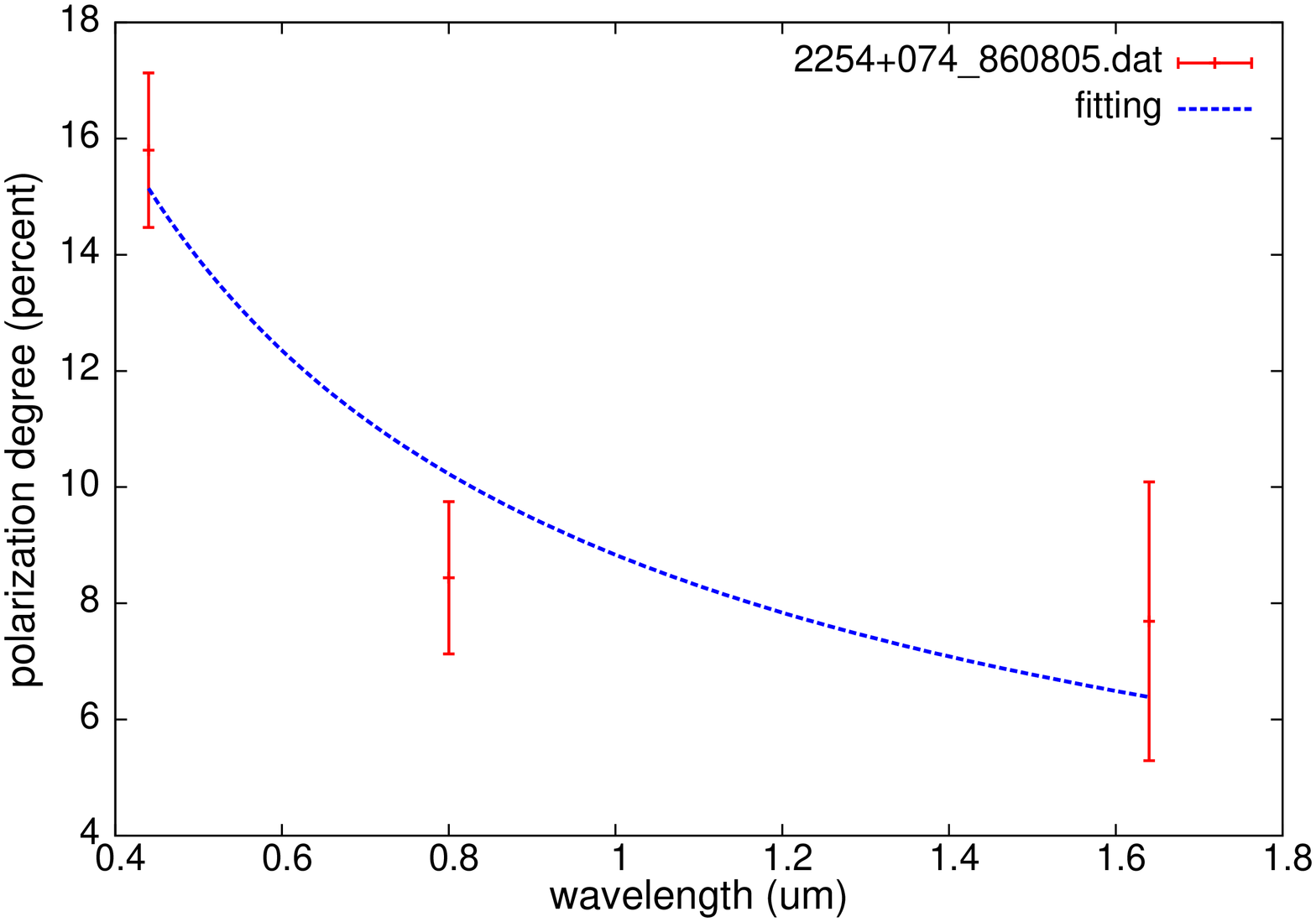} \\
 \includegraphics[scale=0.2,angle=-90]{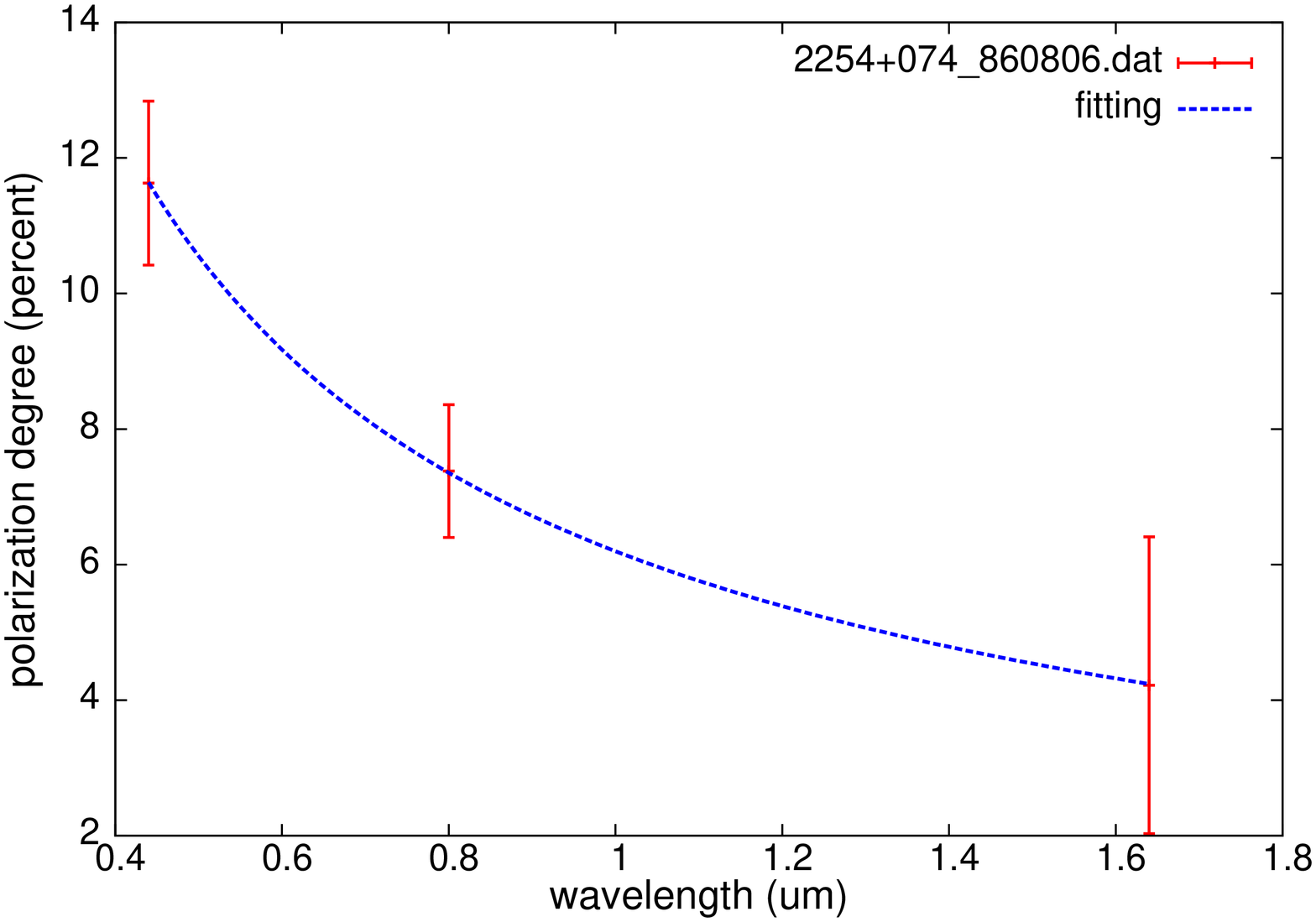}
 \includegraphics[scale=0.2,angle=-90]{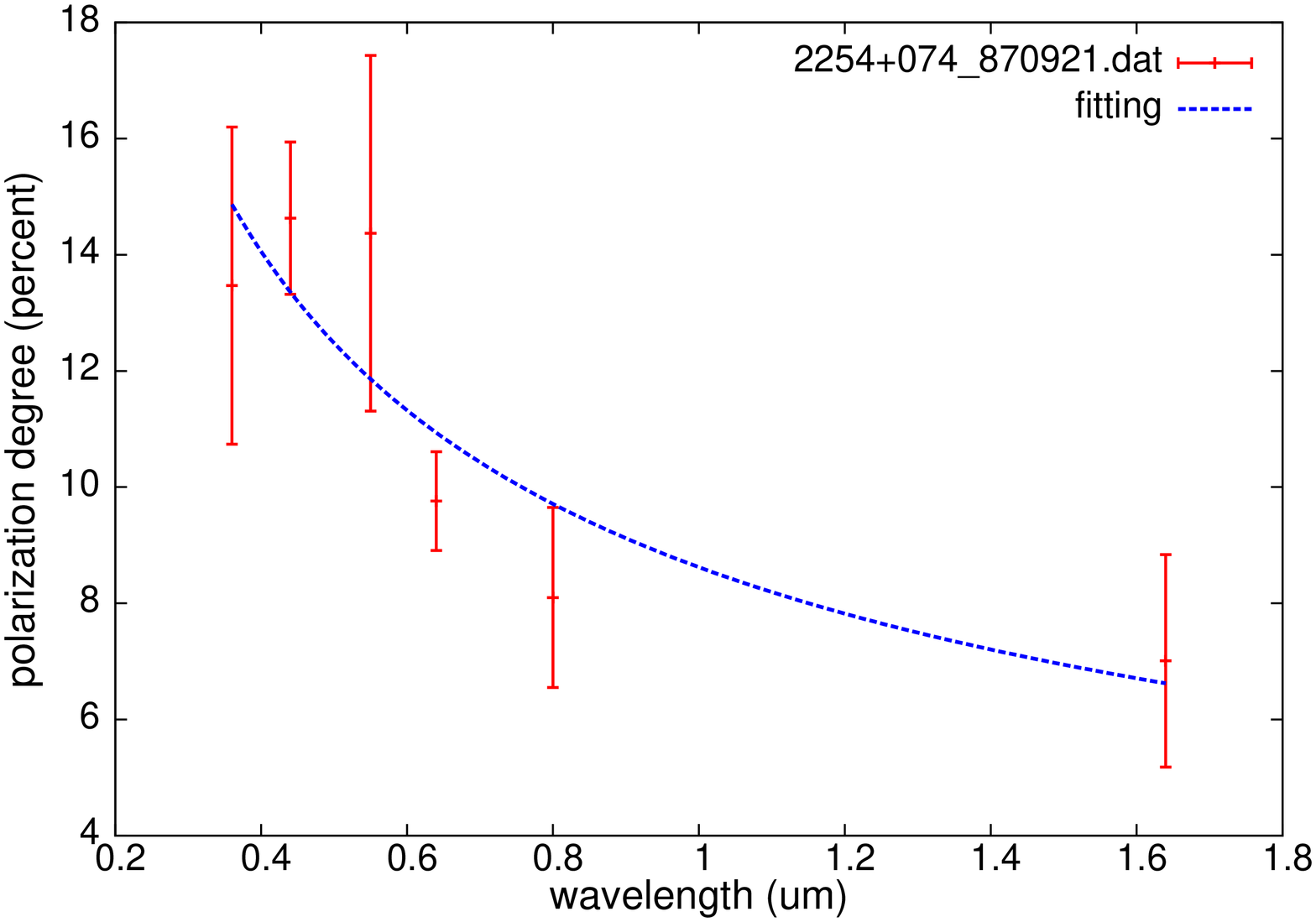}
  \caption{Depolarization fitting of 2254+074 (OY 091). The first panel with the data observed on 1986 July 31 presents the result of $b=0.84\pm0.17$ with
$\chi^2/{\rm d.o.f}=2.26$. The second panel with the data observed on 1986 August 3 presents the result of $b=0.39\pm0.13$ with $\chi^2/{\rm d.o.f}=3.47$.
The third panel with the data observed on 1986 August 5 presents the result of $b=0.66\pm0.27$ with $\chi^2/{\rm d.o.f}=5.33$.
The fourth panel with the data observed on 1986 August 6 presents the result of $b=0.77\pm0.01$ with $\chi^2/{\rm d.o.f}=1.1\times 10^{-3}$.
The fifth panel with the data observed on 1987 September 21 presents the result of $b=0.53\pm0.17$ with $\chi^2/{\rm d.o.f}=3.50$.}
  \label{2254}
\end{figure*}
\clearpage

\begin{table}
\scriptsize 
\caption{\scriptsize{The summary of the fitting results with the FDP feature in the sample of Mead et al. (1990). 
}}
\begin{tabular}{llccccccc}
    \hline
    \hline
    Name & Type & Redshift& Obs. Times of & Obs. Times of & $b$ &$m_{R}$& $m_T$& Remark \\
         &      &         & $dp/d\lambda<0$&$dp/d\lambda>0$&  &       &     &   \\
    \hline
     \multirow{5}*{0048-097}&
     \multirow{5}*{BL-Lac}&
     \multirow{5}*{0.634}&
     \multirow{5}*{5}&
     \multirow{5}*{0}&
     $0.15\pm0.05$ &0.70&0.85&\multirow{5}*{ }\\
     &&&& & $0.17\pm0.04$&0.66&0.83& \\
     &&&& & $0.17\pm0.03$&0.66&0.83& \\
     &&&& & $0.29\pm0.07$&0.42&0.71& \\
     &&&& & $0.26\pm0.02$&0.48&0.74& \\
    \hline
     \multirow{3}*{GC 0109+224}&
     \multirow{3}*{BL-Lac}&
     \multirow{3}*{0.265}&
     \multirow{3}*{3}&
     \multirow{3}*{1}&
     $-0.40\pm0.46^*$ &-&-&\multirow{3}*{high-energy flare\footnotemark[1]} \\
     &&&&  & $1.03\pm0.11$&-&-& \\
     &&&&  & $0.13\pm0.06$&0.74&0.87& \\
    \hline
     \multirow{3}*{PKS 0118-272}&
     \multirow{3}*{BL-Lac}&
     \multirow{3}*{0.559}&
     \multirow{3}*{3}&
     \multirow{3}*{0}&
     $0.19\pm0.04$&0.62&0.81& \\
     &&&&  & $0.11\pm0.02$&0.78&0.89& \\
     &&&&  & $0.09\pm0.05$&0.82&0.91& \\
    \hline
     \multirow{3}*{0138-097}&
     \multirow{3}*{BL-Lac}&
     \multirow{3}*{0.733}&
     \multirow{3}*{2}&
     \multirow{3}*{1}&
     $-0.68\pm0.04^\dag$&-&-& \\
     &&&&  & $0.17\pm0.03$&0.66&0.83& \\
     &&&&  & $0.17\pm0.06$&0.66&0.83& \\
    \hline
     3C 66A & BL-Lac & 0.700& 1 & 0 & $0.19\pm0.01$ &0.62&0.81&$\gamma$-ray flare \footnotemark[2] \\
    \hline
     \multirow{2}*{0754+100}&
     \multirow{2}*{BL-Lac}&
     \multirow{2}*{0.266}&
     \multirow{2}*{2}&
     \multirow{2}*{0}&
     $0.24\pm0.03$&0.52&0.76&\multirow{2}*{NIR flare\footnotemark[3]} \\
     &&&&  & $0.14\pm0.01$&0.72&0.86 \\
    \hline
     \multirow{3}*{0818-128}&
     \multirow{3}*{BL-Lac}&
     \multirow{3}*{0.074}&
     \multirow{3}*{3}&
     \multirow{3}*{0}&
     $0.20\pm0.05$&0.60&0.80& \\
     &&&&  & $0.29\pm0.04$&0.42&0.71& \\
     &&&&  & $0.25\pm0.07$&0.50&0.74& \\
    \hline
     \multirow{2}*{OJ 287}&
     \multirow{2}*{BL-Lac}&
     \multirow{2}*{0.306}&
     \multirow{2}*{2}&
     \multirow{2}*{0}&
     $0.12\pm0.05$&0.76&0.88&\multirow{2}*{high-energy flare\footnotemark[1]} \\
     &&&&  & $0.05\pm0.03$&0.90&0.95& \\
    \hline
     1147+245 & BL-Lac & 0.200& 1 & 0 & $0.74\pm0.10$&-&0.26&\\
    \hline
     \multirow{6}*{3C 279}&
     \multirow{6}*{FSRQ}&
     \multirow{6}*{0.536}&
     \multirow{6}*{6}&
     \multirow{6}*{0}&
     $0.07 \pm 0.01$&0.86&0.93& \multirow{5}*{$\gamma$-ray flare\footnotemark[4]} \\
     &&&& & $0.14 \pm 0.01$&0.72&0.86&\multirow{5}*{$\gamma$-ray Outbursts\footnotemark[5]} \\
     &&&& & $0.12\pm 0.01$&0.76&0.88& \\
     &&&& & $0.13 \pm 0.01$&0.74&0.87& \\
     &&&& & $0.17 \pm 0.02$&0.66&0.83& \\
     &&&& & $0.01 \pm 0.01^*$&-&- \\
    \hline
     \multirow{5}*{1418+546}&
     \multirow{5}*{BL-Lac}&
     \multirow{5}*{0.153}&
     \multirow{5}*{5}&
     \multirow{5}*{0}&
     $0.38 \pm 0.01$&0.24&0.62&\multirow{5}*{high-energy flare\footnotemark[1]} \\
     &&&& & $0.28 \pm 0.11$&0.44&0.72& \\
     &&&& & $0.80 \pm 0.10$&-&0.20& \\
     &&&& & $0.57 \pm 0.37$&-&0.43& \\
     &&&& & $0.32 \pm 0.17$&0.36&0.68 \\
    \hline
     3C 345 & FSRQ & 0.593& 0 & 1 & $-0.47\pm 0.09^\dag$&-&-&high-energy flare\footnotemark[1] \\
    \hline
     1717+178 & BL-Lac & 0.137& 0 & 1 & $-0.47\pm 0.09^\dag$&-&-&high-energy flare\footnotemark[1] \\
    \hline
     I Zw 186&BL-Lac&-&2&0& $3.22 \pm 0.03$&-&-&{$\gamma$-ray flare\footnotemark[6]}\\
    \hline
     \multirow{2}*{1749 + 096}&
     \multirow{2}*{BL-Lac}&
     \multirow{2}*{0.322}&
     \multirow{2}*{2}&
     \multirow{2}*{0}&
     $0.27 \pm 0.12$&0.46&0.73&\multirow{2}*{high-energy flare\footnotemark[1]} \\
     &&&&  & $0.82 \pm0.50$&-&0.18& \\
    \hline
     1921-293 & FSRQ & 0.353& 0 & 1 & $-0.77\pm 0.17^\dag$&-&-&high-energy flare\footnotemark[1] \\
    \hline
     \multirow{2}*{PKS 2155 $-$ 304}&
     \multirow{2}*{BL-Lac}&
     \multirow{2}*{0.116}&
     \multirow{2}*{2}&
     \multirow{2}*{0}&
     $0.41 \pm 0.10$ &0.18&0.59&\multirow{2}*{$\gamma$-ray flare\footnotemark[7]}\\
     &&&&  & $0.10 \pm 0.01$&0.80&0.90 \\
    \hline
     \multirow{3}*{BL Lacertac}&
     \multirow{3}*{BL-Lac}&
     \multirow{3}*{0.069}&
     \multirow{3}*{1}&
     \multirow{3}*{2}&
     $0.13 \pm 0.07$&0.74&0.87&\multirow{3}*{TeV $\gamma$-ray flare\footnotemark[8]} \\
     &&&&  & $-0.12\pm 0.04^\dag$&-&-&\\
     &&&&  & $-0.06\pm 0.08^{\dag *}$&-&-&\\
    \hline
     3C 446 & FSRQ & 1.404& 1 & 0 & $0.26 \pm 0.17$&0.48&0.74 \\
    \hline
     \multirow{5}*{2254+074}&
     \multirow{5}*{BL-Lac}&
     \multirow{5}*{0.190}&
     \multirow{5}*{5}&
     \multirow{5}*{0}&
     $0.84 \pm 0.17$&-&0.16& \\
     &&&& & $0.39 \pm 0.13$&0.22&0.61 \\
     &&&& & $0.66 \pm 0.27$&-&0.34& \\
     &&&& & $0.77 \pm 0.01$&-&0.23& \\
     &&&& & $0.53 \pm 0.17$&-&0.47& \\
    \hline
\end{tabular}
\scriptsize
Notes. $m_R$ indicates the MHD turbulent index
derived in the regular magnetic field dominated case, and $m_T$ indicates the MHD turbulent index derived in the turbulent magnetic field dominated case.\\
 (*){the fitting results with large error-bar}.\\
 (\dag){the notes for the sources with $dp/d\lambda>0$}.\\
 References. (1){\cite{2016arXiv161203165A}}
 (2){\cite{2011ApJ...726...43A, 2009ApJ...693L.104A, 2009arXiv0907.5175R}}
 (3){\cite{2010ATel.2516....1C}}
 (4){\cite{2015ApJ...808L..48P, 2016ApJ...817...61P}}
 (5){\cite{2015ASInC..12..113P}}
 (6){\cite{2015ApJ...808..110A}}
 (7){\cite{2014ATel.6165....1S, 2014ATel.6148....1C, 2013ATel.5248....1K, 2009A&A...502..749A}}
 (8){\cite{1997ApJ...490L.145B,2013ApJ...762...92A}}
 \label{value of b}
\end{table}
\end{document}